\documentclass[aps,prd,reprint,twocolumn,preprintnumbers,floatfix,nofootinbib]{revtex4-1}

\usepackage[utf8]{inputenc}
\usepackage[dvips]{graphicx}
\usepackage[dvipsnames]{xcolor}
\usepackage{color}
\usepackage{relsize}
\usepackage{graphics}
\usepackage{epstopdf}
\usepackage{hyperref}
\usepackage{mathrsfs}
\usepackage{amssymb}
\usepackage{physics}
\usepackage{booktabs}

\usepackage[normalem]{ ulem }
\usepackage{amsthm}
\usepackage{amsmath}
\usepackage{cancel}
\usepackage{fontawesome} 
\usepackage[caption=false]{subfig}
\usepackage{tabularx,ragged2e} 
 \newcolumntype{L}{>{\RaggedRight\arraybackslash}X}

\usepackage{hyperref}

\newcommand{\be}{\begin{equation}}
\newcommand{\ee}{\end{equation}}
\newcommand{\bea}{\begin{eqnarray}}
\newcommand{\eea}{\end{eqnarray}}
\newcommand{\ba}{\begin{aligned}}
\newcommand{\ea}{\end{aligned}}

\newcommand{\mDM}{m_{\rm DM}}

\newcommand{\TBH}{T_{\rm BH}}
\newcommand{\MBH}{M_{\rm BH}}
\newcommand{\TBHi}{T_{\rm BH}^{\rm in}}
\newcommand{\MBHi}{M_{\rm BH}^{\rm in}}
\newcommand{\MPL}{M_p}
\newcommand{\bh}{{\rm BH}}
\newcommand{\as}{a_\star}
\newcommand{\GeV}{\rm GeV}
\newcommand{\g}{\rm g}
\newcommand{\Oh}{\Omega h^2}

\newcommand\myshade{80}
\colorlet{mylinkcolor}{ForestGreen}
\colorlet{mycitecolor}{Red}
\colorlet{myurlcolor}{violet}

\hypersetup{
  linkcolor  = mylinkcolor!\myshade!black,
  citecolor  = mycitecolor!\myshade!black,
  urlcolor   = myurlcolor!\myshade!black,
  colorlinks = true
}

\usepackage{pgfplots}

\pgfplotsset{compat=1.17}

\begin{document}
\sloppy  

\preprint{FERMILAB-PUB-21-304-T, NUHEP-TH/21-06, CP3-21-41, IPPP/21/02}

\vspace*{1mm}

\title{Primordial Black Hole Evaporation and Dark Matter Production:\\
I. Solely Hawking radiation}

\author{Andrew Cheek$^{a}$}
\email{andrew.cheek@uclouvain.be}
\author{Lucien Heurtier$^{b}$}
\email{lucien.heurtier@durham.ac.uk}
\author{Yuber F. Perez-Gonzalez$^{c, d, e}$}
\email{yfperezg@northwestern.edu}
\author{Jessica Turner$^{b}$}
\email{jessica.turner@durham.ac.uk}

\affiliation{$^a$ Centre for Cosmology, Particle Physics and Phenomenology (CP3),
Universit\'e catholique de Louvain, Chemin du Cyclotron 2,
B-1348 Louvain-la-Neuve, Belgium}
\affiliation{$^b$ Institute for Particle Physics Phenomenology, Durham University, South Road, Durham DH1 3LF, U.K.}
\affiliation{$^c$ Theoretical Physics Department, Fermi National Accelerator Laboratory, P.O. Box 500,
Batavia, IL 60510, USA}
\affiliation{$^d$ Department of Physics \& Astronomy, Northwestern University, Evanston, IL 60208, USA}
\affiliation{$^e$ Colegio de Física Fundamental e Interdisciplinaria de las Américas (COFI), 254 Norzagaray street, San Juan, Puerto Rico 00901.}

\begin{abstract}
Hawking evaporation of black holes in the early Universe is expected to copiously produce all kinds of particles, regardless of their charges under the Standard Model  gauge group. For this reason, any fundamental particle, known or otherwise, could be produced during the black hole lifetime. This certainly includes dark matter (DM) particles. This paper improves upon previous calculations of DM production from primordial black holes (PBH) by  consistently including the greybody factors, and by  meticulously tracking a system of coupled Boltzmann equations. We show that the initial PBH densities required to produce the observed relic abundance depend strongly on the DM spin, varying in about $\sim 2$ orders of magnitude between a spin-2 and a scalar DM in the case of non-rotating PBHs. For Kerr PBHs, we have found that the expected enhancement in the production of bosons reduces the initial fraction needed to explain the measurements. We further consider indirect production of DM by assuming the existence of additional and unstable degrees of freedom emitted by the evaporation, which later decay into the DM. For a minimal setup where there is only one heavy particle, we find that the final relic abundance can be increased by at most a factor of $\sim 4$ for a scalar heavy state and a Schwarzschild PBH, or by a factor of $\sim 4.3$ for a spin-2 particle in the case of a Kerr PBH. \href{https://github.com/earlyuniverse/ulysses}{\faGithub}
\end{abstract}

\maketitle

\section{Introduction}

The entire catalogue of experimental evidence for dark matter (DM) comes only from its gravitational effects. Despite this, the particle physics community pins many of its hopes on discovering a DM candidate that has additional interactions with the Standard Model (SM). The three main reasons for this are simple: many well-motivated extensions of the SM include DM candidates with such interactions; there are plausible mechanisms that require interactions to provide the correct DM abundance, and importantly, many such mechanisms are testable by experiments. However, the possibility remains that DM only interacts with the SM gravitationally. If this were the case, the production of DM in the early Universe still requires an explanation. One such explanation is the focus of this paper, namely  that some population of {\em primordial black holes} (PBHs) were abundant and energetic enough to evaporate and produce the relic dark matter we observe today. Notably, such a scenario relies upon particle production via Hawking radiation~\cite{Hawking:1974rv,Hawking:1974sw}, a phenomenon that does not rely on the existence of additional and unobserved interactions. Instead, it arises due to the ambiguity of the definition of the vacuum state in curved spacetime. The disruption of the spacetime resulting from the collapse of some matter generates a thermal flux of particles. Crucially, a black hole (BH) will emanate \emph{all} existing degrees of freedom in nature, without regard to their interactions, and thus constitutes a compelling source of a purely gravitationally interacting DM.

One of the earliest probes of the Universe's history comes from the cosmic microwave background (CMB)~\cite{Ade:2015lrj,Ade:2015xua}. Perhaps the most profound lesson from the CMB is that the observable Universe is remarkably homogeneous. The current scientific consensus is that this is achieved by some early period of cosmic inflation, which also provides the seeds for small matter perturbations that eventually form galaxies. The true model of inflation is far from determined and many of which predict the existence of PBHs. This topic has surged in popularity recently because of the gravitational wave measurements of solar mass black hole binaries. It has been argued that PBHs themselves constitute DM, where their masses are constrained by a large and varied set of experimental searches~\cite{Carr:2016drx,Green:2020jor}. The minimum value for the PBH mass is set by the requirement that they have not evaporated already, determined by Hawking radiation, $M_{\textrm{PBH}}\geq 5\times 10^{14}\,\textrm{g}$~\cite{Carr:2020gox}. 

Even without the requirement that PBHs constitute DM, {\em Big Bang Nucleosynthesis} (BBN) provides serious restrictions on how many PBHs existed in the early Universe for masses $10^9\,\textrm{g}\,\leq M_{\textrm{PBH}}\leq 10^{14}\,\textrm{g}$~\cite{Carr:2009jm,Carr:2020gox,Keith:2020jww}, below which PBHs have evaporated before BBN. A lower limit on the PBH mass comes from constraints on inflation; the Hubble scale during inflation has an upper bound from CMB~\cite{Akrami:2018odb}, which in turn imposes the smallest possible mass to be $M_{\textrm{PBH}}\gtrsim 0.1~\g$~\cite{Carr:2020gox}. Let us stress that such a value is model dependent, specifically on the details of the gravitational collapse and on the features of inflation. One obtains such a minimal value for the PBH mass by assuming a standard slow-roll scenario. For simplicity, we assume such minimal case, and take $M_{\textrm{PBH}}\gtrsim 0.1~\g$~\cite{Carr:2020gox} as the lower limit. This window keeps alive the possibility that PBHs dominated the early Universe and played an important role in its evolution. The consequences of this have been well studied since the discovery of the Hawking radiation~\cite{Carr:1976zz}, and span many different and important aspects, for instance, the generation of Dark Radiation~\cite{Carr:2020xqk,Hooper:2019gtx,Lunardini:2019zob,Inomata:2020lmk,Masina:2020xhk,Masina:2021zpu,Domenech:2021wkk}, matter-antimatter asymmetry production~\cite{Baumann:2007yr,Fujita:2014hha,Hook:2014mla,Hamada:2016jnq,Chaudhuri:2020wjo,Hooper:2020otu,Perez-Gonzalez:2020vnz,Datta:2020bht,JyotiDas:2021shi}, and the implications for the production of DM through evaporation~\cite{Matsas:1998zm,Bell:1998jk,Green:1999yh,Arbey:2021ysg,Khlopov:2004tn,Allahverdi:2017sks,Fujita:2014hha,Lennon:2017tqq,Morrison:2018xla,Hooper:2019gtx,Masina:2020xhk,Gondolo:2020uqv,Bernal:2020kse,Bernal:2020bjf,Bernal:2020ili,Kitabayashi:2021hox,Masina:2021zpu}. Generally, DM particles produced in this way can be very light. However, if they are too light, such DM particles are expected to be relativistic and their free-streaming length will be constrained by observations regarding structure formation~\cite{Baldes:2020nuv, Auffinger:2020afu,Masina:2020xhk, Masina:2021zpu}. 

This is the first paper of a two-part series, where we return to the calculation of DM emission from PBH evaporation to improve existing treatments. We do so by ameliorating the analysis in two different aspects: solving, in detail, the momentum-averaged Boltzmann equations and including consistently the greybody factors, quantities essential for an accurate description of the Hawking evaporation. The code we use for this purpose has been made publicly available\footnote{\url{https://github.com/earlyuniverse/ulysses} \href{https://github.com/earlyuniverse/ulysses}{\faGithub}}. 
In addition, we also provide a semi-analytic solution that is consistent with our numerical analysis. Furthermore, we address the possibility of having \emph{baroque} Dark Sectors, consistent with a large number of degrees of freedom. Since PBH evaporation would produce significant quantities of particles belonging to such sector, one could imagine that, in the scenario, all but one particles are unstable, the generation of the stable DM would be enhanced by such indirect production. In this paper, we assume that this Dark Sector is disconnected from the SM, avoiding thermal production mechanisms such as Freeze-In (FI) or Freeze-Out (FO). In the companion paper~\cite{paperB}, we will consider the situation where there are interactions with the SM sector. We use the infrastructure of {\sc ULYSSES} \cite{Granelli:2020pim}, a publicly available python package that has been typically used to solve Boltzmann equations associated with leptogenesis, to solve the relevant Friedmann and  Boltzmann equations.

This paper is organized as follows. First, we describe  the emission properties of non-rotating (Schwarzschild) and rotating (Kerr) Black Holes in Sec.~\ref{sec:BH}. In each case, we consider the mass and angular momentum loss rate from the BH, the rate of particle emission, and, when possible, the total number of emitted particles. These characteristics will be crucial for the analysis in the subsequent section. Also, we consider the phase-space distribution of emitted particles, which will be helpful to address free-streaming constraints on DM. In Sec.~\ref{sec:evaporation}, we first establish the Friedmann and Boltzmann equations that we solve in the presence of evaporating PBHs. Then, we describe our results for the cases in which the PBHs --- both for Schwarzschild and Kerr --- are the only source of DM. We then focus on the next-to-minimal case which consists of a dark sector containing only DM together with one heavy metastable state. Finally, we make our concluding remarks in Sec.~\ref{sec:conclus}. We have included two  appendices: App.~\ref{ap:A} provides useful formulae related to the BH evaporation properties and derive some specific quantities used in the main text, and App.~\ref{ap:A1}, which contains the decay width of scalars, vectors and massive tensors into a fermion-antifermion pair. We use natural units where $\hbar = c = k_{\rm B} = 1$ throughout this manuscript. 

\section{Black Hole Evaporation}
\label{sec:BH}

Black holes were initially thought to be eternal and were expected to increase their mass by accreating additional matter or even other black holes. Nevertheless, when the BH quantum properties were inspected, it was shown that they also emit particles with a thermal spectrum related to BH surface gravity~\cite{Hawking:1974rv,Hawking:1974sw}, making the BHs lose mass and angular momentum in the process. Hence, the properties of the emitted particles depend only on the specific characteristics of the BH, which, according to the no-hair conjecture, are its mass, angular momentum, and charge. We focus here on two distinct cases, Schwarzschild (non-rotating) and Kerr (rotating) PBHs. Next, we discuss the emission properties and the BH evaporation rates for each case separately.

\subsection{Schwarzschild Black Holes}

\begin{figure}[t!]
 \includegraphics[width=0.475\textwidth]{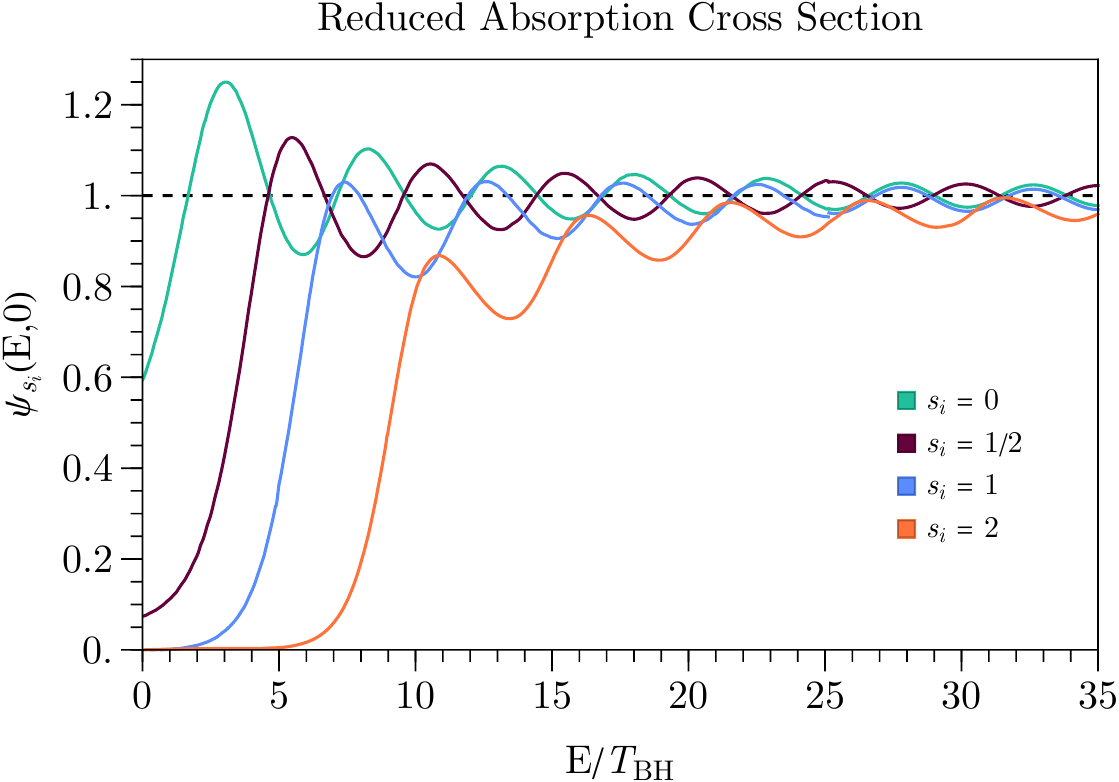}
 \caption{Ratio of the greybody factors to the geometric optics limit for massless particles and different spins, $s_i=0$ (\-e\-me\-rald),
 $s_i=1/2$ (purple), $s_i=1$ (orange), $s_i=2$ (light blue), as function of $E/\TBH$. \label{fig:greyb}}
\end{figure}

Schwarzschild BHs correspond to the simplest scenario, where the BHs are solely described by their mass, $M_{\rm BH}$. As Hawking demonstrated in his seminal papers~\cite{Hawking:1974rv,Hawking:1974sw}, the emitted particles from the evaporation process have a thermal spectrum with temperature related to the mass as ($G$ the gravitational constant)
\begin{align}
 T_{\rm BH} = \frac{1}{8\pi G M_{\rm BH}}\sim 1.06~{\rm GeV}\left(\frac{10^{13}~{\rm g}}{M_{\rm BH}}\right).
\end{align}
The emission rate of any particle species $i$ of mass $\mu_i$, spin $s_i$, and number of degrees of freedom $g_i$ from the evaporation of a BH, within time $\dd{t}$ and momentum $[p,p+\dd{p}]$ interval, is given by
\be\label{eq:BHrate}
\frac{\dd^2 \mathcal{N}_{i}}{\dd p\,\dd t}=\frac{g_i}{2\pi^2}\frac{\sigma_{s_i}(\MBH,\mu_i,p)}{\exp\left[E_i(p)/\TBH\right]-( -1)^{2s_i}}\frac{p^3}{E_i(p)}\,,
\ee
where $E_i(p)=\sqrt{\mu_i^2+p^2}$, and $\sigma_{s_i}$ stands for the absorption cross-section. From this emission rate, we will be able to obtain the time evolution of the BH mass and the phase-space distribution of the different particles evaporated. The absorption cross-section $\sigma_{s_i}$ -- or the related greybody factor, $\Gamma_{s_i}\equiv\sigma_{s_i}p^2/\pi$ -- is a crucial characteristic of the Hawking emission rate as it describes the possible back-scattering of particles due to the gravitational or centrifugal potentials~\cite{Hawking:1974rv,Hawking:1974sw,Page:1976df,Page:1977um}. We note that in the literature this factor is sometimes neglected. However, recent works such as Refs.~\cite{Auffinger:2020afu, Masina:2021zpu} provide the most comprehensive inclusion of these greybody factors. Here, in a similar fashion, we include these factors as consistently as possible, given the results in the literature. For instance, we incorporate the the absorption cross-section for \emph{massive} fermions emitted from Schwarzschild BHs, obtained in Refs.~\cite{Unruh:1976fm,Doran:2005vm}. For massive bosons, we will only include the cross-section obtained by assuming a massless field~\cite{Page:1976df}. Since particle emission is only possible when $E_i \geq \mu_i$, while the correction to the greybody factors due to the finite mass is not large for such values of energy~\cite{MacGibbon:1990zk}, we do not expect a significant effect from such an approximation. For values $G\MBH p\gg 1$, and independently of the particle's spin, the greybody factors tend to the \emph{geometrical-optics} limit, $\sigma_{s_i}(E,\mu)|_{\rm GO}=27\pi G^2\MBH^2$~\cite{Page:1976df,Page:1977um,MacGibbon:1990zk,MacGibbon:1991tj}. Hence, it is convenient to define the ratio of the full greybody factors to the geometrical-optics limit\footnote{For sake of clarity, we do not write the dependence of the absorption cross section on the particle's mass from now on. Let us stress, however, that for fermions emitted from non-rotating BHs, we do include the modifications due to the finite mass~\cite{Doran:2005vm}.}~\cite{Ukwatta:2015iba}
\begin{align}
    \psi_{s_i}(E)\equiv\frac{\sigma_{s_i}(E)}{27\pi G^2\MBH^2}.
\end{align}
In Fig.~\ref{fig:greyb} we present the reduced greybody factors, $\psi_{s_i}(E)$, for the case of massless particles and different spins, $s_i=0$ (emerald), $s_i=1/2$ (purple), $s_i=1$ (orange), $s_i=2$ (light blue), as function of $E/\TBH$. The oscillations present in such quantities are related to the different contributions of the partial waves, each having a different value of the total angular momentum quantum number. Moreover, we observe that the low energy contributions are suppressed from higher particle spin values. This crucial characteristic will play an important role in the accurate determination of the relic abundance.

\begin{figure}
    \includegraphics[width=0.45\textwidth]{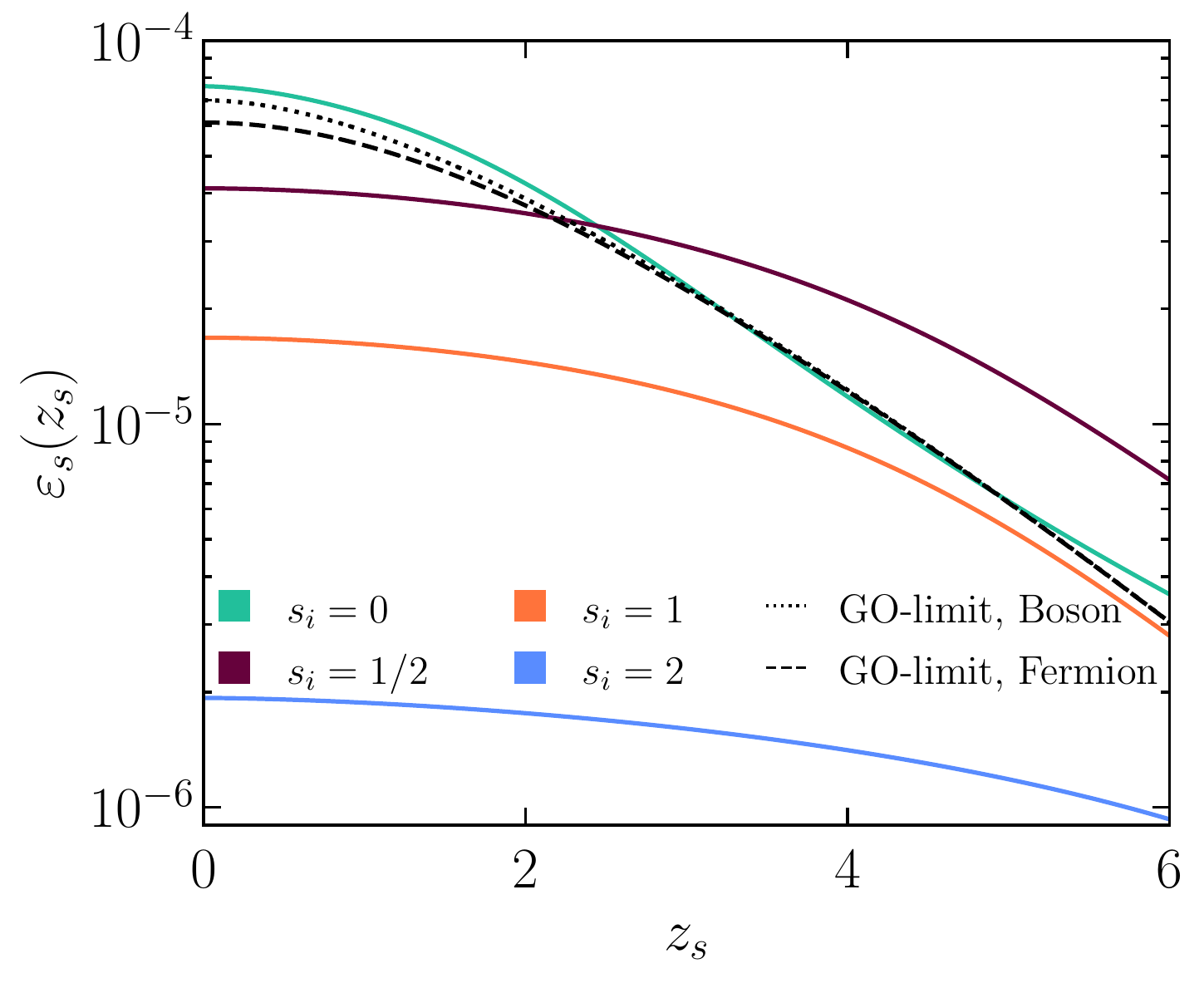}
    \caption{Evaporation function, GO-limit refers to the geometric optics limit.}
    \label{fig:evaporation_func}
\end{figure}

BHs lose their mass over time because of the evaporation process.  The reduction in mass can be obtained by summing Eq.~\eqref{eq:BHrate} over the different species and integrating over the phase space, to obtain \cite{PhysRevD.41.3052,PhysRevD.44.376}
\begin{align}\label{eq:dMBHdt}
\frac{\dd \MBH}{\dd t}&\equiv \sum_i \left.\frac{
\dd \MBH}{\dd t}\right|_i = -\sum_i \int_0^\infty E_i \frac{\dd^2 \mathcal{N}_{i}}{\dd p\dd t} \dd p\,,\nonumber\\ 
&=-\varepsilon(\MBH)\frac{\MPL^4}{\MBH^2}\,,
\end{align}
where $M_p=G^{-1/2}$ denotes the Planck mass. Here, we have defined $\varepsilon(\MBH)$ as the evaporation function  which is dependent on the BH instantaneous mass,
\begin{align}
    \varepsilon(\MBH) \equiv \sum_i g_i\varepsilon_i(z_i),
\end{align}
with the functions $\varepsilon_i(z_i)$ given by
\begin{align}
   \varepsilon_i(z_i) = \frac{27}{8192\pi^5}\int_{z_i}^\infty\frac{\psi_{s_i}(x)(x^2-z_i^2)}{\exp(x)-(-1)^{2s_i}}\,x\dd{x}\,,
\end{align}
where the integration is performed over the dimensionless parameter $x=E_i/\TBH$, and $z_i=\mu_i/\TBH$. The spin-dependent expressions of $\varepsilon_i (z_i)$ for massless particles, in the geometrical-optics limit, and a fitted form obtained after integrating over the full greybody factors are explicitly given in the App.~\ref{ap:A}. In Fig.~\ref{fig:evaporation_func} we present the different contributions to the evaporation function for particles with different spins, together with the results in the geometrical-optics limit as function of $z_i$. As we  observe in this figure, the geometrical-optics limits closely resembles the expected evaporation function for scalars where for bosons with non-zero spin, the approximated forms overestimate the mass loss rate, while for fermions there is a underestimation when $z_i\gtrsim 4$.

Let us determine the momentum-integrated emission rate, $\Gamma_{{\rm BH} \to i}$, and the total number of emitted particles per BH, $\mathcal{N}_i$. Integrating the Hawking rate, Eq.~\eqref{eq:BHrate}, over the momentum, we obtain
\begin{align}\label{eq:Hwtotrate}
   \Gamma_{{\rm BH} \to i} &\equiv\int \dd{p} \frac{\dd^2 \mathcal{N}_{i}}{\dd p\,\dd t}\notag\,,\\
   &=\frac{27 g_i}{1024\pi^4}\frac{1}{G\MBH} \Psi_i (z_i)\notag\,,\\
    &\sim 9.802\times 10^{29} g_i \left(\frac{10^5 {\rm~g}}{\MBH}\right)\left(\frac{\Psi_i (z_i)}{0.897}\right){\rm~s^{-1}}\,,
\end{align}
where
\begin{align*}
    \Psi_i (z_i) \equiv \int_{z_i}^\infty \frac{\psi_{s_i}(x)(x^2-z_i^2)}{\exp(x)-(-1)^{2s_i}} \,\dd{x}\,.
\end{align*}
In the massless case $\mu_i=0$, $\Psi$ simply takes a numerical value which depends on the particle's spin~\cite{Ukwatta:2015iba}
\begin{equation}\label{eq:psi0}
    \Psi_i (0) = \begin{cases} 
    2.45 & s=0 \\
    0.897 & s=1/2 \\
    0.273 & s=1 \\
    0.026 & s = 2
    \end{cases}\,.
\end{equation}
We provide useful analytic expressions for $\Psi_i (z_i)$ in App.~\ref{ap:A}.
The total number of emitted particles of the species $i$ during the BH existence is simply computed by integrating the total rate over time,
\begin{align}
    \mathcal{N}_i&=\int_0^{\tau}\dd{t}\Gamma_i(\MBH)\notag\\
    &= \eta_i(z_i^{\rm in})\frac{g_i}{g_{\star}(\TBH^{\rm in})} \left(\frac{\MBH^{\rm in}}{\MPL}\right)^2 ,
    \label{eq:semi_analytic}
\end{align}
where $\tau$ is the BH lifetime, and 
\begin{align}
    \eta_i(z_i^{\rm in})=\frac{27}{1024\pi^4}\frac{g_{\star}(\TBH^{\rm in})}{\left(z_i^{\rm in}\right)^2}\int_0^{z_i^{\rm in}}\frac{\Psi_i(z_i)}{\sum_j g_j\varepsilon_j(\mathfrak{m}_j z_i)}\,z_i\dd{z_i}\,,
\end{align}
with $z_i^{\rm in} = \mu_i/\TBH^{\rm in}$ the ratio of the particle's mass to the \emph{initial} BH temperature, and $\mathfrak{m}_j\equiv \mu_j/\mu_i$ the ratio of each existing particle mass to the mass of the species $i$. The derivation of $\eta_i(z_i^{\rm in})$ is presented in App.~\ref{ap:A}. Differently from what has been previously done in the literature, we have not assumed any relation between the particle mass and the BH temperature. Instead, $\mathcal{N}_i$ is general: the Boltzmann suppression present when $\TBH < \mu_i$ is automatically included in it. 
Let us compare the total number of emitted particles including the greybody factors to the geometric optics limit, $\mathcal R_{\cal N} = \left.\mathcal N_i\right|_{\rm w}/\left.\mathcal N_i\right|_{\rm w.o.}$ for a particle with $\mu_i\ll \TBHi$ we have
\begin{equation}
    \mathcal{R}_{\cal N}=\begin{cases}
    0.84 & s_i=0\\
    0.61 & s_i=1/2 \\
    0.28 & s_i=1\\
    0.02 & s_i=2
    \end{cases}\,,
\end{equation}
we therefore observe that by not including correctly the greybody factors, there is a significant overestimation of the number of produced particles by a BH.

\subsection{Kerr Black Holes}\label{subs:KBH}

Another possibility is that the evaporating BHs have some non-zero angular momentum. Such rotating BHs, also known as Kerr BHs, could have formed with some initial spin or acquired their angular momenta via some specific mechanisms, such as mergers~\cite{Buonanno:2007sv,Kesden:2008ga,Tichy:2008du}. The BH temperature for the Kerr scenario is modified due to the presence of the angular momentum, 
\begin{align}
 T_{\rm BH} = \frac{1}{4\pi G M_{\rm BH}}\frac{\sqrt{1-a_\star^2}}{1+\sqrt{1-a_\star^2}}\,,
\end{align}
where the dimensionless parameter $a_\star$ is related to $J$, the BH angular momentum, as $a_\star = J \MPL^2/M^2$. Such parameter can have values $a_\star\in [0,1]$, so that for the case of close-to-maximally rotating BHs, the temperature tends to be zero.

The spectra of emitted particles have an additional dependence on the BH angular momentum,
\begin{align}\label{eq:KBHrate}
\frac{\dd^2 \mathcal{N}_{i}}{\dd p\dd t}&=\frac{g_i}{2\pi^2} \sum_{l=s_i}\sum_{m=-l}^l\frac{\dd^2 \mathcal{N}_{ilm}}{\dd p\dd t}\,,
\end{align}
with
\begin{align}\label{eq:KBHrate_lm}
\frac{\dd^2 \mathcal{N}_{ilm}}{\dd p\dd t}&=\frac{\sigma_{s_i}^{lm}(\MBH,p,a_\star)}{\exp\left[(E_i - m\Omega)/\TBH\right]-( -1)^{2s_i}}\frac{p^3}{E_i}\,,
\end{align}
where $\Omega = (a_\star/2G\MBH)(1/(1+\sqrt{1-a_\star^2}))$ is the angular velocity of the horizon and $l,m$ the total and axial angular momentum quantum numbers, respectively. From the emission rate in Eq.~\eqref{eq:KBHrate} it is clear that the absorption cross-section also depends on $a_\star$. In what follows we will use the procedure established in Refs.~\cite{Chandrasekhar:1975zz,10.2307/79115,Chandrasekhar:1977kf} in order to compute the cross-sections $\sigma_{s_i}^{lm}$ appearing in Eq.~\eqref{eq:KBHrate_lm} in the case of scalar, fermion, and vector particles in the Kerr scenario\footnote{For consistency, we have checked that our numerical results are similar to those contained in the code {\tt BlackHawk}~\cite{Arbey:2019mbc}, finding an agreement at the levels of $\sim 1.37\%$ ($\sim 0.44\%$) for massless scalars, $\sim 1.39\%$ ($\sim 10\%$) for massless fermions, and $\sim 0.55\%$ ($\sim 1.8\%$) for massless vectors in the case of $a_\star=0$ ($a_\star=0.99$).}. For the spin-2 case \cite{Chandrasekhar:1976zz} we use the greybody factors from {\tt BlackHawk} as a numerical input when using Eq.~\eqref{eq:KBHrate_lm}. Interestingly, the emission of higher-spin particles is enhanced for BHs with a non-zero angular momentum. Thus, we could expect an enhanced emission of spin-2 DM particles, such that it would be possible to increase the relic density. This will be explored in more detail in the next section.

Similarly to the mass depletion in the Schwarzschild case, for Kerr black holes the angular momentum decreases in time because of particle emission. The equation for the angular momentum is obtained by integrating the rate multiplied by the axial angular momentum number in Eq.~\eqref{eq:KBHrate_lm}~\cite{Page:1976df},
\begin{align}
 \frac{\dd J}{\dd t} &= -\sum_i \int_0^\infty \sum_{lm} m \frac{\dd^2 \mathcal{N}_{ilm}}{\dd p\dd t} dp\notag\,,\\
 &=-a_\star\frac{\MPL^2}{\MBH}\gamma(\MBH,a_\star)\,,
\end{align}
with $\gamma(\MBH,a_\star) = \sum_i\gamma_i(\MBH,a_\star)$ the angular momentum evaporation function.
Substituting the definition of $a_\star$ in  Eq.~\eqref{eq:KBHrate_lm}, one finds the evolution equations as function of time for both mass and spin,
\begin{subequations}\label{eq:KerrMA}
\begin{align}
 \frac{d\MBH}{dt} &= - \varepsilon(\MBH, a_\star)\frac{\MPL^4}{\MBH^2}\,,\\
 \frac{da_\star}{dt} &= - a_\star[\gamma(\MBH,a_\star) - 2\varepsilon(\MBH,a_\star)]\frac{\MPL^4}{\MBH^3}\,.
\end{align}
\end{subequations}
The functions, $\gamma_i(\MBH,a_\star)$ and $\varepsilon_i(\MBH,a_\star)$, for the different spins can be parametrized in a similar fashion as in the Schwarzschild case,
\begin{subequations}
\begin{align}
   \varepsilon_i(z_i,\as) &= \frac{27}{8192\pi^5}\int_{z_i}^\infty\sum_{lm}\frac{\psi_{s_i}^{lm}(x,\as)(x^2-z_i^2)x\dd{x}}{\exp(x^\prime/2f(\as))-(-1)^{2s_i}}\,,\\
   \gamma_i(z_i,\as) &= \frac{27}{1024\pi^4}\int_{z_i}^\infty\sum_{lm}\frac{m\,\psi_{s_i}^{lm}(x,\as)(x^2-z_i^2)\dd{x}}{\exp(x^\prime/2f(\as))-(-1)^{2s_i}}\,,
\end{align}
\end{subequations}
where now $x = 8\pi G\MBH E_i$, $z_i = 8\pi G\MBH \mu_i$, $x^\prime = x - m\Omega^\prime$, with $\Omega^\prime = 8\pi G\MBH\Omega$, and
\begin{align*}
    f(\as) = \frac{\sqrt{1-a_\star^2}}{1+\sqrt{1-a_\star^2}}\,.
\end{align*}
The previous definitions were chosen in order to have a smooth transition to the Schwarzschild case when $\as\to 0$. We have determined fitted forms for these factors from explicit integration of the greybody factors in the Kerr case, see App.~\ref{ap:A}. We parametrize the emission rate for spinning BHs similarly to the Schwarzschild case,
\begin{align}\label{eq:HwtotrateK}
    \Gamma_{{\rm BH} \to i}(\MBH,\as)= \frac{27 g_i}{1024\pi^4}\frac{1}{G\MBH} \Psi_i (z_i,\as),
\end{align}
where, analogously, we have
\begin{align}
    \Psi_i (z_i,\as) \equiv \int_{z_i}^\infty\sum_{lm}\frac{\,\psi_{s_i}^{lm}(x,\as)(x^2-z_i^2)}{\exp(x^\prime/2f(\as))-(-1)^{2s_i}}\dd{x}\,.
\end{align}
Obtaining a closed form for the total number of particles in the Kerr case is not straightforward. It is not possible to take as an independent variable the BH mass, as done in the Schwarzschild case since the angular momentum also changes with time. 

Finally, note that in the limit of an initial $a_\star = 0$, one readily recovers the Schwarzschild functions. Thus, in our simulations, we solve the Eq.~\eqref{eq:KerrMA} in the cosmological context and impose $a_\star=0$ as an initial condition when analyzing the specific scenario of Schwarzschild BHs.  

\subsection{Phase-space Distribution of Evaporated Particles}
\label{sec:psd}
The phase-space distribution of particles emitted from BHs has a significant impact on the evolution of the Universe. For the simple setup explored in this study, the mean free path of DM is the quantity of most consequence, limiting the formation of small-scale structures.    

The mean free path of the emitted particles strongly depends on the evolution of their respective phase-space distributions. In the usual FO and FI cases, such distributions are dictated by the Boltzmann distributions already present in the thermal bath. In the presence of BH evaporation, such phase-space distributions may be significantly distorted. Indeed, when they evaporate, BHs produce particles with a typical momentum $\langle p(t)\rangle\sim \TBH(t)$. Because $\TBH$ is an increasing function of time when BHs evaporate, the momentum of the particles they produce is directly related to the dynamics of the Hawking evaporation. For a particle of mass $\mu_i$, this typically leads to two major production regimes:
\begin{itemize}
  \item $\mu_i\lesssim \TBH^{\rm in}$:  most of the particles produced via evaporation are relativistic, as they carry a momentum $p\gtrsim \TBH^{\rm in}$.
  \item $\mu_i\gtrsim \TBH^{\rm in}$:  the production is statistically suppressed until the BH temperature increases above the particle mass. Therefore, most of the production occurs when $\TBH\sim \mu_i$ producing a population of non-relativistic evaporated products.
\end{itemize}
\begin{figure*}
 \centering
 \includegraphics[width=\linewidth]{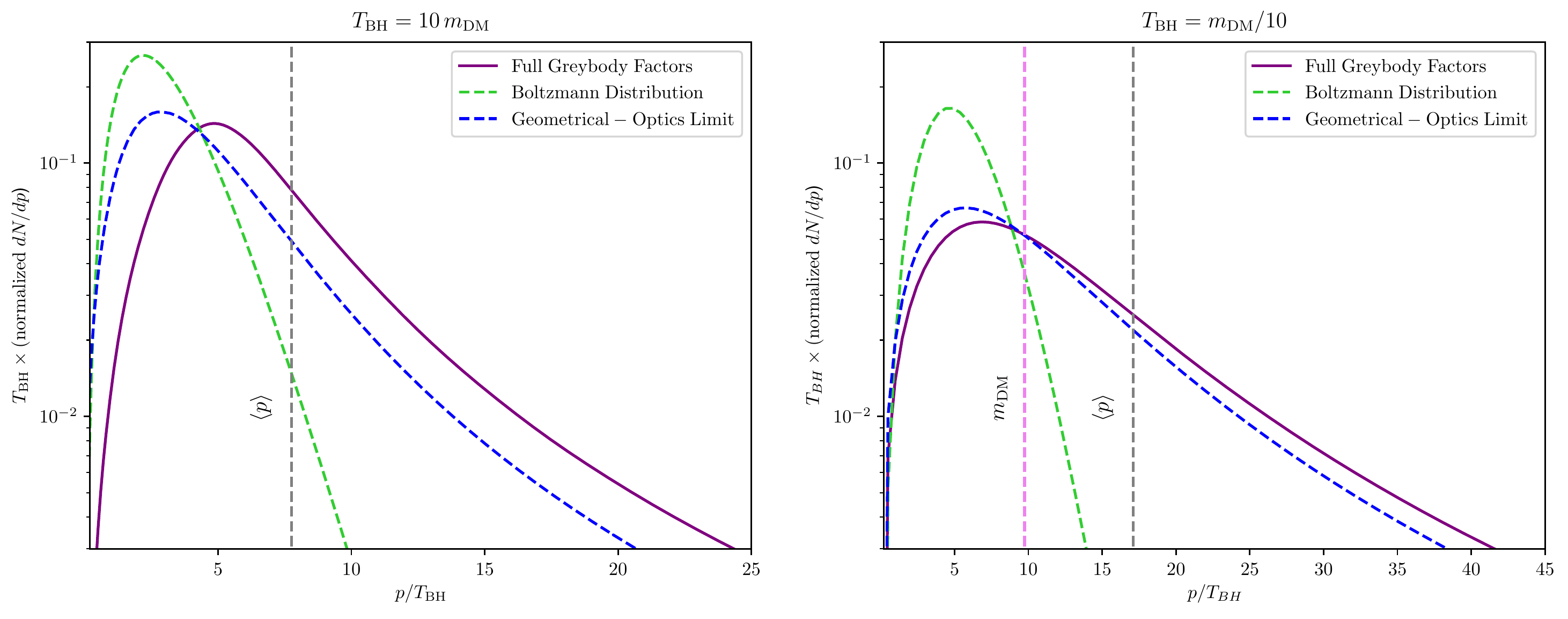}
 \caption{\footnotesize\label{fig:phasespace} Phase-space distribution of dark matter particles produced via BH evaporation, in the two cases where $\TBH=10 \mDM$ \textbf{(left)} and $\TBH=\mDM/10$ \textbf{(right)}. We compare the distribution from the full calculation \textbf{(violet)}, with the Boltzmann distribution \textbf{(green dashed)} and the Geometrical-Optics limit (blue dashed). We also indicate the average momentum, $\langle p\rangle$ \textbf{(grey dashed)} as calculated in \eqref{eq:ave_mom}.}
\end{figure*}
Given the expression of evaporation rate per unit of time and momentum in Eq.~\eqref{eq:BHrate}, we can derive the phase-space distribution of the different particles produced through BHs evaporation. 
In Ref.~\cite{Baldes:2020nuv} such a distribution was derived in the geometrical-optics limit in the case where the DM mass, $\mDM$, verifies $\mDM\ll \TBH$.
Note, however, that in Refs.~\cite{Auffinger:2020afu,Masina:2021zpu}, the phase-space distribution was first computed including the greybody factors, showing the crucial impact of incorporating such factors correctly.
We also go beyond the geometrical-optics limit and solve those phase-space distributions using our expressions for the greybody factors by simply integrating Eq.~\eqref{eq:BHrate} over time~\footnote{In principle, taking into account the expansion of the Universe during the evaporation process may slightly alter this result. However, it was shown in Ref.~\cite{Baldes:2020nuv} that such an effect is negligible.}
\be
\frac{\dd \mathcal N_{i,s_i}}{\dd p}=\int_{t_i}^{t_{\rm ev}}\frac{\dd^2 \mathcal{N}_{i, s_i}}{\dd p\dd t}\dd t\,.
\ee
Extending the results of Ref.~\cite{Baldes:2020nuv} to the massive DM case, we can compare our results to the geometrical-optics limit of such a formula 
\be\label{eq:geodistri}
\frac{\dd \mathcal N_{i,s_i}}{\dd p}=\frac{15 g_i M_p^2}{8\pi^5 g^\star_{\rm BH}}\frac{p}{(p^2+\mu_i^2)^2}f_{s_i}\left(\frac{\sqrt{p^2+\mu_i^2}}{\TBHi}\right)\,,
\ee
where the function $f_{s_i}$ is an integral that can be computed analytically
\bea
f_{s_i}(x) &\equiv& \int_0^x\frac{y^4\dd y}{\exp(y)-(-1)^{2s_i}}\,,\nonumber\\
&=&\frac{1}{5 \epsilon_i} \left[120 \left( \text{Li}_5\left[\epsilon_i\right]-\text{Li}_5\left[\epsilon_i e^x\right]\right)\right.\nonumber\\
&+&\left.20 x \left(x^2 \text{Li}_2\left[\epsilon_i e^x\right]-3 x \text{Li}_3\left[\epsilon_i e^x\right]+6 \text{Li}_4\left[\epsilon_i e^x\right]\right)\right.\nonumber\\
&-&\left.x^4 \left(x-5 \log \left[1- e^x/\epsilon_i \right]\right)\right]\,,\nonumber\\
\eea
and ${\rm Li}_n$ are the polylog functions of order $n$ and $\epsilon_i=(-1)^{2 s_i}$ . In Fig.~\ref{fig:phasespace} we depict the phase-space distribution of a fermionic DM particle produced by evaporation in two representative cases where $\mDM \ll \TBH$ (left panel) and $\mDM\gg \TBH$ (right panel). We indicate in violet the phase-space distribution of DM particles that we obtain using the full greybody factors in Eq.~\eqref{eq:BHrate}. As expected, such a distribution is peaked around the BH temperature, similarly to what was obtained in Ref.~\cite{Baldes:2020nuv}. We compare our results with the distribution of Eq.~\eqref{eq:geodistri} obtained in the geometrical-optics limit and find that our distribution is slightly shifted towards larger values of the momenta.
Such a shift is related to the suppression of the low momenta present in the greybody factors,
similar to what was observed in Ref.~\cite{Auffinger:2020afu}.
We also indicate (dashed green line) the corresponding Boltzmann distribution evaluated at the temperature $\TBH$ as well as the value of the typical momentum of evaporated particles (grey dashed line). In the right panel of Fig.~\ref{fig:phasespace} one can see that the DM phase-space distribution instead peaks at $p\sim \mDM$, since BHs mainly produce DM particles after their temperature rises above $\mDM$. Again we can notice a significant shift between our findings and the geometrical-optics limit obtained using the prescription of Ref.~\cite{Baldes:2020nuv}. Finally, the authors of Ref.~\cite{Baldes:2020nuv} evaluated the Boltzmann distribution at $\sim 3 \TBH$ to make the distribution peaks match. We can see that such a prescription must be modified to match a Boltzmann distribution with the full distribution we obtained because of the aforementioned shift towards larger momenta. 

An important constraint that the purely gravitational production via Hawking evaporation is subject to corresponds to the warm DM bound. From our discussion above, we have found that the emitted particles could have a large average momenta depending on their masses. In such a case, the redshift resulting from the expansion of the Universe might not be large enough to make the DM non-relativistic at the moment of structure formation, hence contradicting observations. Following previous treatments~\cite{Baldes:2020nuv,Masina:2020xhk,Masina:2021zpu}, we compute the average DM velocity today $v_{0}$ from the expected average momentum,
\begin{align}
    v_{0}=\frac{a_{\rm ev}}{a_0}\frac{\langle p_i \rangle}{m_{\rm DM}},
    \label{eq:wdm}
\end{align}
with $a_{\rm ev}(a_0)$ the scale factors at evaporation (today). We will impose that such a velocity should be smaller than the maximum value allowed from Lyman-$\alpha$ constraints, assuming all DM coming from PBH evaporation, to have a sufficiently cold DM~\cite{Bode:2000gq,Boyarsky:2008xj,Baldes:2020nuv,Baur:2017stq}. 

The average momentum of an emitted particle will be computed for spinning BHs in a simple and general manner. Reversing the integration order, that is, integrating the Hawking rate first over the momentum and using the definitions of the evaporation functions, Eq.~\eqref{eq:KerrMA}, and the momentum integrated Hawking rates, Eq.~\eqref{eq:HwtotrateK}, and then integrating over time, we have
\begin{align}\label{eq:ave_mom}
    \langle p_i \rangle = \frac{\int \dd{t}\,\varepsilon_i(z_i,\as) \MBH^{-2}}{\int\, \dd{t}\Gamma_{{\rm BH} \to i}(\MBH,\as)}\,.
\end{align}
This complementary approach will be used in our numerical procedure to enforce the warm DM constrain in our results. It it worth noting that more accurate determinations of the WDM constraint have been undertaken by the authors of Refs.~\cite{Baldes:2020nuv, Auffinger:2020afu, Masina:2021zpu} where the DM phase space has been used as an input to the cosmic linear perturbation solver CLASS~\cite{CLASSI, CLASSII, CLASSIV}.

\section{Production of Dark Matter via Primordial Black Hole Evaporation}\label{sec:evaporation}

Several mechanisms lead to the formation of PBHs in the early Universe after inflation~\cite{Carr:2020xqk,Carr:2020gox,Khlopov:2008qy}. For simplicity, we assume that a population of PBHs was formed with a monochromatic mass spectrum.
Let us stress that assuming such a simple spectrum allows us to give more generic statements, since it decouples the particle production from the details of the PBH formation. 
Clearly, the PBH mass spectrum obtained from a given mechanism will depend on specific parameters related to the model.
For instance, PBHs formed from collapse of inhomogeneties relies upon the critical value of the overdensities that enter the horizon~\cite{Carr:2020gox,Carr:2020xqk}. 
Other specific models, such as collapses from multi-field inflatons, cosmic strings, bubble collisions, domain walls, or even the PBH formation in an early matter dominated era, will produce distinct mass spectra.
Note, however, that the assumption of a monochromatic spectrum is not totally unrealistic, as the PBHs could have formed at very specific time, thus having a rather narrow spectrum.
We leave the extension of our results to more realistic mass distributions for future work.
Moreover, we consider that the initial PBH mass is proportional to the particle horizon mass at the moment of formation in a radiation-dominated era~\cite{Carr:2020xqk}
\begin{align}\label{eq:Min}
 \MBH^{\rm in} = \frac{4\pi}{3} \gamma\frac{\rho_i}{H_i^3}\,,
\end{align}
where $\gamma$ is a factor related to the gravitational collapse, assumed here to be equal to $(1/\sqrt{3})^3\approx 0.2$. The initial PBH population is characterized by the initial fraction of the PBH energy density, $\rho^{\rm in}_{\rm PBH}$, with respect to the total energy density $\rho^{\rm in}$, which can be expressed through the parameter $\beta\equiv \rho^{\rm in}_{\rm PBH}/\rho^{\rm in}$, or, more commonly, using the definition
\begin{align}\label{eq:betap}
 \beta^\prime \equiv \gamma^{1/2}\left(\frac{g_\star (T_{\rm in})}{106.75}\right)^{-1/4}\frac{\rho^{\rm in}_{\rm PBH}}{\rho^{\rm in}}\,,
\end{align}
where $T_{\rm in}$ is the plasma temperature at the time of the PBH formation, and the additional factors are included as the initial PBH fraction always appears corrected by them~\cite{Carr:2020xqk}. Since the PBH energy density scales as $a^{-3}$, it is possible to have a PBH-dominated era depending on the initial value of $\beta^\prime$. Such a possibility will play an important role when we consider the effects of the evaporation on the DM production. 
Furthermore, for the case of Kerr PBHs, we assume a monochromatic angular momentum distribution, similarly to the mass, such that all BHs had the same initial value of the angular momentum. As with the mass spectrum, this simplification allows us to remain moderately independent of the PBHs formation mechanisms.
For the specific case of Kerr BHs, PBHs can acquire a non zero spin via different models, such as mergers, accretion, or even the formation mechanism mentioned before could produce BHs with some spin
(see, e.g.~\cite{Hooper:2020evu,Flores:2021tmc}, 
and Ref.~\cite{Arbey:2021ysg} for a first computation of $\Delta N_{\rm eff}$ from Dark Radiation considering an extended spin distribution).

Therefore, the early Universe will be comprised of three different energy density components, the PBH population plus radiation related to the SM and, possibly, a Dark Sector (DS).
The Hubble parameter, therefore, should take into account these three elementary contributions,
\be\label{eq:Hubble}
\frac{3H^2M_p^2}{8\pi}=\rho_{\rm SM} + \rho_{\rm DS} + \rho_{\rm PBH}\,.
\ee
By means of Hawking evaporation, PBHs will not only change the evolution of the Universe but also emit a large number of particles, regardless of their possible interactions. The set of produced particles will affect the Universe's energy budget and, as we have mentioned before, could lead to the generation of the observed DM.

The capacity of PBHs to produce DM particles when they evaporate strongly depends on two factors: $(i)$ whether the temperature of the black holes is smaller or larger than the DM mass, and $(ii)$ whether PBHs evaporate in a matter or radiation dominated era~\cite{Lennon:2017tqq, Morrison:2018xla, Hooper:2019gtx, Gondolo:2020uqv, Masina:2020xhk,  Bernal:2020bjf, Bernal:2020ili, Masina:2021zpu}. In order to track effectively the number of DM particles produced by a PBH population in the early Universe, we must specify how the phase-space distribution of such states changes over time. We \emph{define} for the species $i$\footnote{We include in the definition the factor of $p^2/(2\pi^2)$ because the integration of the Hawking rate over momentum and time directly gives the total number of emitted particles.}
\begin{align}
\left.g_i\frac{p^2}{2\pi^2}\frac{\partial f_i}{\partial t}\right|_\bh(t,p)=n_\bh \frac{\dd^2 \mathcal{N}_{i}}{\dd p\dd t}\,,
\end{align}
where $n_\bh$ is the PBH number density. Hence, it is possible to write a Boltzmann equation for such a species in a FLRW Universe,
\begin{align}
\frac{\partial f_i}{\partial t}-Hp\frac{\partial f_i}{\partial p}=C[f_i]+\left.\frac{\partial f_{i}}{\partial t}\right|_\bh\,,
\end{align}
where we have included possible interactions via a collision term $C[f_i]$. In the following, however, we assume that the DM does not interact with the SM thermal plasma, so that such a collision term will be absent. We can obtain the usual equation for number densities after integrating over the phase space,
\begin{align}\label{eq:NumEq}
\dot{n}_{i} + 3H n_{i} &=  g_i\int\left.\frac{\partial f_{i}}{\partial t}\right|_\bh\frac{p^2\dd p}{2\pi^2}\,\notag\,,\\
&=n_\bh\, \Gamma_{{\rm BH} \to i}(\MBH,\as)\,.
\end{align}
\begin{figure*}[t!]
 \centering
 \includegraphics[width=0.475\linewidth]{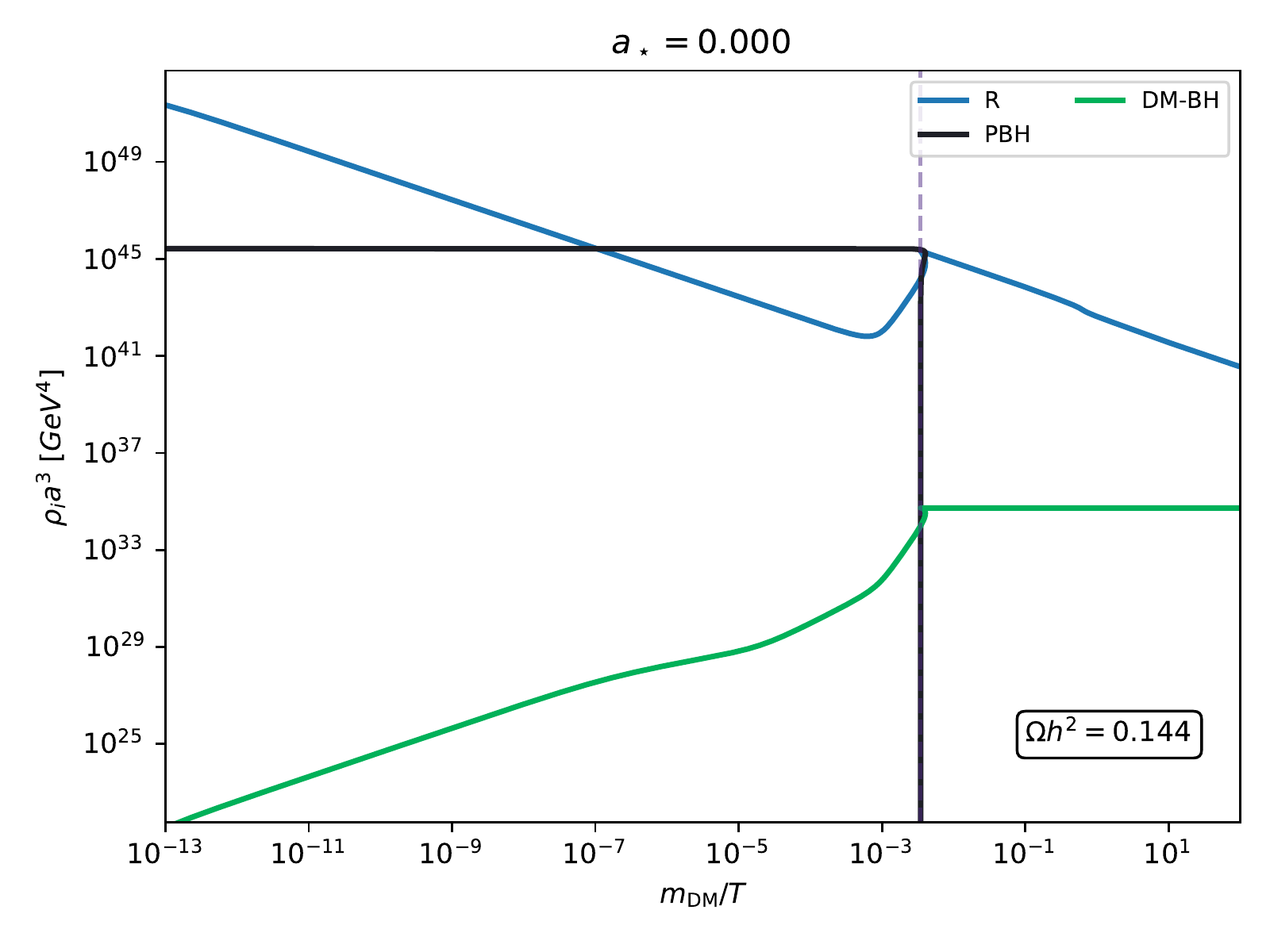}
 \includegraphics[width=0.475\linewidth]{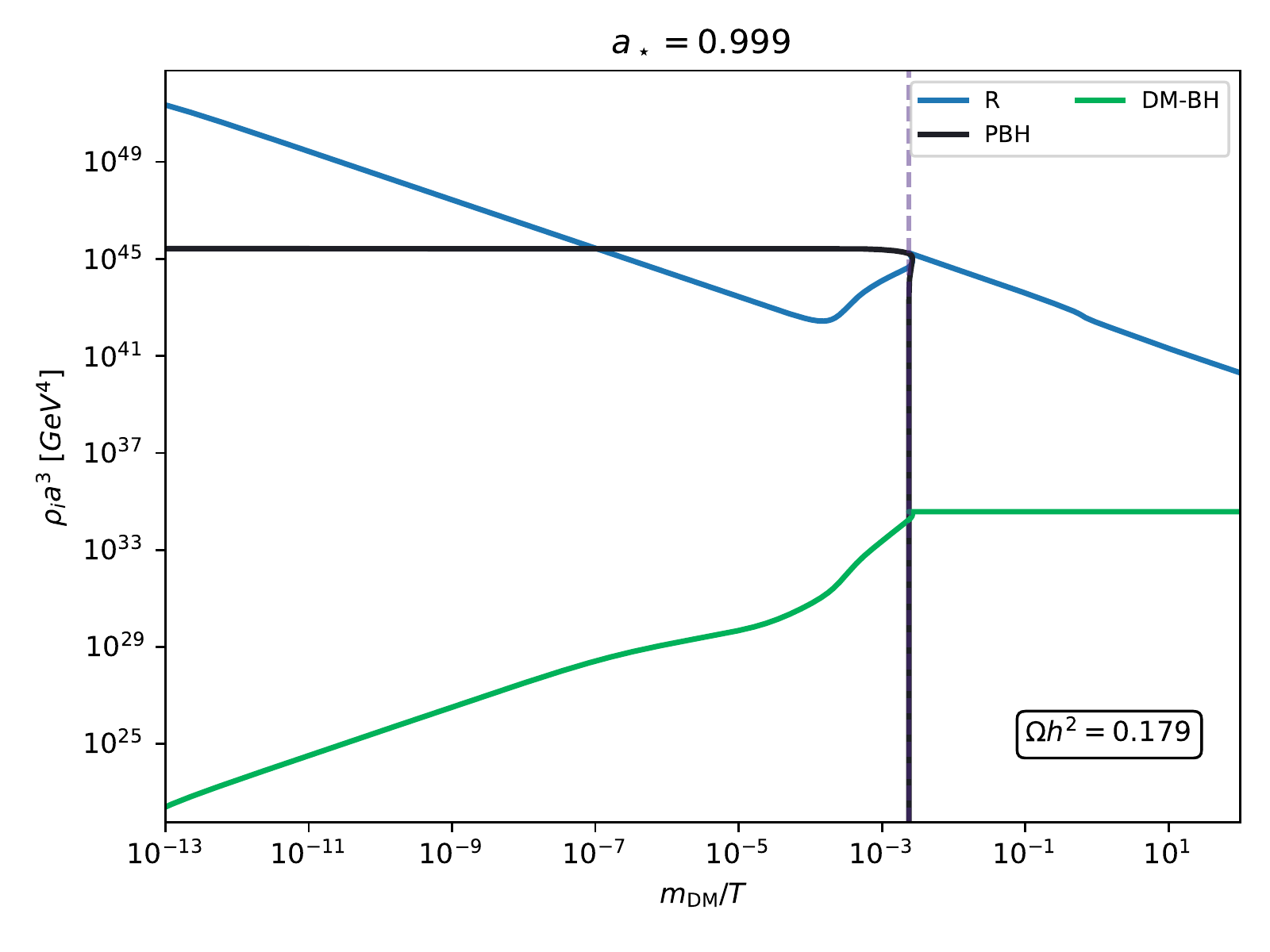}
 \caption{\footnotesize\label{fig:sol} Solutions of the Friedmann-Boltzmann equations for $\mDM = 0.1~\GeV$, $\beta^\prime = 10^{-7}$ and $\MBHi = 10^6~\g$ Schwarzschild (left) and Kerr (right) PBHs. We present $\rho_i a^3$ as function of $\mDM/T$ for the SM radiation (blue), PBH (black), and DM (green) energy densities. In each case, we record the final relic abundance. }
\end{figure*}
The Friedmann equations for the $\rho_{\rm PBH}$, $\rho_{\rm SM}$ PBH and SM radiation energy densities, respectively, are given by
\begin{subequations}\label{eq:FBEqs}
\begin{align}
    \dot{\rho}_{\rm SM} + 4H \rho_{\rm SM} &= -\frac{1}{\MBH}\left.\frac{\dd \MBH}{\dd t}\right|_{\rm SM}\rho_{\rm PBH}\,,\\
    \dot{\rho}_{\rm PBH} + 3H \rho_{\rm PBH} &= \frac{1}{\MBH}\frac{\dd \MBH}{\dd t}\rho_{\rm PBH}\,,
\end{align}
\end{subequations}
where the energy produced by the evaporation depends on the mass loss rate since
\begin{align}
\frac{\dd \rho_{\rm ev}}{\dd t}=\sum_i \int_0^\infty E_i \left.\frac{\partial f_i}{\partial t}\right|_\bh\frac{p^2\dd p}{2\pi^2}=-\frac{\rho_{\rm PBH}}{\MBH} \frac{\dd \MBH}{\dd t}\,,
\end{align}
where we used that $\rho_{\rm PBH}=\MBH n_\bh$. The set of Friedmann equations includes two different effects related to the presence of a PBH population. First, PBHs behave as matter, $\rho_{\rm PBH}\propto a^{-3}$, enabling the possibility of early matter domination, as mentioned above. Second, the evaporation produces SM particles that reheat the Universe. Thus, to determine the DM generation consistently, we solve the system of equations, Eq.~\eqref{eq:FBEqs}, together with the mass and angular momentum PBH loss rates, Eq.~\eqref{eq:KerrMA}, and an equation for the DM number density in the lines of Eq.~\eqref{eq:NumEq}. The solution is found using the {\tt ULYSSES} python package~\cite{Granelli:2020pim}, which allows for a rapid determination of the resulting DM relic abundance, including the PBH evaporation. 

Some words are in order about the numerical procedure. In general, it is not possible to na\"ively apply a differential equation solver to the full system of equations, especially when the DM mass is much larger than the initial PBH temperature because of the stiffness present in the mass loss rate. Such stiffness is a consequence of the explosive nature of the particle emission in the final stages of the BH lifetime. Starting with a relatively large PBH mass, $\MBH^{\rm in} \gg 1~{\rm g}$, it is not possible to reach $\MPL$, a value which we aim to attain when we solve the equations, with direct use of a numerical solver. Instead, we use a \emph{zoom-in} procedure: We iteratively solve the Boltzmann equation on smaller and smaller time scales until the PBH mass reaches the Planck mass, $M_p$. We have checked that the solutions are stable and correctly account for the case when the particle emission only occurs during the final moments of the BH existence. 

Once our coupled equations have reached a stable point, where the Universe is radiation dominated and there is no longer any production of DM, we can use the temperature at which the evaporation occurs, $T_{\rm ev}$
and entropy conservation to obtain today's dark matter density parameter ($T_0$ is the present temperature),
\begin{equation}
    \Omega_{\rm DM}=  \frac{1}{\rho^0_{\rm crit}}\frac{g_{\star s}(T_0)T_0^3}{g_{\star s}(T_{\rm ev})T_{\rm ev}^3} \rho_{\rm DM}^{\rm sim}.
\end{equation}
We present in Fig~\ref{fig:sol} prototypical solutions of the Friedmann-Boltzmann equations for $\mDM = 0.1~\GeV$, $\beta^\prime = 10^{-7}$ and $\MBHi = 10^6~\g$ Schwarzschild (left) and Kerr (right) PBHs. The time evolution of $\rho_i a^{3}$ is displayed for the SM, PBH and DM energy densities. The value of the relic abundance is also shown. We observe in both cases, PBHs modify the evolution of the Universe and generate DM. After a radiation-dominated phase, the PBH density, in this case, leads to an early matter dominated era, which ends when the PBHs evaporate. During the final states of the evaporation, a large entropy injection into the SM takes place, while DM production is accelerated. Such entropy injection is modified if the PBHs had a non-zero $\as$. We will return to these solutions in more detail in the next subsections.

Similarly to our semi-analytic expression in  Eq.~\eqref{eq:semi_analytic}, for the total number of DM particles produced per Schwarzchild BH, $\mathcal{N}_{\rm DM}$, we can obtain the same parameter, 
\begin{equation}
    \Omega_{\rm DM}=  \frac{1}{\rho^0_{\rm crit}}\frac{g_{\star s}(T_0)T_0^3}{g_{\star s}(T_{\rm ev})T_{\rm ev}^3} n_{\rm BH}^{\rm eva}\mathcal{N}_{\rm DM} m_{\rm DM}\,,
\end{equation}
where $n_{\rm BH}^{\rm ev}$ is the BH number density at the evaporation, which for a monochromatic mass spectrum can be related to the initial number density $n_{\rm BH}^{\rm in}$ by $n_{\rm BH}^{\rm ev}(a^{\rm ev})^3=n_{\rm BH}^{\rm in}(a^{\rm in})^3$ and thus
\begin{align}
   \Omega_{\rm DM}=  \frac{1}{\rho^0_{\rm crit}}\frac{g_{s \star}(T_0)T_0^3}{g_{s \star}(T_{\rm ev})T_{\rm ev}^3} \left(\frac{a^{\rm in}}{a^{\rm ev}}\right)^3\frac{\rho_{\rm BH}^{\rm in}}{M_{\rm BH}^{\rm in}}\mathcal{N}_{\rm DM} m_{\rm DM}\,.
\end{align}
In general, it is difficult to get a good approximation for all the above values at evaporation (see, however,~\cite{paperB}). Nevertheless, in the case where the populations of PBHs remain a negligible component of the Universe's energy density, entropy conservation can be assumed, leading to the simpler form of the relic density
\begin{align}
    \Omega_{\rm DM}&=  \frac{1}{\rho^0_{\rm crit}}\frac{g_{\star S}(T_0)T_0^3}{g_{\star s}(T_{\rm in})T_{\rm in}^3} \frac{\rho_{\rm BH}^{\rm in}}{M_{\rm BH}^{\rm in}}\mathcal{N}_{\rm DM} m_{\rm DM}\,,
\end{align}
which is fully calculable using Eq.~\eqref{eq:semi_analytic} and the initial conditions, Eq.~\eqref{eq:Min} - \eqref{eq:betap}, leading to 
\begin{widetext}
\begin{align}
    \Omega_{\rm DM}h^2
    &=\frac{\pi^2}{30}\left(\frac{45}{16 \pi^3}\right)^{1/4} \left(\frac{g_{\star S}(T_0)T_0^3}{\rho^0_{\rm crit} h^{-2}}\right)\left(\frac{\MPL}{\MBH^{\rm in}}\right)^{3/2}\beta^\prime \mathcal{N}_{\rm DM} m_{\rm DM}\notag\\
    &\simeq 1.595 \left(\frac{\gamma}{0.2}\right)^{1/2}\left(\frac{g_\star(\TBHi)}{106.75}\right)^{-1/4}\left(\frac{1~{\rm g}}{\MBH^{\rm in}}\right)^{3/2}\left(\frac{m_{\rm DM}}{1~\GeV}\right)\,\beta\,\mathcal{N}_{\rm DM}\,.
\label{eq:relic_approx}
\end{align}
\end{widetext}
Where we apply this method, we find agreement with the fully numerical method to the level below $1\%$. In the case where PBHs play a much greater role in the cosmological evolution, we use the approximations in~\cite{paperB} and obtain values that agree up to some $\mathcal{O}(1)$ multiplicative factor. This gives us a high degree of confidence in the accuracy of our calculation. By numerically solving the Boltzmann equations and including the greybody factors as accurately as possible, we believe that this work constitutes a step forward in the work connecting DM production and PBHs. As previously mentioned, Refs.~\cite{Auffinger:2020afu, Masina:2021zpu} take great care in consistently using the greybody factors but use approximate analytic solutions to obtain $\Omega_{\rm DM}h^2$. These approximations are most appropriate when the PBH population does not affect the thermal history of the universe as seen in our validation of the code. Moving to the numerical framework for solving these systems allows for a more sophisticated analysis to be performed, where dark sectors for have non-gravitational interactions with the Standard Model. 

Next, we describe our results regarding the DM production from Schwarzschild and Kerr PBHs, and then we analyze the effects of having a \emph{baroque} dark sector composed of a large number of particles, whose lightest particle is stable and thus constitutes the perfect candidate to be the DM present in the Universe.

\subsection{Direct Production}

\begin{figure}[t!]
 \includegraphics[width=0.475\textwidth]{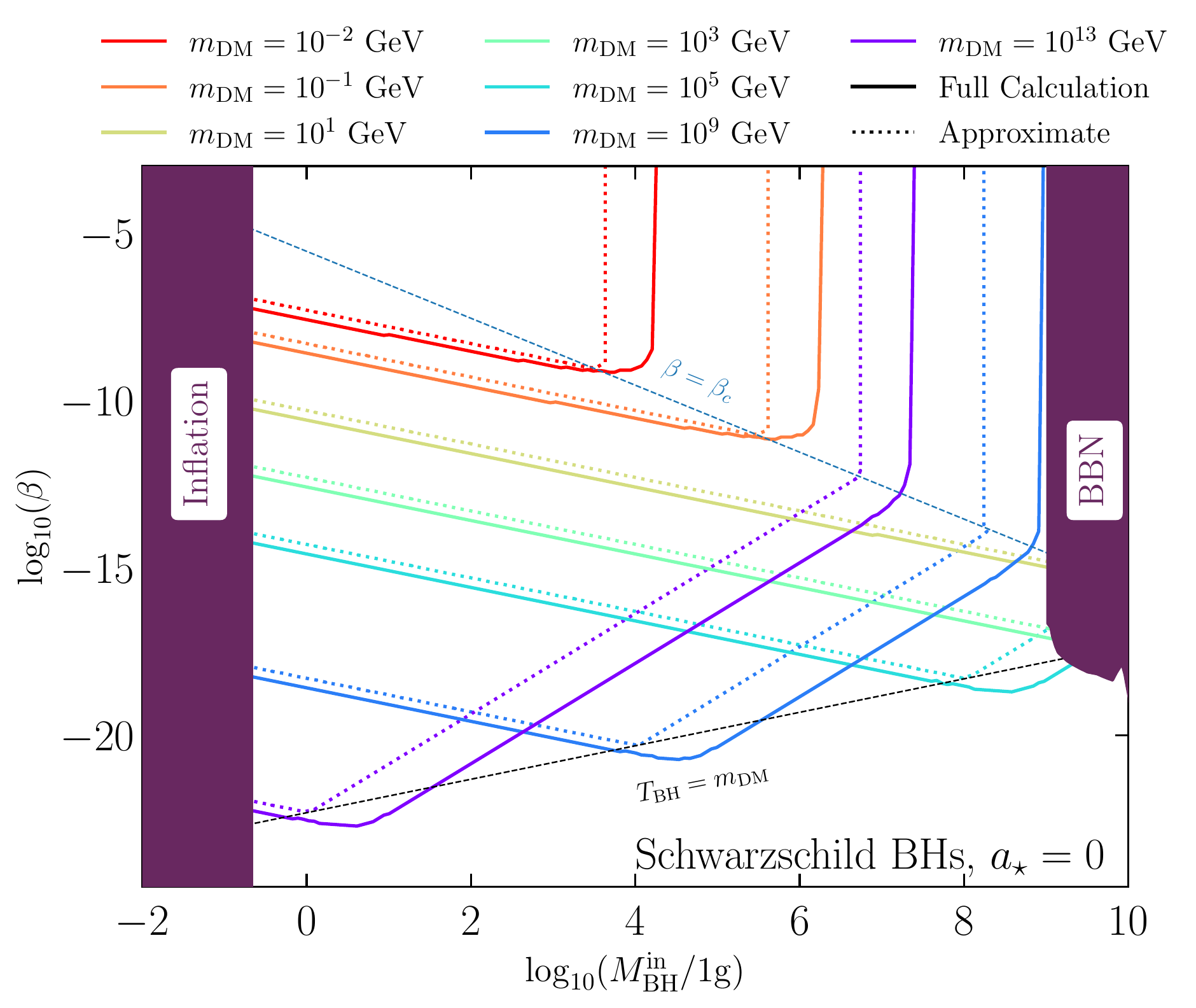}
 \caption{PBH energy fraction $\beta$ as a function of the PBH mass leading to the observed relic abundance $\Omega h^2=0.11$ for different values of the DM mass (in GeV). The dashed contours show the analytical estimations derived in previous works in the Geometric Optics limit, specifically from Ref.~\cite{Gondolo:2020uqv}, whereas the plain lines were derived in this work, including the full greybody factors. We assume the DM to be a fermion.}\label{fig:DM_PBH}
\end{figure}

In the case where PBHs are the only source of DM, the values of $\beta$ and $\MBHi$ leading to the correct relic abundance are indicated in Fig.~\ref{fig:DM_PBH} for various values of the DM mass. For any of those masses, a point above the corresponding coloured contours leads to an overproduction of DM ($\Omega h^2 > 0.11$) while DM is underproduced in points below the coloured contour.

In the limit where $\MBHi\to 0$, the Hawking temperature $\TBHi \propto (\MBHi)^{-1}$ is always larger than the DM mass. Therefore PBHs produce DM particles during the entire evaporation process. 
In that limit, the relic density of DM particles produced from of evaporation is linearly related to the fraction of PBHs $\beta$. 
A too-large value of this fraction leads to an overabundance of DM, which sets an upper bound on $\beta$. For larger PBH masses, $\TBHi$ might be smaller than the DM mass while PBHs still evaporate during a radiation-dominated era (this is typically the case for DM masses above $10^9~\GeV$). 
In that case, the larger $\MBHi$, the fewer DM particles are being produced during evaporation, which explains why the relic density contours go up after crossing the $\TBHi=\mDM$ line in Fig.~\ref{fig:DM_PBH}. 
For even larger $\MBHi$, PBHs dominate the universe energy density before they evaporate and reheat the SM bath at a temperature $T_{\rm ev}$. This is the case if their energy fraction $\beta$ at the time of PBH formation $T=T_{\rm in}$ is larger than $\beta_c\equiv T_{\rm ev}/T_{\rm in}$. 
In that case, the relic abundance of DM particles does not depend on the PBH fraction anymore but rather only on the PBH mass, this is reflected by the contours being vertical past the line $\beta=\beta_c$.
Interestingly, on the right of those vertical lines, PBHs can significantly reheat the Universe, and therefore modify the evolution of the SM thermal bath while not overproducing DM particles. 
Note that in most of the previous works, the contours depicted in Fig.~\ref{fig:DM_PBH} were derived analytically, ignoring the greybody factors~\cite{Gondolo:2020uqv} and/or fully tracking the Boltzmann equations, we indicate such a result with dashed lines. 
Our studies used the evaporation rates, including the full greybody factors, leading to significantly shifted contours towards larger PBH masses (plain coloured lines), assuming the DM to be fermionic. 
Since the Hawking rate departs from being a full blackbody spectrum because of the absorption probabilities, the number of emitted particles is larger than expected in the approximated purely-Planckian form. 
Moreover, the evaporation temperature is greater when including the greybody factors. 
Thus, we observe that smaller values of $\beta$ are required to give the correct relic abundance.

Let us notice that our results coincide qualitatively with those from Ref.~\cite{Auffinger:2020afu} for the case of a PBH dominated Universe, when fully including the greybody factors for the different spins. 
For a Universe where there was not PBH domination, but the BHs constitute an important contribution to the energy budget of the Universe, our results differ from those in Ref.~\cite{Auffinger:2020afu}.
Such a difference arises because in our code we always include the PBH contribution to the evolution of the Universe, which can alter the final relic density.

\begin{figure}[t!]
 \includegraphics[width=0.475\textwidth]{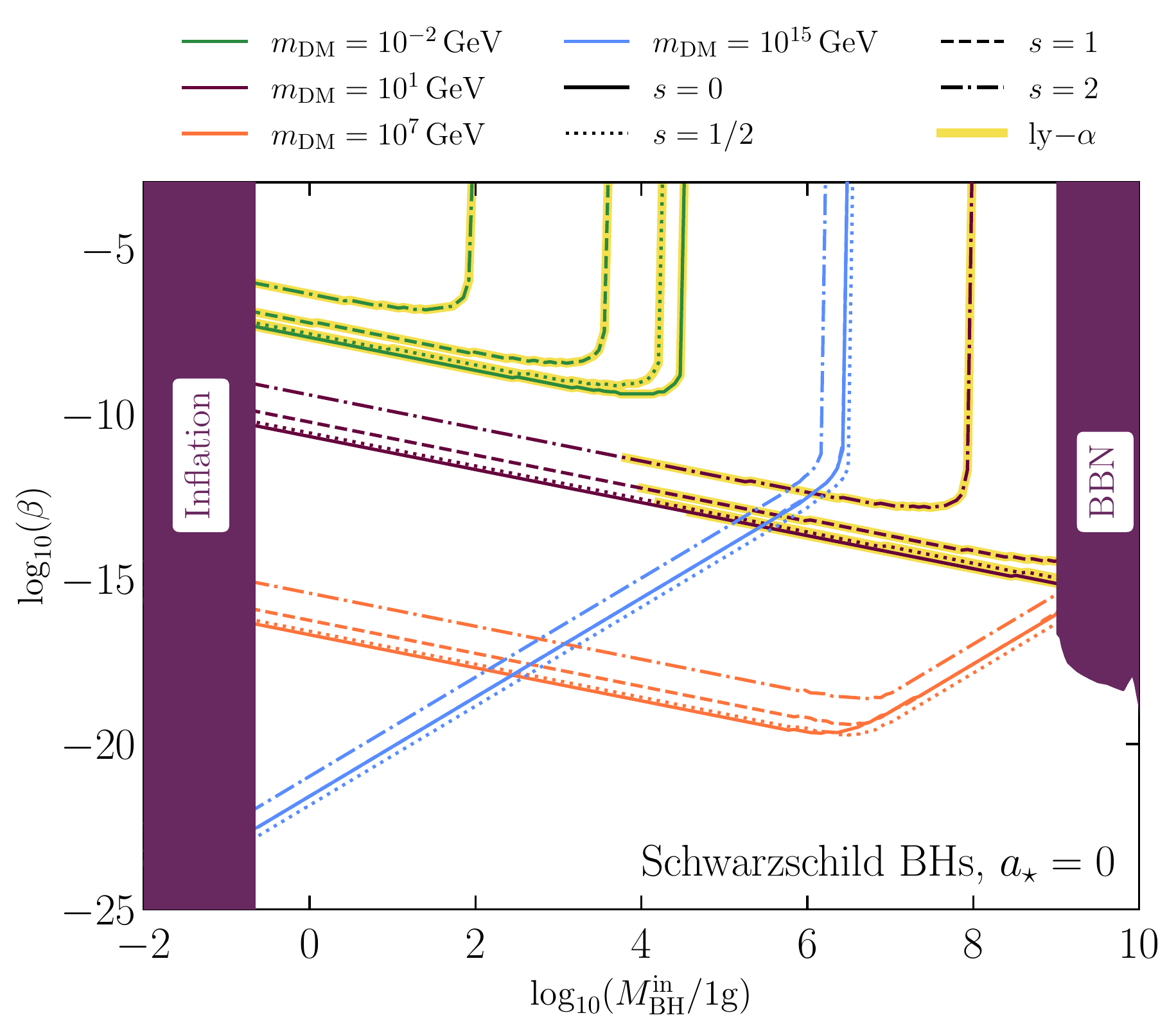}
 \includegraphics[width=0.475\textwidth]{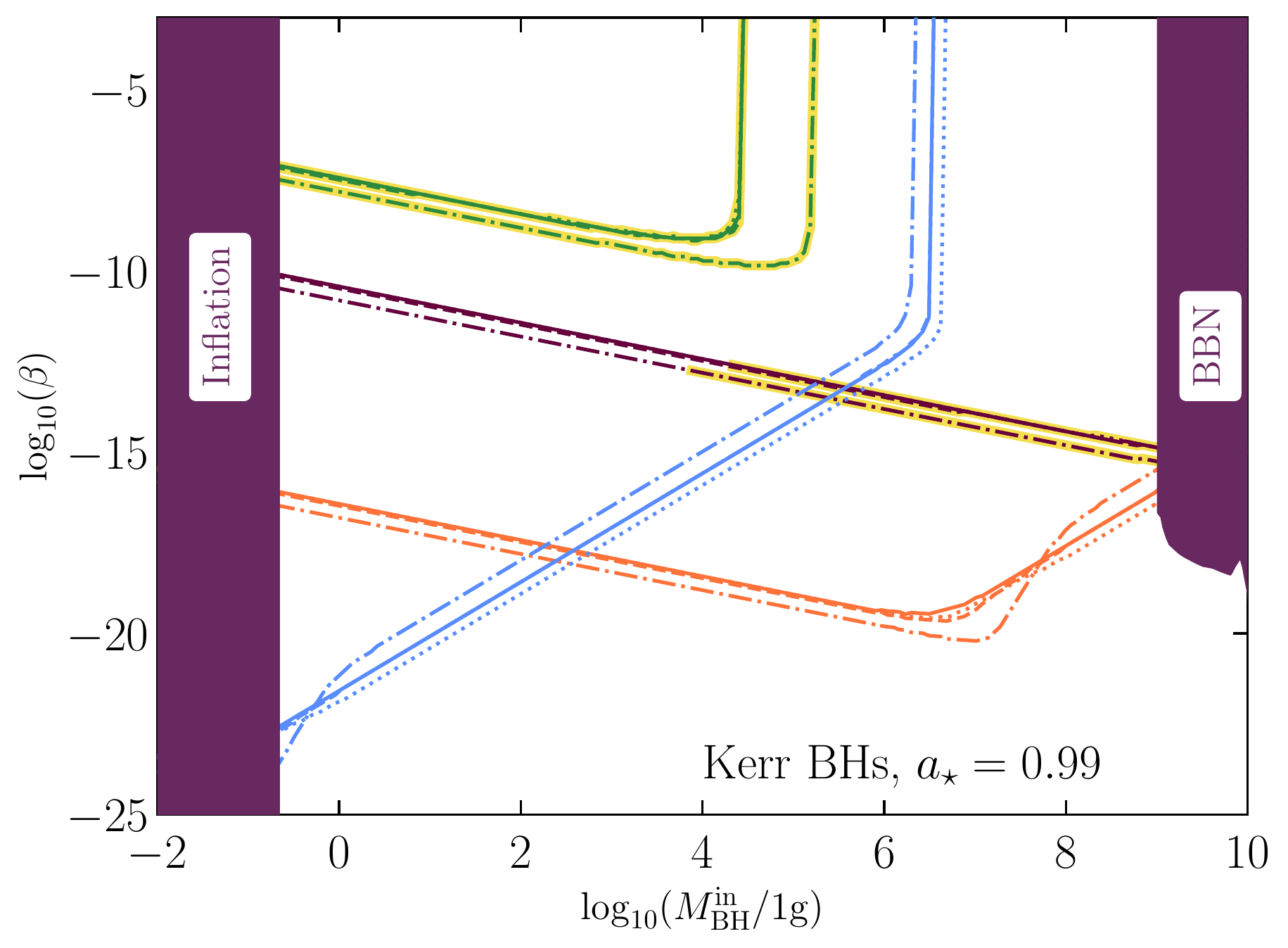}
 \caption{Similar to Figure \ref{fig:DM_PBH} but now showing only the results from the full numerical evaluation of the coupled Boltzmann equations. We show the results for four different intrinsic spins of the DM. \textbf{Upper panel} assumes all PBHs are Schwarzchild, $a_{\star}=0$. \textbf{Lower panel} assumes all PBHs are Kerr and approaching the maximal angular momentum, $a_{\star}=0.99.$ Parameters that do not fulfil structure formation constraints are indicated by marking the line with yellow. }
 \label{fig:DM_PBH_type_schw}
\end{figure}

Yet another effect of including the greybody factors correctly is that the relic abundance depends on the spin of the DM particle. In the top panel of Fig.~\ref{fig:DM_PBH_type_schw}, we show such dependence for several values of the DM mass and for spin $s_i=0,1/2,1,2$. For lighter DM masses relative to the initial BH mass, $\TBHi \gtrsim m_{\rm DM}$, we observe that larger values of $\beta$ are required to produce the correct $\Omega h^2$ for larger spins. Similar conclusions were found in Refs.~\cite{Lennon:2017tqq, Auffinger:2020afu} and it is a direct result of the suppression of the number of emitted particles for higher spins due to the greybody factors, see Eq.~\eqref{eq:psi0} and Fig.~\ref{fig:greyb}. On the other hand, when $\TBHi \lesssim m_{\rm DM}$, the cutoff induced by the non-zero mass affects the scalar case especially, such that the number of emitted particles is reduced in comparison to the case where the PBH temperature is higher than the DM mass. Hence, the required value of $\beta$ necessary to obtain the observed $\Omega h^2$ is larger and becomes similar to the values needed in the case of a Vector DM. As specified in the previous section, we have applied the warm dark matter constraint in the same figure. We indicate where the DM would violate the cold dark matter condition by marking the line with yellow. From this, we observe that for light masses, $\mDM \lesssim 10~\GeV$ for $\MBH^{\rm in}\gtrsim 10^4~\g$, the DM particles emitted from the evaporation are too hot, thus in tension with the observations~\cite{Baldes:2020nuv,Masina:2020xhk,Gondolo:2020uqv}. We find that our method proves to be more conservative than is reported in Refs.~\cite{Baldes:2020nuv,Masina:2020xhk,Gondolo:2020uqv,Auffinger:2020afu}. It is true that our procedure described in Sec.~\ref{sec:psd} is certainly more approximate than that of Refs.~\cite{Baldes:2020nuv, Auffinger:2020afu} and we leave the implementation of such methods in our code to future work. The warm dark matter constraint becomes less relevant for heavier masses, and for $\mDM \gtrsim 100~\GeV$, the full parameter space would obey such a limit. 

\subsection{Effect of the BH spin}

As mentioned in Section~\ref{subs:KBH}, Kerr PBHs could have a unique impact on the DM generation given the peculiar features present in such a case. Specifically, the enhanced emission of spin-2 particles that can compensate for the large initial fractions required to account for all the DM. In Fig.~\ref{fig:DM_PBH_type_schw}, bottom panel, we present the energy fraction $\beta$ as a function of the initial PBH mass for the Kerr case, assuming a value of $\as = 0.99$. Interestingly, we observe that, when $\TBHi \gtrsim m_{\rm DM}$ is valid in all the parameter space, the values of $\beta$ that give the correct relic abundance coincide for scalars, fermions and vectors. Such agreement is related to the increase of high-spin emission reflected in the greybody factors.
Moreover, the initial PBH fraction that gives the correct relic abundance for spin-2 DM is reduced by $\sim 2$ orders of magnitude with respect to the Schwarzschild case. For the case where $\TBHi \lesssim m_{\rm DM}$, a similar behavior to the non-rotating case is present; the emission cutoff due to the DM mass diminishes the overall particle production, specifically for scalars and fermions. Nevertheless, for tensor DM, there is an interesting effect when $\TBHi \lesssim m_{\rm DM}$. Even though a large Boltzmann suppression is still present, the enhanced emission of tensor particles~\cite{Dong:2015yjs} is a significant countervailing effect, which leads to enhanced particle production. In Fig.~\ref{fig:DM_PBH_type_schw}, such amplification is responsible for the structure observable for $m_{\rm DM} = 10^7~\GeV\,(10^{15}~\GeV)$ for BH masses of $\MBH^{\rm in} \sim 1 - 10~{\g}~(10^{8} - 10^{9}~{\g})$. Such an effect is also present for scalars, fermions, and vectors, although much less conspicuously\footnote{In the first version of this manuscript the tensor results saw a larger deviation. This was due to an numerical error which has been corrected.}. Ref.~\cite{Masina:2021zpu} also investigated the affect of Kerr BHs on DM production, focusing predominantly on light $m_{\rm DM}$, however one can observe the amplification of Tensor particle production at high $a_{\star}$ in their Fig.~10. Finally, regarding the warm DM constraint, we observe that the BH spin increases the parameter space that is excluded by such a limit in comparison to the non-rotating case, this is also in agreement with Ref.~\cite{Masina:2021zpu}. Still, we have demonstrated that the DM production from Kerr BHs has many compelling features not encountered before.

\subsection{Indirect Production: Presence of additional dark sector particles}\label{subsec:ind}

The DM could be part of a much larger \emph{dark sector}, containing a large quantity of particles. Such a baroque scenario should not be inconceivable from what we have learnt about the SM sector. Indeed, supersymmetric (SUSY) models constitute the perfect example of UV complete scenarios that are expected to contain many additional degrees of freedom~\cite{Green:1999yh}. Let us assume that the DM particle belongs to an extended sector that does not interact with the SM. Moreover, for simplicity, let us consider that just one particle is stable, just like the lightest superpartner in SUSY with some R-parity. Suppose that there are $i$ copies of $X$ particles, where $X=\{S,F,V,G\}$ indicates whether the particles are scalars, fermions, vectors or tensors, respectively. The total number of final DM particles produced via PBH evaporation $N_{\rm DM}^{\rm tot}$ will be the sum of all emitted particles 
\begin{align}
    {\cal N}_{\rm DM}^{\rm tot} = {\cal N}_{\rm DM} + \sum_i\sum_X \mathfrak{n}_{X_i\to\rm DM}\, {\cal N}_{X_i}\,,
\end{align}
being $\mathfrak{n}_{X_i\to\rm DM}$ the number of DM particles resulting from the decay of $X_i$, such that $\mathfrak{n}_{X_i\to\rm DM}\geq 2$. Following our previous analytical estimation of the final relic abundance, we can examine the enhancement of $\Omega h^2$ with respect to the case where there is just the DM particles,
\begin{align}
    \frac{\left.\Omega_{\rm DM} h^2\right|_{\rm DM+X_i}}{\left.\Omega_{\rm DM} h^2\right|_{\rm DM}}=\frac{g_{\star S}(T_{\rm ev})}{g_{\star S}(T_{\rm ev}^t)}\left(\frac{T_{\rm ev} a_{\rm ev}}{ T_{\rm ev}^t a_{\rm ev}^t}\right)^3\frac{{\cal N}_{\rm DM}^{\rm tot}}{{\cal N}_{\rm DM}}\,,
\end{align}
where $T_{\rm ev}^t, a_{\rm ev}^t$ are the Universe temperature and scale factor at evaporation in the extended dark sector case. From this, we observe that the effect of having additional dark sector particles is twofold. First, the increase of DM particles evidently enlarges the final relic density. Second, since the emission of the additional particles affects the BH lifetime, the Universe properties when the PBHs evaporate are changed, and thus $\Oh$.

Let us be more specific and consider the situation in which the dark sector is only composed by the DM particle and a heavier state $X$, assuming for simplicity one decay channel, $X\to {\rm DM + DM}$, such that $\mathfrak{n}_{X\to\rm DM} = 2$. In order to be consistent in our treatment, we solve the same set of Eqs.~\eqref{eq:FBEqs}, plus the following equations for the number density of $X$ and DM including the exchange terms
\begin{subequations}
\begin{align}
\dot{n}_{\rm DM} + 3H n_{\rm DM} &= n_\bh\, \Gamma_{\rm BH \to DM} + 2\, \left\langle\Gamma_{X\to {\rm DM}}\right\rangle_{\rm ev} n_X\,,\\
    \dot{n}_{X} + 3H n_{X} &= - \left\langle\Gamma_{X\to {\rm DM}}\right\rangle_{\rm ev} n_X + n_\bh\, \Gamma_{{\rm BH} \to X}\,,
\end{align}
\end{subequations}
with the thermally averaged decay width of $X$ given by
\begin{align}
    \left\langle\Gamma_{X\to {\rm DM}}\right\rangle_{\rm ev} = \Gamma_{X\to {\rm DM}}\left\langle \frac{m_X}{E_X}\right\rangle_{\rm ev}\,,
\end{align}
where ``\rm{ev}'' indicates that the average is taken with respect to the BH temperature, and $\Gamma_{X\to {\rm DM}}$ the decay width of X in vacuum\footnote{Clearly, the decay width depends on the particle nature of X, that is, on whether it is a scalar, vector or massive tensor. We provide the specific decay widths assumed here in the App.~\ref{ap:A1}.} (for further details, see the companion paper~\cite{paperB}).

\begin{figure}[t!]
 \includegraphics[width=0.475\textwidth]{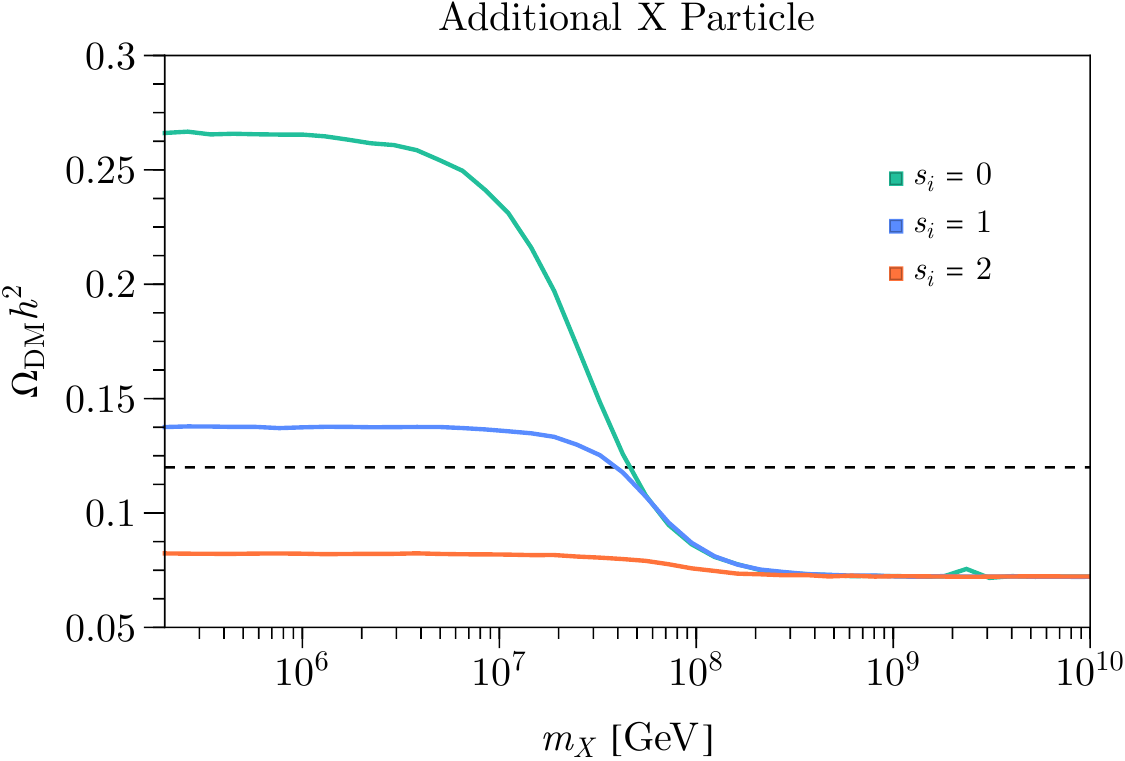}
 \caption{Relic abundance as a function of the mass of an additional $X$ heavy state decaying into DM for different assumed values of the spin of such particle, scalar (emerald), vector (light blue), and massive tensor (orange). We assume $\MBH^{\rm in} = 10^6~\g, m_{\rm DM} = 10^5~\GeV, $ and $\beta^\prime = 10^{-17.75}$. The horizontal dashed line indicates the observed value of $\Oh$. \label{fig:Xfig}}
\end{figure}

In Fig.~\ref{fig:Xfig}, we present $\Omega h^2$ for a Fermionic DM as a function of the mass of $X$. We show the result for different types of $X$, scalar (emerald), vector (light blue), and massive tensor (orange), for $m_{\rm DM} = 10^5~\GeV$, $\beta^\prime = 10^{-17.75}$, and $\MBH^{\rm in} = 10^6~\g$. Without accounting for greybody factors, one could expect that $\Oh$ should increase by a factor of 3 since the $X$ would decay into two DM particles. However, the more accurate calculation leads to enhancements of $\sim 3.7$, $\sim 1.9$ and $\sim 1.3$ for a scalar, vector or tensor $X$ respectively. Once more, we are seeing the greybody factors affect the emission of higher spin particles more significantly, reducing the contribution of $X$ to the total. We note that $\MBHi=10^6\,\rm{g}$ means that $\TBHi\sim 10^7\,\rm{GeV}$, suggesting that the suppression of $X$ particle emission should occur when $m_{X}\gtrsim 10^7\,\rm{GeV}$. This is corroborated by Fig.~\ref{fig:Xfig} and we see that by $m_{X}\sim 2\times 10^8\,\rm{GeV}$ enhancements of $\Omega h^2$ from $X$ decay is negligible. Of course this suppression is independent of the particle's spin, so we see the behaviour across the three cases in the figure. 

Interestingly, once there is a large enough separation of scales between $X$ and DM, the warm DM bounds need to be considered once more. Unlike previously, we now can have a mix of cold and warm DM from Hawking emission and dark sector decay respectively. According to Ref. \cite{Baur:2017stq, Auffinger:2020afu} when the fraction of warm/cold DM is less than $\sim 0.2$, constraints from structure formation do not apply. Despite this, even when the parent dark sector particles do not contribute much, such as the $X=G$, Schwarzschild PBHs case, the fraction of DM particles which could be warm is sufficiently large $\sim 0.3$. This has implications for the coexistence of dark sectors that contain a dark matter candidate as its lightest member and PBHs. For the decay products, the average momentum is given by
\begin{align}\label{eq:decay_mom}
    \langle p_{\rm DM}^{\rm dec.prod.}\rangle = \left(\frac{m_X^2 + \langle p_X \rangle^2}{\mathfrak{n}_{X_i\to\rm DM}^2} -m_{\rm DM}\right)^{1/2}
\end{align}
where $\langle p_X \rangle$ is calculated following Eq.~\eqref{eq:ave_mom} but for the parent particle. Plugging Eq.~\eqref{eq:decay_mom} into Eq.~\eqref{eq:wdm} gives the velocity today which can be compared with Lyman-$\alpha$ constraints\cite{Baur:2017stq, Baldes:2020nuv}. By taking the setup in Fig. \ref{fig:Xfig}, we find that $m_X$ would have to be above $10^{10}$ GeV to produce warm dark matter with mass $10^5$ GeV. At which point, the constraint would be irrelevant because the contribution decayed DM has on the relic abundance is negligible. The scale separation that leads to a heavy warm dark matter component is highly dependent on the time of BH evaporation, the later evaporation occurs the less the particles will red-shift. For example, taking  the Fig. \ref{fig:Xfig} setup but $M_{\rm PBH}^{\rm in}=10^8$ g, now $m_{X}\geq 10^8$ GeV produces warm DM when $m_{\rm{DM}}=10^5$ GeV.

\begin{figure}[t!]
 \includegraphics[width=0.475\textwidth]{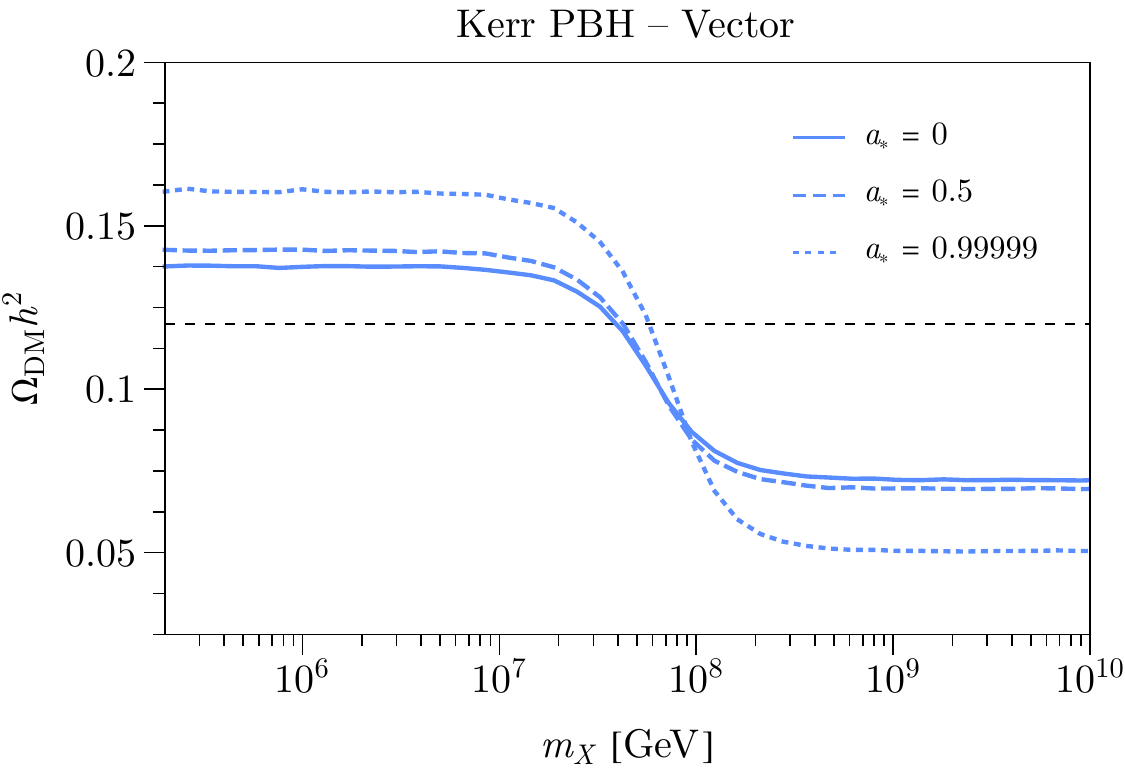}
 \caption{Relic abundance as a function of the mass of an additional heavy vector $X$ decaying into DM for different assumed values of the PBH spin parameter, $\as = 0.$ (full), $\as = 0.5$ (dashed), and $\as = 0.99999$ (dotted). We assume $\MBH^{\rm in} = 10^6~\g, m_{\rm DM} = 10^5~\GeV, $ and $\beta^\prime = 10^{-17.75}$. The horizontal dashed line indicates the observed value of $\Oh$. \label{fig:KXfig}}
\end{figure}

If the PBH population had an initial non-zero angular momentum, we find that the spin of $X$ plays a crucial role in the final $\Oh$. We present the relic abundance as a function of $m_X$ in Fig.~\ref{fig:KXfig} for three different values of $\as=\{0, 0.5, 0.99999\}$ corresponding to full, dashed and dotted lines, respectively, assuming the heavier state to be a vector and considering the same parameters as in Fig.~\ref{fig:Xfig}. We find two different effects at play here; when the particle $X$ is kinematically accessible by the evaporation, the indirect DM production is largely enhanced because of the PBH spin. Meanwhile, the relic density is decreased when $m_{X}\gtrsim \TBHi$ in comparison to the Schwarzschild case since Kerr PBHs inject much more entropy to the early Universe due to the amplified production of SM boson states. From such effects, we have that the increase in the final relic density is $\sim\{1.9, 2.0, 2.2\}$ for $\as=\{0, 0.5, 0.99999\}$ with reference to the Schwarzschild value without any additional state, respectively. Such an augmentation is more stringent if the $X$ particle has a spin of 2, reaching a value of $\sim 4.3$ for $\as=0.99999$. Thus, one can see how having a rich dark sector at high masses can quickly overclose the Universe even if there is a tiny number of PBHs in the early Universe.

\section{Conclusions}\label{sec:conclus}

Black holes are one of the most fascinating objects predicted by General Relativity. Initially thought to be everlasting, we have learnt that they instead evaporate by emitting a thermal flux of particles, losing simultaneously their mass and angular momentum. Such evaporation in the early Universe could have critical consequences on our understanding of how the Universe came to be what we observe. In particular, since the Hawking radiation is \emph{democratic} in nature, i.~e., BHs emit all existing degrees of freedom in nature, the observed relic abundance could be the result of the evaporation of PBHs, even in the case that the DM only interacts gravitationally.

In this paper, we have addressed distinct effects that impact the DM production by PBHs in the purely gravitationally interacting scenario thoroughly. We have solved the system of Friedmann - Boltzmann equations, and investigated systematically the distinct features present in this scenario for both Schwarzschild and Kerr PBHs.  Especially, after including consistently the greybody factors in the description, we have demonstrated how the DM relic abundance can depend on the particle's spin, in such a way that the initial PBH fraction necessary to obtain the observed values is $\sim 2$ orders of magnitude larger for massive tensors than for scalar DM. Besides, by correctly including the mass cutoff due to Boltzmann suppression, we have identified the modifications of the required fractions when the initial PBH temperature is smaller than the DM mass. In such a case, the emission only occurs in the last stages of the BH lifetime. Regarding the warm DM bounds that affect this scenario, we have computed the average momenta of the emitted particles, finding it to be larger than estimated before because of the energy dependence present in the greybody factors. Light DM masses are thus in tension with small scale structure measurements, similarly to the previous results present in the literature. We have also illustrated the properties of DM production in the case that PBHs had an initial non-zero angular momentum. The enhancement of the emission expected for bosons, particularly for spin-2 particles, reduces the initial fractions needed to generate the observed DM. Interestingly, we have identified the regions in the parameter space where the enhanced Hawking emission due to the BH spin plays a significant role in the particle production. 

Finally, we have analysed the impact of having a large dark sector containing a unique stable particle, the DM candidate. For such models, PBHs would also emit the additional unstable particles of the dark sector during its evaporation, will would produce an additional surplus of DM particles. Such indirect production alters not only the number of DM particles during the PBH evaporation but also the PBH lifetime and impacts the Universe's evolution via entropy injection. We scrutinized a minimal scenario where there exist just one additional heavier particle that decays into the DM. In this case, we found that the increase on the relic abundance can be as large as a factor $\sim 4$ in the case that the heavy particle is a scalar. For other types of spins, the factor is smaller. This dependence on the spins is simply understood as the effect due to the greybody factors. In this regard, we also investigated the indirect mechanisms for Kerr PBHs, finding, as expected, an enhancement by a factor of $\sim 4.3$ for the tensor case when the PBHs initially had a close-to-maximal angular momentum. Assuredly, an extended dark sector can lead to a rich phenomenology. Moreover, if we assume the existence of interactions with the SM, there could be significant modifications to the results presented here. Such a treatment is left for the second part of this series~\cite{paperB}.

\appendix

\section*{Acknowledgments}
The authors would like to thank Robert Bird for useful discussions. This manuscript has been authored by Fermi Research Alliance, LLC under Contract No. DE-AC02-07CH11359 with the U.S. Department of Energy, Office of Science, Office of High Energy Physics. AC is supported by the F.R.S.-FNRS under the Excellence of Science EOS be.h project n. 30820817. Computational resources have been provided by the supercomputing facilities of the Université Catholique de Louvain (CISM/UCL) and the Consortium des Équipements de Calcul Intensif en Fédération Wallonie Bruxelles (CÉCI) funded by the Fond de la Recherche Scientifique de Belgique (F.R.S.-FNRS) under convention 2.5020.11 and by the Walloon Region. The work of LH is funded by the UK Science and Technology Facilities Council (STFC) under grant ST/P001246/1.

\section{Analytic derivation of PBH emission properties}\label{ap:A}


\subsection{Schwarzschild case}

Here we go through the analytic derivation of the emission rates and total number of particles of Schwarzschild BHs including greybody factors. The Hawking spectrum, for a given particle species, $i$ is parametrized as
\begin{equation}
    \frac{\dd[2]{\mathcal N_i}}{\dd{E}\dd{t}} = \frac{27g_i G^2\MBH^2}{2\pi}\frac{\psi_{s_i}(E)(E^2-\mu_i^2)}{\exp(E/\TBH)-(-1)^{2s_i}}\,,
\end{equation}
where $g_i$ is the internal d.o.f, $s_i$ is the spin and $\psi_{s_i}(E)$ is the absorption cross section normalized to the geometric optics limit, and $G$ and $\MBH$ are the gravitational constant and the mass of the BH respectively. Introducing the dimensionless parameters $x\equiv E/\TBH$ and $z_i=\mu_i/\TBH$, the total emission rate per particle species is
\begin{equation}\label{eq:GamSBH}
    \Gamma_{{\rm BH} \to i}=  \frac{27 g_s}{1024\pi^4}\frac{1}{G\MBH} \underbrace{\int_{z_i}^\infty \frac{\psi_{s_i}(x)(x^2-z_i^2)}{\exp(x)-(-1)^{2s_i}} \dd{x}}_{\Psi_i (z_i)}\,,
\end{equation}
where for now we simply take the integral result unspecified as $\Psi_i(z_i)$. An assumption that is often made is that $\psi(x,z)=1$, that is, take the greybody factors equal to the geometrical-optics limit, which allows one to perform the integral analytically, 
\begin{equation}
    \Psi_i (z)=2\epsilon_i\left[z\text{Li}_2(\epsilon_i e^{-z})+\text{Li}_3(\epsilon_i e^{-z})\right]\,,
\end{equation}
being ${\rm Li}_n$ the polylog functions of order $n$, and $\epsilon_i=(-1)^{2 s_i}$. We then have
\begin{align}
     \Gamma_{{\rm BH} \to i}
     &=\frac{27 g_s}{512\pi^4}\frac{\epsilon_i}{G\MBH}\left[z\text{Li}_2(\epsilon_i e^{-z})+\text{Li}_3(\epsilon_i e^{-z})\right]\,.
\end{align}
Therefore, under this assumption and taking $\mu_i=0$
\begin{equation}
    \Gamma_{{\rm BH} \to i}=\frac{27g_s}{32\pi^3}\zeta(3)\begin{cases}
    1 & \text{for Bosons}.\\
    3/2 & \text{for Fermions}
    \end{cases};
\end{equation}
this allows one to make a comparison between the calculation with the full greybody factors in the massless limit. 

We can carry out a similar procedure for the evaporation function $\varepsilon_i(z_i)$ per particle species, defined by
\begin{align}
    \varepsilon_i(z_i)&\equiv-\frac{\MBH^2}{\MPL^4}\frac{\dd{\MBH}}{\dd{t}},\notag\\ 
    &=\frac{27 g_i}{8192\pi^5}\int_{z_i}^\infty\frac{\psi_{s_i}(x^2-z^2)}{\exp(x)-(-1)^{2s_i}}x\dd{x}\,,
\end{align}
where we now have defined the function $\varepsilon_i(z_i)$ in a similar fashion to $\Psi_i(z)$. Its fairly straightforward to obtain the massless geometric optics limit for $\varepsilon_i(0)$,
\begin{equation}
    \varepsilon_i(0)=\frac{27 g_i}{8192\pi^5}\begin{cases}
    \frac{\pi^4}{15} & \text{for Bosons}\\
    \frac{7 \pi^4}{120} & \text{for Fermions}\,,
    \end{cases}
\end{equation}
so 
\begin{equation}
    \dv{M}{t}=-\frac{27}{4}\frac{1}{30720\pi}\frac{\MPL^4}{M^2}g_{\star}(\TBH),
\end{equation}
in agreement with Ref.~\cite{Baldes:2020nuv}.

\begin{figure*}[t!]
 \includegraphics[width=0.75\textwidth]{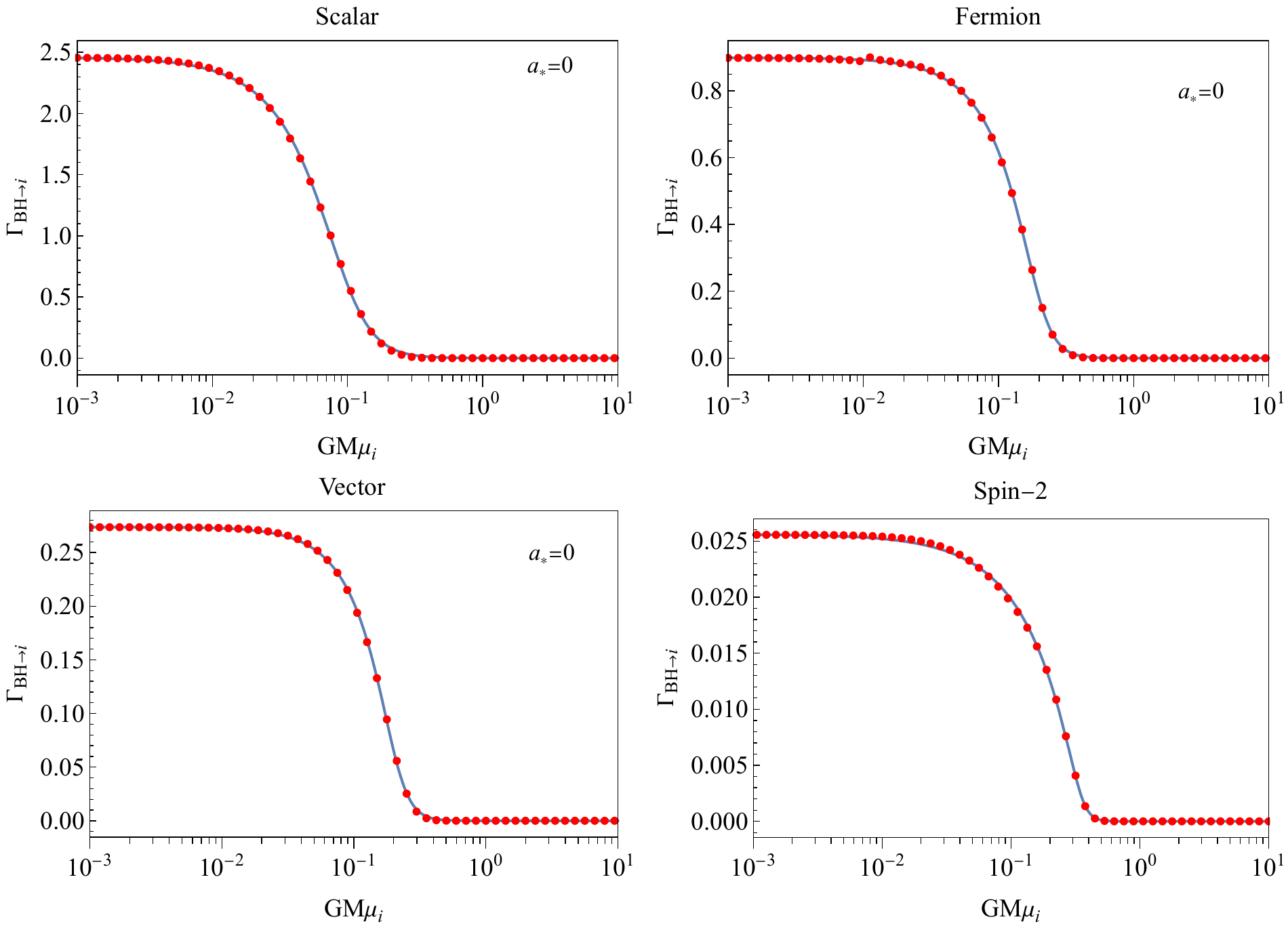}
 \caption{Total emission rate as function of $GM\mu_i$ for the different types of particles, scalars, fermions, vectors and spin-2 for a Schwarzschild BH. The red points correspond to the values obtained directly by integration of Eq.~\eqref{eq:GamSBH}, while the blue lines are our fitted forms. \label{fig:fitS}}
\end{figure*}
To find the total number of emitted particle, we need to integrate over the lifetime, $\tau$ of the BH. 
\begin{equation}
    \mathcal{N}_i=\int_0^{\tau}\dd{t}\dv{\mathcal{N}_i}{t},
\end{equation}
where we have chosen the time the BHs are formed to be $t=t_{\rm in}=0$. Using the mass loss rate eq.\eqref{eq:dMBHdt} we can make a change of variables 
\begin{equation}
    \mathcal{N}_i = \frac{27g_s}{1024\pi^4} G \int_{0}^{\MBH^{\rm in}}\frac{\Psi_i(z)}{\varepsilon(M)} M\dd{M}\,,
\end{equation}
where $\varepsilon\equiv \sum_i g_{s_i}\varepsilon(\MBH)$. For $z=0$, and taking the Geometric Optics limits, 
\begin{equation}
    \Psi_i(0)=2\zeta(3),\quad\quad \varepsilon(M)=\frac{27}{4}\frac{1}{30720\pi}g_{\star}(T),
\end{equation}
one recovers the results from Refs.~\cite{Baldes:2020nuv,Gondolo:2020uqv}
\begin{align}
    \mathcal{N}_i=\frac{120\zeta(3)}{\pi^3}\frac{g_s}{g_{\star}(\TBH^{\rm in})}\left(\frac{\MBH^{\rm in}}{\MPL}\right)^2\,.
\end{align}
To keep the greybody factors in the equation we can rewrite $\varepsilon(\MBH)$ as
\begin{equation}
    \varepsilon(\MBH)= \sum_j g_{s_i}\varepsilon_j\left(\frac{\mu_j}{\TBH}\right)=\sum_j g_{s_j}\varepsilon_{j}\left(\underbrace{\frac{\mu_j}{\mu_i}}_{\mathfrak{m}_i}\frac{\mu_i}{\TBH}\right).
\end{equation}
Writing $\varepsilon(z_i) = \sum_j g_{s_j}\varepsilon_j(\mathfrak{m}_j z_i)$ and using that $z_i=8\pi G M \mu_i$, we obtain
\begin{align}
    \mathcal{N}_i 
    &= \eta(z_i^{\rm in})\frac{g_s}{g_{\star}(\TBH^{\rm in})} \left(\frac{\MBH^{\rm in}}{\MPL}\right)^2\,,
\end{align}
where 
\begin{equation}
     \eta(z_i^{\rm in})=\frac{27}{1024\pi^4}\frac{1}{\left(z_i^{\rm in}\right)^2} g_{\star}(\TBH^{\rm in})\int_0^{z^{\rm in}}\frac{\Psi(z)}{\sum_i g_{j}\varepsilon_j(\mathfrak{m}_j z)}z\dd{z}\,.
\end{equation}
\begin{table*}[t]
\caption{Fitting parameters for our analytical form, Eq.~\eqref{eq:fitf}, in the Schwarzschild case.  \label{tab:pars}}
    \vspace{1mm}
    \centering
    \begin{tabular}{c|cccc|cccc}
        \toprule
         & \multicolumn{4}{c}{$\Psi_i$} & \multicolumn{4}{c}{$\varepsilon_s$} \\
         & $A_s$ & $B_s$ &  $C_s$ & $\nu_s$ & $A_s$ & $B_s$ &  $C_s$ & $\nu_s$ \\ \colrule
        Scalar & $2.457$ & $7.50218$ & $2.9437$ & $0.4208$ & $7.61\times 10^{-5}$ & $7.79884$ & $3.80742$ & $0.4885$\\ 
        Fermion & $0.897$ & $12.3573$ & $8.7436$ & $0.3045$ & $4.12\times 10^{-5}$ & $13.0496$ & $9.91178$ & $0.3292$\\ 
        Vector & $0.2736$ & $13.465$ & $9.8134$ & $0.3049$ & $1.68\times 10^{-5}$ & $14.0361$ & $10.7138$ & $0.3072$\\ 
        Graviton & $0.0259$ & $22.325$ & $21.232$ & $0.1207$ & $1.93\times 10^{-6}$ & $21.5094$ & $20.5135$ & $0.1734$\\\botrule
    \end{tabular}
\end{table*}
In order to obtain a semi-analytic approximation all functions, $\Psi_i(z)$ and $\varepsilon_i(z)$ have been fitted to a generalized logistic form
\begin{equation}\label{eq:fitf}
    A_s\left\{1-(1+\exp\{-B_s\log_{10}(z)+C_s\})^{-\nu_s}\right\}\,,
\end{equation}
where the parameters $\{A_s,B_s,C_s,\nu_s\}$ depend on the spin of the particle. We give such parameters in Tab.~\ref{tab:pars}, and we present an example for the fit function for $\Gamma_{{\rm BH} \to i}$ for all particle types together with the values obtained by direct integration in Fig.~\ref{fig:fitS}.\\

\subsection{Kerr case}

\begin{figure*}[t!]
 \includegraphics[width=0.75\textwidth]{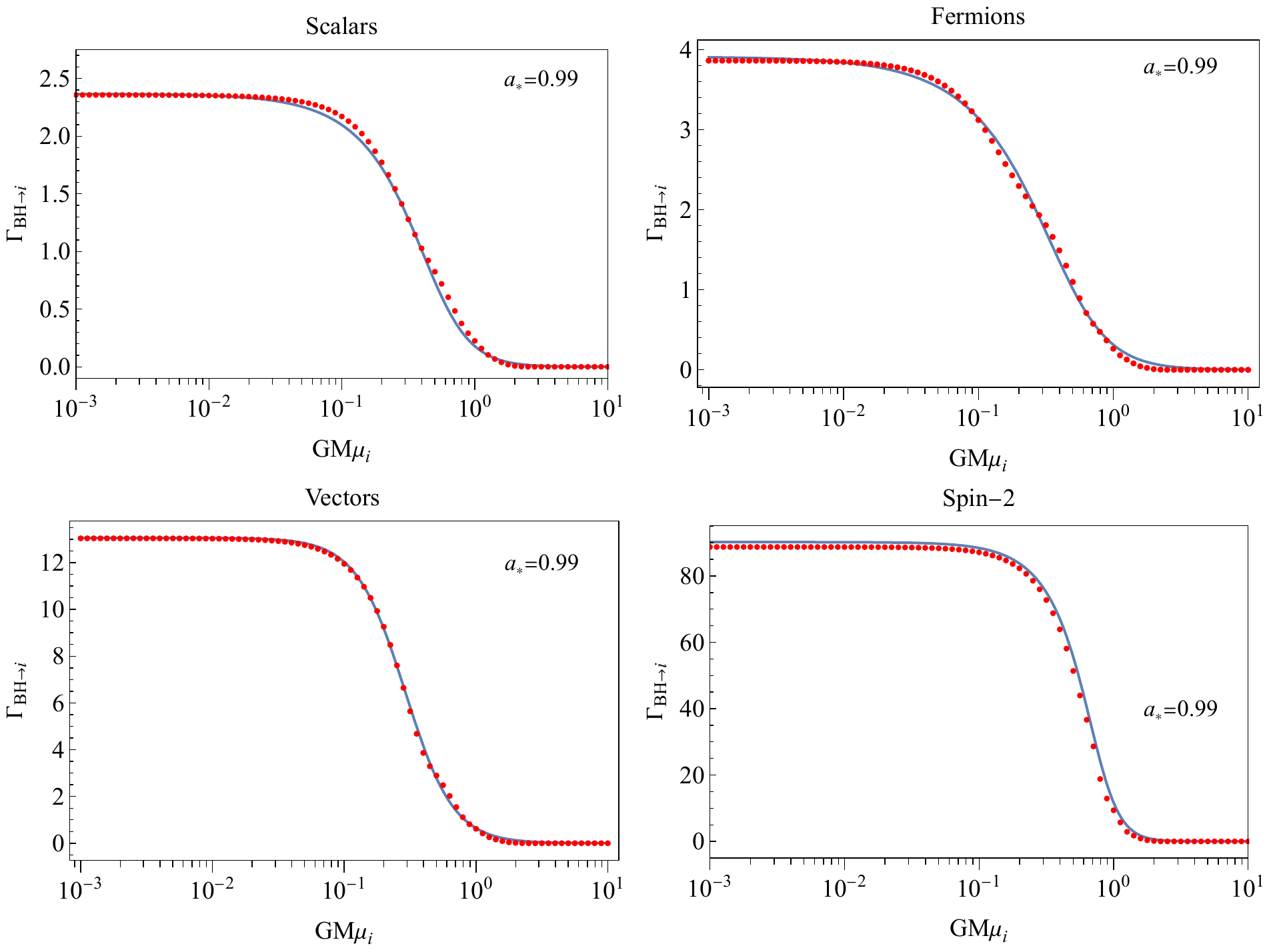}
 \caption{Total emission rate as function of $GM\mu_i$ for the different types of particles, scalars, fermions, vectors and spin-2 for a Kerr BH with $a_\star=0.99$. The red points correspond to the values obtained directly by integration of Eqs.~\eqref{eq:GamKBHA} and \eqref{eq:GamKBHB}, while the blue lines are our fitted forms. \label{fig:fitK}}
\end{figure*}

For initially rotating BHs we can perform a similar analysis. The Hawking rate is modified because of the presence of the non-zero angular momentum, 
\begin{widetext}
\begin{align}
\frac{\dd^2 \mathcal{N}_{i}}{\dd E_i\dd t}&=\frac{27g_i G^2\MBH^2}{2\pi} \sum_{l=s_i}\sum_{m=-l}^l\frac{\psi_{s_i}^{lm}(\MBH,p,a_\star)(E_i^2-\mu_i^2)}{\exp\left[(E_i - m\Omega)/\TBH\right]-( -1)^{2s_i}}\,,
\end{align}
being $\psi_{s_i}^{lm}(\MBH,p,a_\star)$ the greybody factor dependent associated to the partial wave with quantum numbers $l,m$, and normalized to $27\pi^2 G^2\MBH^2$. The total emission rate, obtained after integration over the energy, is
\begin{align}\label{eq:GamKBHA}
    \Gamma_{{\rm BH} \to i}= \frac{27 g_i}{1024\pi^4}\frac{1}{G\MBH}\Psi_i (z_i,\as)\,,
\end{align}
being
\begin{align}\label{eq:GamKBHB}
    \Psi_i (z_i,\as)= \int_{z_i}^\infty\sum_{lm}\frac{\,\psi_{s_i}^{lm}(x,\as)(x^2-z_i^2)}{\exp[(x(1+\sqrt{1-\as^2})-4\pi m\as)/2\sqrt{1-\as^2}]-(-1)^{2s_i}}\dd{x},
\end{align}
where $x = E_i/\TBH^S$ and $z_i=\mu_i/\TBH^S$, being $\TBH^S=(8\pi G \MBH)^{-1}$ the temperature for \emph{Schwarzschild} BHs. After considering the greybody factors for each spin type, and integrating numerically, we have performed a fit to $\Psi_i (z_i,\as)$ in the form 
\begin{equation}\label{eq:fitfK} A_s(\as)\left\{1-(1+\exp\{B_s(\as)\log_{10}(z_i/8\pi)+C_s(\as)\})^{-\nu_s(\as)}\right\},
\end{equation}
where now $A_s(\as),B_s(\as),C_s(\as),\nu_s(\as)$ are functions of $\as$, and are fitted according to the functions,
\begin{subequations}\label{eq:ABCn-Kerr}
\begin{align}
    \log_{10} A_s(\as) &= \frac{\alpha^s_5 \as^2}{(\as^2-1.025)^2} + \sum_{j=0}^4 \alpha^s_j\as^j,\\
    B_s(\as) &= \frac{\beta^s_5 \as^2}{(\as^2-1.025)^2} + \sum_{j=0}^4 \beta^s_j\as^j,\\
    \log_{10} C_s(\as) &= \frac{\eta^s_5 \as^2}{(\as^2-1.025)^2} + \sum_{j=0}^4 \eta^s_j\as^j,\\
    \log_{10} \nu_s(\as) &= \frac{\delta^s_5 \as^2}{(\as^2-1.025)^2} + \sum_{j=0}^4 \delta^s_j\as^j\,.
\end{align}
\end{subequations}
The fitting parameters $\{\alpha^s_j,\beta^s_j,\eta^s_j,\delta^s_j\}$, $j=1,\ldots,5$ for each spin are given in Tables~\ref{tab:parsKS}-\ref{tab:parsKG}. We present an example for the fit function for $\Gamma_{{\rm BH} \to i}$ for all particle types together with the values obtained by direct integration in Fig.~\ref{fig:fitK}.

\begin{table*}
\subfloat[$A_s(\as)$\label{tab:table1_aS}]{
\centering
\begin{tabular}{c|r|r|r}
        \toprule
         & \multicolumn{1}{c|}{$\Psi$} & \multicolumn{1}{c|}{$\varepsilon$} &  \multicolumn{1}{c}{$\gamma$} \\ \colrule
        $\alpha_0$ & $3.89166$ & $-4.11848$ & $-4.04521$\\ 
        $\alpha_1$ & $-0.03924$ & $-0.41827$ & $-0.25175$\\ 
        $\alpha_2$ & $0.59957$ & $2.58436$ & $2.31410$\\ 
        $\alpha_3$ & $-2.30988$ & $-5.76425$ & $-3.47358$\\ 
        $\alpha_4$ & $1.55282$ & $4.01628$ & $2.20081$\\ 
        $\alpha_5$ & $0.00023$ & $0.00008$ & $0.00007$\\ 
        \botrule
    \end{tabular}
}
\hspace{1cm}
\subfloat[$B_s(\as)$\label{tab:table1_bS}]{
\centering
\begin{tabular}{c|r|r|r}
        \toprule
         & \multicolumn{1}{c|}{$\Psi$} & \multicolumn{1}{c|}{$\varepsilon$} &  \multicolumn{1}{c}{$\gamma$} \\ \colrule
        $\beta_0$ & $0.90067$ & $0.86256$ & $1.14795$\\ 
        $\beta_1$ & $-0.28757$ & $1.06174$ & $-0.18821$\\ 
        $\beta_2$ & $2.06242$ & $-6.40438$ & $0.95797$\\ 
        $\beta_3$ & $-6.0310$ & $10.38130$ & $-2.36396$\\ 
        $\beta_4$ & $4.34910$ & $-5.12991$ & $1.16129$\\ 
        $\beta_5$ & $0.00020$ & $0.00011$ & $0.00014$\\ 
        \botrule
    \end{tabular}
}
\\
\subfloat[$C_s(\as)$\label{tab:table1_cS}]{
\centering
\begin{tabular}{c|r|r|r}
        \toprule
         & \multicolumn{1}{c|}{$\Psi$} & \multicolumn{1}{c|}{$\varepsilon$} &  \multicolumn{1}{c}{$\gamma$} \\ \colrule
        $\eta_0$ & $7.68412$ & $7.02688$ & $8.02772$\\ 
        $\eta_1$ & $-1.19450$ & $2.99615$ & $0.80777$\\ 
        $\eta_2$ & $3.42557$ & $-25.1091$ & $-10.13620$\\ 
        $\eta_3$ & $-19.2999$ & $31.0490$ & $3.33735$\\ 
        $\eta_4$ & $11.6408$ & $-14.56991$ & $-0.49068$\\ 
        $\eta_5$ & $-0.00076$ & $-0.00145$ & $0.00141$\\ 
        \botrule
    \end{tabular}
}
\hspace{1cm}
\subfloat[$\nu_s(\as)$\label{tab:table1_dS}]{
\centering
\begin{tabular}{c|r|r|r}
        \toprule
         & \multicolumn{1}{c|}{$\Psi$} & \multicolumn{1}{c|}{$\varepsilon$} &  \multicolumn{1}{c}{$\gamma$} \\ \colrule
        $\delta_0$ & $-0.43895$ & $-0.28079$ & $-0.55216$\\ 
        $\delta_1$ & $-0.57066$ & $-1.87129$ & $0.47187$\\ 
        $\delta_2$ & $2.32570$ & $11.5739$ & $-2.55669$\\ 
        $\delta_3$ & $-0.98160$ & $-20.6905$ & $5.12681$\\ 
        $\delta_4$ & $-0.97489$ & $11.1745$ & $-2.65038$\\ 
        $\delta_5$ & $0.00035$ & $-0.00024$ & $-0.00022$\\ 
        \botrule
    \end{tabular}
}
\caption{Fitting parameters of our parametrized from Eqs.~\eqref{eq:fitfK}, \eqref{eq:ABCn-Kerr} for scalars.}
\label{tab:parsKS}
\end{table*}
\begin{table*}
\subfloat[$A_s(\as)$\label{tab:table1_aF}]{
\centering
\begin{tabular}{c|r|r|r}
        \toprule
         & \multicolumn{1}{c|}{$\Psi$} & \multicolumn{1}{c|}{$\varepsilon$} &  \multicolumn{1}{c}{$\gamma$} \\ \colrule
        $\alpha_0$ & $-0.040863$ & $-4.38503$ & $-3.51098$\\ 
        $\alpha_1$ & $-0.01122$ & $-0.01683$ & $-0.05455$\\ 
        $\alpha_2$ & $0.79019$ & $1.18529$ & $0.81172$\\ 
        $\alpha_3$ & $-0.80843$ & $-1.212644$ & $-1.62004$\\ 
        $\alpha_4$ & $0.53561$ & $0.80341$ & $1.23503$\\ 
        $\alpha_5$ & $0.00017$ & $0.00026$ & $0.00015$\\ 
        \botrule
    \end{tabular}
}
\hspace{1cm}
\subfloat[$B_s(\as)$\label{tab:table1_bF}]{
\centering
\begin{tabular}{c|r|r|r}
        \toprule
         & \multicolumn{1}{c|}{$\Psi$} & \multicolumn{1}{c|}{$\varepsilon$} &  \multicolumn{1}{c}{$\gamma$} \\ \colrule
        $\beta_0$ & $1.02775$ & $1.05952$ & $1.00516$\\ 
        $\beta_1$ & $0.25174$ & $-0.54908$ & $-0.30863$\\ 
        $\beta_2$ & $-1.91938$ & $3.00617$ & $1.35856$\\ 
        $\beta_3$ & $3.71237$ & $-6.69030$ & $-3.82659$\\ 
        $\beta_4$ & $-2.57412$ & $3.97656$ & $2.60927$\\ 
        $\beta_5$ & $0.00020$ & $0.00010$ & $0.00007$\\ 
        \botrule
    \end{tabular}
}
\\
\subfloat[$C_s(\as)$\label{tab:table1_cF}]{
\centering
\begin{tabular}{c|r|r|r}
        \toprule
         & \multicolumn{1}{c|}{$\Psi$} & \multicolumn{1}{c|}{$\varepsilon$} &  \multicolumn{1}{c}{$\gamma$} \\ \colrule
        $\eta_0$ & $8.64208$ & $8.19678$ & $7.50414$\\ 
        $\eta_1$ & $0.64604$ & $2.07543$ & $1.55438$\\ 
        $\eta_2$ & $-11.8172$ & $-19.0044$ & $-18.3266$\\ 
        $\eta_3$ & $18.2938$ & $22.0031$ & $23.8847$\\ 
        $\eta_4$ & $-13.6375$ & $-12.3065$ & $-12.7473$\\ 
        $\eta_5$ & $-0.00101$ & $-0.00102$ & $-0.00084$\\ 
        \botrule
    \end{tabular}
}
\hspace{1cm}
\subfloat[$\nu_s(\as)$\label{tab:table1_dF}]{
\centering
\begin{tabular}{c|r|r|r}
        \toprule
         & \multicolumn{1}{c|}{$\Psi$} & \multicolumn{1}{c|}{$\varepsilon$} &  \multicolumn{1}{c}{$\gamma$} \\ \colrule
        $\delta_0$ & $-0.49451$ & $-0.47427$ & $-0.43234$ \\ 
        $\delta_1$ & $-0.16979$ & $1.15307$ & $0.31938$ \\ 
        $\delta_2$ & $1.60784$ & $-6.77883$ & $-1.43802$ \\ 
        $\delta_3$ & $-2.20497$ & $15.0118$ & $5.66093$ \\ 
        $\delta_4$ & $1.33059$ & $-9.39296$ & $-4.65577$ \\ 
        $\delta_5$ & $-0.00039$ & $-1.40672\times 10^{-6}$ & $0.00004$ \\ 
        \botrule
    \end{tabular}
}
\caption{Fitting parameters of our parametrized from Eqs.~\eqref{eq:fitfK}, \eqref{eq:ABCn-Kerr} for Fermions.}
\label{tab:parsKF}
\end{table*}
\begin{table*}
\subfloat[$A_s(\as)$\label{tab:table1_aV}]{
\centering
\begin{tabular}{c|r|r|r}
        \toprule
         & \multicolumn{1}{c|}{$\Psi$} & \multicolumn{1}{c|}{$\varepsilon$} &  \multicolumn{1}{c}{$\gamma$} \\ \colrule
        $\alpha_0$ & $-0.55674$ & $-4.77544$ & $-3.63003$\\ 
        $\alpha_1$ & $-0.19425$ & $0.06299$ & $0.26289$\\ 
        $\alpha_2$ & $3.85503$ & $3.15186$ & $0.08464$\\ 
        $\alpha_3$ & $-4.67469$ & $-3.52187$ & $0.61410$\\ 
        $\alpha_4$ & $2.56983$ & $1.99381$ & $0.00797$\\ 
        $\alpha_5$ & $0.00019$ & $0.00019$ & $0.00018$\\ 
        \botrule
    \end{tabular}
}
\hspace{1cm}
\subfloat[$B_s(\as)$\label{tab:table1_bV}]{
\centering
\begin{tabular}{c|r|r|r}
        \toprule
         & \multicolumn{1}{c|}{$\Psi$} & \multicolumn{1}{c|}{$\varepsilon$} &  \multicolumn{1}{c}{$\gamma$} \\ \colrule
        $\beta_0$ & $1.12764$ & $1.14383$ & $1.12718$\\ 
        $\beta_1$ & $-0.00975$ & $-0.00734$ & $0.00040$\\ 
        $\beta_2$ & $0.17278$ & $-0.05991$ & $-0.02018$\\ 
        $\beta_3$ & $-0.15243$ & $0.13639$ & $-0.26672$\\ 
        $\beta_4$ & $-0.36279$ & $-0.57533$ & $-0.18571$\\ 
        $\beta_5$ & $-0.00006$ & $0.00010$ & $0.00007$\\ 
        \botrule
    \end{tabular}
}
\\
\subfloat[$C_s(\as)$\label{tab:table1_cV}]{
\centering
\begin{tabular}{c|r|r|r}
        \toprule
         & \multicolumn{1}{c|}{$\Psi$} & \multicolumn{1}{c|}{$\varepsilon$} &  \multicolumn{1}{c}{$\gamma$} \\ \colrule
        $\eta_0$ & $8.99996$ & $8.88373$ & $8.61971$ \\ 
        $\eta_1$ & $-1.07481$ & $-1.45578$ & $-0.00904$ \\ 
        $\eta_2$ & $3.30235$ & $2.35702$ & $-3.23898$ \\ 
        $\eta_3$ & $-12.0108$ & $-14.6024$ & $-7.78225$\\ 
        $\eta_4$ & $3.96199$ & $7.09857$ & $4.85716$\\ 
        $\eta_5$ & $-0.00060$ & $-0.00111$ & $-0.00109$\\ 
        \botrule
    \end{tabular}
}
\hspace{1cm}
\subfloat[$\nu_s(\as)$\label{tab:table1_dV}]{
\centering
\begin{tabular}{c|r|r|r}
        \toprule
         & \multicolumn{1}{c|}{$\Psi$} & \multicolumn{1}{c|}{$\varepsilon$} &  \multicolumn{1}{c}{$\gamma$} \\ \colrule
        $\delta_0$ & $-0.52989$ & $-0.51754$ & $-0.52112$ \\ 
        $\delta_1$ & $-0.00885$ & $0.03176$ & $0.09238$ \\ 
        $\delta_2$ & $0.02979$ & $0.07427$ & $-0.47192$ \\ 
        $\delta_3$ & $-0.67753$ & $-0.40853$ & $0.81145$ \\ 
        $\delta_4$ & $1.14249$ & $0.94323$ & $0.16319$ \\ 
        $\delta_5$ & $-0.00004$ & $-0.00026$ & $-0.00022$ \\ 
        \botrule
    \end{tabular}
}
\caption{Fitting parameters of our parametrized from Eqs.~\eqref{eq:fitfK}, \eqref{eq:ABCn-Kerr} for Vectors.}
\label{tab:parsKV}
\end{table*}
\begin{table*}
\subfloat[$A_s(\as)$\label{tab:table1_aG}]{
\centering
\begin{tabular}{c|r|r|r}
        \toprule
         & \multicolumn{1}{c|}{$\Psi$} & \multicolumn{1}{c|}{$\varepsilon$} &  \multicolumn{1}{c}{$\gamma$} \\ \colrule
        $\alpha_0$ & $-1.71000$ & $-5.71338$ & $-4.26363$\\ 
        $\alpha_1$ & $0.60445$ & $0.55086$ & $-0.08493$\\ 
        $\alpha_2$ & $5.87217$ & $7.51779$ & $5.63412$\\ 
        $\alpha_3$ & $-6.26968$ & $-9.50996$ & $-6.57782$\\ 
        $\alpha_4$ & $3.33794$ & $5.34276$ & $3.89968$\\ 
        $\alpha_5$ & $0.00025$ & $0.00033$ & $0.00028$\\ 
        \botrule
    \end{tabular}
}
\hspace{1cm}
\subfloat[$B_s(\as)$\label{tab:table1_bG}]{
\centering
\begin{tabular}{c|r|r|r}
        \toprule
         & \multicolumn{1}{c|}{$\Psi$} & \multicolumn{1}{c|}{$\varepsilon$} &  \multicolumn{1}{c}{$\gamma$} \\ \colrule
        $\beta_0$ & $1.27648$ & $1.29608$ & $1.27466$\\ 
        $\beta_1$ & $0.23464$ & $0.06644$ & $0.481525$\\ 
        $\beta_2$ & $-0.96709$ & $-0.07557$ & $-2.48397$\\ 
        $\beta_3$ & $1.59757$ & $0.06767$ & $4.3418$\\ 
        $\beta_4$ & $-1.05198$ & $-0.30298$ & $-2.68619$\\ 
        $\beta_5$ & $-0.00010$ & $-0.00006$ & $0.000005$\\ 
        \botrule
    \end{tabular}
}
\\
\subfloat[$C_s(\as)$\label{tab:table1_cG}]{
\centering
\begin{tabular}{c|r|r|r}
        \toprule
         & \multicolumn{1}{c|}{$\Psi$} & \multicolumn{1}{c|}{$\varepsilon$} &  \multicolumn{1}{c}{$\gamma$} \\ \colrule
        $\eta_0$ & $9.06958$ & $9.06618$ & $8.91067$ \\ 
        $\eta_1$ & $2.94595$ & $1.29693$ & $2.77640$ \\ 
        $\eta_2$ & $-25.4011$ & $-16.2621$ & $-21.98310$ \\ 
        $\eta_3$ & $39.1391$ & $21.5632$ & $28.0939$\\ 
        $\eta_4$ & $-24.1023$ & $-14.4600$ & $-16.1920$\\ 
        $\eta_5$ & $-0.00125$ & $-0.00051$ & $-0.000732$\\ 
        \botrule
    \end{tabular}
}
\hspace{1cm}
\subfloat[$\nu_s(\as)$\label{tab:table1_dG}]{
\centering
\begin{tabular}{c|r|r|r}
        \toprule
         & \multicolumn{1}{c|}{$\Psi$} & \multicolumn{1}{c|}{$\varepsilon$} &  \multicolumn{1}{c}{$\gamma$} \\ \colrule
        $\delta_0$ & $-0.64634$ & $-0.64402$ & $-0.63890$ \\ 
        $\delta_1$ & $-0.05947$ & $-0.25035$ & $-0.39021$ \\ 
        $\delta_2$ & $0.01428$ & $1.23193$ & $1.97792$ \\ 
        $\delta_3$ & $-0.01704$ & $-2.46744$ & $-3.60446$ \\ 
        $\delta_4$ & $0.25369$ & $1.85603$ & $2.40479$ \\ 
        $\delta_5$ & $0.00009$ & $0.00002$ & $-1.06\times 10^{-6}$ \\ 
        \botrule
    \end{tabular}
}
\caption{Fitting parameters of our parametrized from Eqs.~\eqref{eq:fitfK}, \eqref{eq:ABCn-Kerr} for spin-2 particles.}
\label{tab:parsKG}
\end{table*}
The mass $\varepsilon_i(z_i,\as)$ and angular momentum $\gamma_i(z_i,\as)$ evaporation functions per spin defined by
\begin{subequations}
\begin{align}
    \varepsilon_i(z_i,\as)&\equiv-\frac{\MBH^2}{\MPL^4}\frac{\dd{\MBH}}{\dd{t}}\,,\\ 
    \gamma_i(z_i,\as)&\equiv-\frac{1}{\as}\frac{\MBH}{\MPL^2}\frac{\dd{J}}{\dd{t}}\,,
\end{align}
\end{subequations}
being $J$ the BH angular momentum. These evaporation functions are parametrized as
\begin{subequations}
\begin{align}
   \varepsilon_i(z_i,\as) &= \frac{27}{8192\pi^5}\int_{z_i}^\infty\sum_{lm}\frac{\psi_{s_i}^{lm}(x,\as)(x^2-z_i^2)}{\exp[(x(1+\sqrt{1-\as^2})-4\pi m\as)/2\sqrt{1-\as^2}]-(-1)^{2s_i}}\,x\dd{x}\,,\\
   \gamma_i(z_i,\as) &= \frac{27}{1024\pi^4}\int_{z_i}^\infty\sum_{lm}\frac{m\,\psi_{s_i}^{lm}(x,\as)(x^2-z_i^2)\dd{x}}{\exp[(x(1+\sqrt{1-\as^2})-4\pi m\as)/2\sqrt{1-\as^2}]-(-1)^{2s_i}}\,,
\end{align}
\end{subequations}
\end{widetext}
with $x$ and $z_i$ defined as before. Similar to the particle emission rate, we fit these evaporation parameters according to a general logistic form, Eq.~\eqref{eq:fitfK}, where the parameters are also given in Tables~\ref{tab:parsKS}-\ref{tab:parsKG}.

\section{Decay Widths}\label{ap:A1}

In this appendix we quote the decay widths of scalars, vectors and massive tensors into a fermion-antifermion pair, used in the subsection \ref{subsec:ind}. For $X$ having a mass $m_X$ and a coupling $g_D$ with the fermions $\chi$, we have
\begin{align}
    \Gamma_{X\to {\rm DM}}^S = \frac{m_X}{8\pi}g_D^2\left(1-\frac{4m_{\rm DM}^2}{m_X^2}\right)^\frac{3}{2}\,,
\end{align}
for $X$ being a scalar. In the case that $X$ is a massive vector, we have
\begin{align}
    \Gamma_{X\to {\rm DM}}^V= \frac{m_X}{12\pi}g_D^2\left(1+\frac{2m_{\rm DM}^2}{m_X^2}\right)\left(1-\frac{4m_{\rm DM}^2}{m_X^2}\right)^\frac{1}{2}\,,
\end{align}
and, finally, for a massive spin-2 particle~\cite{Han:1998sg,Lee:2013bua,Falkowski:2016glr}
\begin{align}
    \Gamma_{X\to {\rm DM}}^G = \frac{m_X}{160\pi}g_D^2\left(1+\frac{8}{3}\frac{m_{\rm DM}^2}{m_X^2}\right)\left(1-\frac{4m_{\rm DM}^2}{m_X^2}\right)^\frac{3}{2}\,.
\end{align}

\bibliographystyle{apsrev4-1}
\bibliography{main.bib}

\begin{thebibliography}{74}%
\makeatletter
\providecommand \@ifxundefined [1]{%
 \@ifx{#1\undefined}
}%
\providecommand \@ifnum [1]{%
 \ifnum #1\expandafter \@firstoftwo
 \else \expandafter \@secondoftwo
 \fi
}%
\providecommand \@ifx [1]{%
 \ifx #1\expandafter \@firstoftwo
 \else \expandafter \@secondoftwo
 \fi
}%
\providecommand \natexlab [1]{#1}%
\providecommand \enquote  [1]{``#1''}%
\providecommand \bibnamefont  [1]{#1}%
\providecommand \bibfnamefont [1]{#1}%
\providecommand \citenamefont [1]{#1}%
\providecommand \href@noop [0]{\@secondoftwo}%
\providecommand \href [0]{\begingroup \@sanitize@url \@href}%
\providecommand \@href[1]{\@@startlink{#1}\@@href}%
\providecommand \@@href[1]{\endgroup#1\@@endlink}%
\providecommand \@sanitize@url [0]{\catcode `\\12\catcode `\$12\catcode
  `\&12\catcode `\#12\catcode `\^12\catcode `\_12\catcode `\%12\relax}%
\providecommand \@@startlink[1]{}%
\providecommand \@@endlink[0]{}%
\providecommand \url  [0]{\begingroup\@sanitize@url \@url }%
\providecommand \@url [1]{\endgroup\@href {#1}{\urlprefix }}%
\providecommand \urlprefix  [0]{URL }%
\providecommand \Eprint [0]{\href }%
\providecommand \doibase [0]{http://dx.doi.org/}%
\providecommand \selectlanguage [0]{\@gobble}%
\providecommand \bibinfo  [0]{\@secondoftwo}%
\providecommand \bibfield  [0]{\@secondoftwo}%
\providecommand \translation [1]{[#1]}%
\providecommand \BibitemOpen [0]{}%
\providecommand \bibitemStop [0]{}%
\providecommand \bibitemNoStop [0]{.\EOS\space}%
\providecommand \EOS [0]{\spacefactor3000\relax}%
\providecommand \BibitemShut  [1]{\csname bibitem#1\endcsname}%
\let\auto@bib@innerbib\@empty
\bibitem [{\citenamefont {Hawking}(1974)}]{Hawking:1974rv}%
  \BibitemOpen
  \bibfield  {author} {\bibinfo {author} {\bibfnamefont {S.~W.}\ \bibnamefont
  {Hawking}},\ }\href {\doibase 10.1038/248030a0} {\bibfield  {journal}
  {\bibinfo  {journal} {Nature}\ }\textbf {\bibinfo {volume} {248}},\ \bibinfo
  {pages} {30} (\bibinfo {year} {1974})}\BibitemShut {NoStop}%
\bibitem [{\citenamefont {Hawking}(1975)}]{Hawking:1974sw}%
  \BibitemOpen
  \bibfield  {author} {\bibinfo {author} {\bibfnamefont {S.~W.}\ \bibnamefont
  {Hawking}},\ }\bibfield  {booktitle} {\emph {\bibinfo {booktitle} {{Euclidean
  quantum gravity}}},\ }\href {\doibase 10.1007/BF02345020, 10.1007/BF01608497}
  {\bibfield  {journal} {\bibinfo  {journal} {Commun. Math. Phys.}\ }\textbf
  {\bibinfo {volume} {43}},\ \bibinfo {pages} {199} (\bibinfo {year} {1975})},\
  \bibinfo {note} {[,167(1975)]}\BibitemShut {NoStop}%
\bibitem [{\citenamefont {Ade}\ \emph {et~al.}(2016{\natexlab{a}})\citenamefont
  {Ade} \emph {et~al.}}]{Ade:2015lrj}%
  \BibitemOpen
  \bibfield  {author} {\bibinfo {author} {\bibfnamefont {P.~A.~R.}\
  \bibnamefont {Ade}} \emph {et~al.} (\bibinfo {collaboration} {Planck}),\
  }\href {\doibase 10.1051/0004-6361/201525898} {\bibfield  {journal} {\bibinfo
   {journal} {Astron. Astrophys.}\ }\textbf {\bibinfo {volume} {594}},\
  \bibinfo {pages} {A20} (\bibinfo {year} {2016}{\natexlab{a}})},\ \Eprint
  {http://arxiv.org/abs/1502.02114} {arXiv:1502.02114 [astro-ph.CO]}
  \BibitemShut {NoStop}%
\bibitem [{\citenamefont {Ade}\ \emph {et~al.}(2016{\natexlab{b}})\citenamefont
  {Ade} \emph {et~al.}}]{Ade:2015xua}%
  \BibitemOpen
  \bibfield  {author} {\bibinfo {author} {\bibfnamefont {P.~A.~R.}\
  \bibnamefont {Ade}} \emph {et~al.} (\bibinfo {collaboration} {Planck}),\
  }\href {\doibase 10.1051/0004-6361/201525830} {\bibfield  {journal} {\bibinfo
   {journal} {Astron. Astrophys.}\ }\textbf {\bibinfo {volume} {594}},\
  \bibinfo {pages} {A13} (\bibinfo {year} {2016}{\natexlab{b}})},\ \Eprint
  {http://arxiv.org/abs/1502.01589} {arXiv:1502.01589 [astro-ph.CO]}
  \BibitemShut {NoStop}%
\bibitem [{\citenamefont {Carr}\ \emph {et~al.}(2016)\citenamefont {Carr},
  \citenamefont {Kuhnel},\ and\ \citenamefont {Sandstad}}]{Carr:2016drx}%
  \BibitemOpen
  \bibfield  {author} {\bibinfo {author} {\bibfnamefont {B.}~\bibnamefont
  {Carr}}, \bibinfo {author} {\bibfnamefont {F.}~\bibnamefont {Kuhnel}}, \ and\
  \bibinfo {author} {\bibfnamefont {M.}~\bibnamefont {Sandstad}},\ }\href
  {\doibase 10.1103/PhysRevD.94.083504} {\bibfield  {journal} {\bibinfo
  {journal} {Phys. Rev. D}\ }\textbf {\bibinfo {volume} {94}},\ \bibinfo
  {pages} {083504} (\bibinfo {year} {2016})},\ \Eprint
  {http://arxiv.org/abs/1607.06077} {arXiv:1607.06077 [astro-ph.CO]}
  \BibitemShut {NoStop}%
\bibitem [{\citenamefont {Green}\ and\ \citenamefont
  {Kavanagh}(2021)}]{Green:2020jor}%
  \BibitemOpen
  \bibfield  {author} {\bibinfo {author} {\bibfnamefont {A.~M.}\ \bibnamefont
  {Green}}\ and\ \bibinfo {author} {\bibfnamefont {B.~J.}\ \bibnamefont
  {Kavanagh}},\ }\href {\doibase 10.1088/1361-6471/abc534} {\bibfield
  {journal} {\bibinfo  {journal} {J. Phys. G}\ }\textbf {\bibinfo {volume}
  {48}},\ \bibinfo {pages} {4} (\bibinfo {year} {2021})},\ \Eprint
  {http://arxiv.org/abs/2007.10722} {arXiv:2007.10722 [astro-ph.CO]}
  \BibitemShut {NoStop}%
\bibitem [{\citenamefont {Carr}\ \emph {et~al.}(2020)\citenamefont {Carr},
  \citenamefont {Kohri}, \citenamefont {Sendouda},\ and\ \citenamefont
  {Yokoyama}}]{Carr:2020gox}%
  \BibitemOpen
  \bibfield  {author} {\bibinfo {author} {\bibfnamefont {B.}~\bibnamefont
  {Carr}}, \bibinfo {author} {\bibfnamefont {K.}~\bibnamefont {Kohri}},
  \bibinfo {author} {\bibfnamefont {Y.}~\bibnamefont {Sendouda}}, \ and\
  \bibinfo {author} {\bibfnamefont {J.}~\bibnamefont {Yokoyama}},\ }\href@noop
  {} {\  (\bibinfo {year} {2020})},\ \Eprint {http://arxiv.org/abs/2002.12778}
  {arXiv:2002.12778 [astro-ph.CO]} \BibitemShut {NoStop}%
\bibitem [{\citenamefont {Carr}\ \emph {et~al.}(2010)\citenamefont {Carr},
  \citenamefont {Kohri}, \citenamefont {Sendouda},\ and\ \citenamefont
  {Yokoyama}}]{Carr:2009jm}%
  \BibitemOpen
  \bibfield  {author} {\bibinfo {author} {\bibfnamefont {B.~J.}\ \bibnamefont
  {Carr}}, \bibinfo {author} {\bibfnamefont {K.}~\bibnamefont {Kohri}},
  \bibinfo {author} {\bibfnamefont {Y.}~\bibnamefont {Sendouda}}, \ and\
  \bibinfo {author} {\bibfnamefont {J.}~\bibnamefont {Yokoyama}},\ }\href
  {\doibase 10.1103/PhysRevD.81.104019} {\bibfield  {journal} {\bibinfo
  {journal} {Phys. Rev. D}\ }\textbf {\bibinfo {volume} {81}},\ \bibinfo
  {pages} {104019} (\bibinfo {year} {2010})},\ \Eprint
  {http://arxiv.org/abs/0912.5297} {arXiv:0912.5297 [astro-ph.CO]} \BibitemShut
  {NoStop}%
\bibitem [{\citenamefont {Keith}\ \emph {et~al.}(2020)\citenamefont {Keith},
  \citenamefont {Hooper}, \citenamefont {Blinov},\ and\ \citenamefont
  {McDermott}}]{Keith:2020jww}%
  \BibitemOpen
  \bibfield  {author} {\bibinfo {author} {\bibfnamefont {C.}~\bibnamefont
  {Keith}}, \bibinfo {author} {\bibfnamefont {D.}~\bibnamefont {Hooper}},
  \bibinfo {author} {\bibfnamefont {N.}~\bibnamefont {Blinov}}, \ and\ \bibinfo
  {author} {\bibfnamefont {S.~D.}\ \bibnamefont {McDermott}},\ }\href@noop {}
  {\  (\bibinfo {year} {2020})},\ \Eprint {http://arxiv.org/abs/2006.03608}
  {arXiv:2006.03608 [astro-ph.CO]} \BibitemShut {NoStop}%
\bibitem [{\citenamefont {Akrami}\ \emph {et~al.}(2020)\citenamefont {Akrami}
  \emph {et~al.}}]{Akrami:2018odb}%
  \BibitemOpen
  \bibfield  {author} {\bibinfo {author} {\bibfnamefont {Y.}~\bibnamefont
  {Akrami}} \emph {et~al.} (\bibinfo {collaboration} {Planck}),\ }\href
  {\doibase 10.1051/0004-6361/201833887} {\bibfield  {journal} {\bibinfo
  {journal} {Astron. Astrophys.}\ }\textbf {\bibinfo {volume} {641}},\ \bibinfo
  {pages} {A10} (\bibinfo {year} {2020})},\ \Eprint
  {http://arxiv.org/abs/1807.06211} {arXiv:1807.06211 [astro-ph.CO]}
  \BibitemShut {NoStop}%
\bibitem [{\citenamefont {Carr}(1976)}]{Carr:1976zz}%
  \BibitemOpen
  \bibfield  {author} {\bibinfo {author} {\bibfnamefont {B.~J.}\ \bibnamefont
  {Carr}},\ }\href {\doibase 10.1086/154351} {\bibfield  {journal} {\bibinfo
  {journal} {Astrophys. J.}\ }\textbf {\bibinfo {volume} {206}},\ \bibinfo
  {pages} {8} (\bibinfo {year} {1976})}\BibitemShut {NoStop}%
\bibitem [{\citenamefont {Carr}\ and\ \citenamefont
  {Kuhnel}(2020)}]{Carr:2020xqk}%
  \BibitemOpen
  \bibfield  {author} {\bibinfo {author} {\bibfnamefont {B.}~\bibnamefont
  {Carr}}\ and\ \bibinfo {author} {\bibfnamefont {F.}~\bibnamefont {Kuhnel}},\
  }\href {\doibase 10.1146/annurev-nucl-050520-125911} {\bibfield  {journal}
  {\bibinfo  {journal} {Ann. Rev. Nucl. Part. Sci.}\ }\textbf {\bibinfo
  {volume} {70}},\ \bibinfo {pages} {355} (\bibinfo {year} {2020})},\ \Eprint
  {http://arxiv.org/abs/2006.02838} {arXiv:2006.02838 [astro-ph.CO]}
  \BibitemShut {NoStop}%
\bibitem [{\citenamefont {Hooper}\ \emph {et~al.}(2019)\citenamefont {Hooper},
  \citenamefont {Krnjaic},\ and\ \citenamefont {McDermott}}]{Hooper:2019gtx}%
  \BibitemOpen
  \bibfield  {author} {\bibinfo {author} {\bibfnamefont {D.}~\bibnamefont
  {Hooper}}, \bibinfo {author} {\bibfnamefont {G.}~\bibnamefont {Krnjaic}}, \
  and\ \bibinfo {author} {\bibfnamefont {S.~D.}\ \bibnamefont {McDermott}},\
  }\href {\doibase 10.1007/JHEP08(2019)001} {\bibfield  {journal} {\bibinfo
  {journal} {JHEP}\ }\textbf {\bibinfo {volume} {08}},\ \bibinfo {pages} {001}
  (\bibinfo {year} {2019})},\ \Eprint {http://arxiv.org/abs/1905.01301}
  {arXiv:1905.01301 [hep-ph]} \BibitemShut {NoStop}%
\bibitem [{\citenamefont {Lunardini}\ and\ \citenamefont
  {Perez-Gonzalez}(2020)}]{Lunardini:2019zob}%
  \BibitemOpen
  \bibfield  {author} {\bibinfo {author} {\bibfnamefont {C.}~\bibnamefont
  {Lunardini}}\ and\ \bibinfo {author} {\bibfnamefont {Y.~F.}\ \bibnamefont
  {Perez-Gonzalez}},\ }\href {\doibase 10.1088/1475-7516/2020/08/014}
  {\bibfield  {journal} {\bibinfo  {journal} {JCAP}\ }\textbf {\bibinfo
  {volume} {08}},\ \bibinfo {pages} {014} (\bibinfo {year} {2020})},\ \Eprint
  {http://arxiv.org/abs/1910.07864} {arXiv:1910.07864 [hep-ph]} \BibitemShut
  {NoStop}%
\bibitem [{\citenamefont {Inomata}\ \emph {et~al.}(2020)\citenamefont
  {Inomata}, \citenamefont {Kawasaki}, \citenamefont {Mukaida}, \citenamefont
  {Terada},\ and\ \citenamefont {Yanagida}}]{Inomata:2020lmk}%
  \BibitemOpen
  \bibfield  {author} {\bibinfo {author} {\bibfnamefont {K.}~\bibnamefont
  {Inomata}}, \bibinfo {author} {\bibfnamefont {M.}~\bibnamefont {Kawasaki}},
  \bibinfo {author} {\bibfnamefont {K.}~\bibnamefont {Mukaida}}, \bibinfo
  {author} {\bibfnamefont {T.}~\bibnamefont {Terada}}, \ and\ \bibinfo {author}
  {\bibfnamefont {T.~T.}\ \bibnamefont {Yanagida}},\ }\href {\doibase
  10.1103/PhysRevD.101.123533} {\bibfield  {journal} {\bibinfo  {journal}
  {Phys. Rev. D}\ }\textbf {\bibinfo {volume} {101}},\ \bibinfo {pages}
  {123533} (\bibinfo {year} {2020})},\ \Eprint
  {http://arxiv.org/abs/2003.10455} {arXiv:2003.10455 [astro-ph.CO]}
  \BibitemShut {NoStop}%
\bibitem [{\citenamefont {Masina}(2020)}]{Masina:2020xhk}%
  \BibitemOpen
  \bibfield  {author} {\bibinfo {author} {\bibfnamefont {I.}~\bibnamefont
  {Masina}},\ }\href {\doibase 10.1140/epjp/s13360-020-00564-9} {\bibfield
  {journal} {\bibinfo  {journal} {Eur. Phys. J. Plus}\ }\textbf {\bibinfo
  {volume} {135}},\ \bibinfo {pages} {552} (\bibinfo {year} {2020})},\ \Eprint
  {http://arxiv.org/abs/2004.04740} {arXiv:2004.04740 [hep-ph]} \BibitemShut
  {NoStop}%
\bibitem [{\citenamefont {Masina}(2021)}]{Masina:2021zpu}%
  \BibitemOpen
  \bibfield  {author} {\bibinfo {author} {\bibfnamefont {I.}~\bibnamefont
  {Masina}},\ }\href@noop {} {\  (\bibinfo {year} {2021})},\ \Eprint
  {http://arxiv.org/abs/2103.13825} {arXiv:2103.13825 [gr-qc]} \BibitemShut
  {NoStop}%
\bibitem [{\citenamefont {Dom\`enech}\ \emph {et~al.}(2021)\citenamefont
  {Dom\`enech}, \citenamefont {Takhistov},\ and\ \citenamefont
  {Sasaki}}]{Domenech:2021wkk}%
  \BibitemOpen
  \bibfield  {author} {\bibinfo {author} {\bibfnamefont {G.}~\bibnamefont
  {Dom\`enech}}, \bibinfo {author} {\bibfnamefont {V.}~\bibnamefont
  {Takhistov}}, \ and\ \bibinfo {author} {\bibfnamefont {M.}~\bibnamefont
  {Sasaki}},\ }\href@noop {} {\  (\bibinfo {year} {2021})},\ \Eprint
  {http://arxiv.org/abs/2105.06816} {arXiv:2105.06816 [astro-ph.CO]}
  \BibitemShut {NoStop}%
\bibitem [{\citenamefont {Baumann}\ \emph {et~al.}(2007)\citenamefont
  {Baumann}, \citenamefont {Steinhardt},\ and\ \citenamefont
  {Turok}}]{Baumann:2007yr}%
  \BibitemOpen
  \bibfield  {author} {\bibinfo {author} {\bibfnamefont {D.}~\bibnamefont
  {Baumann}}, \bibinfo {author} {\bibfnamefont {P.~J.}\ \bibnamefont
  {Steinhardt}}, \ and\ \bibinfo {author} {\bibfnamefont {N.}~\bibnamefont
  {Turok}},\ }\href@noop {} {\  (\bibinfo {year} {2007})},\ \Eprint
  {http://arxiv.org/abs/hep-th/0703250} {arXiv:hep-th/0703250 [HEP-TH]}
  \BibitemShut {NoStop}%
\bibitem [{\citenamefont {Fujita}\ \emph {et~al.}(2014)\citenamefont {Fujita},
  \citenamefont {Kawasaki}, \citenamefont {Harigaya},\ and\ \citenamefont
  {Matsuda}}]{Fujita:2014hha}%
  \BibitemOpen
  \bibfield  {author} {\bibinfo {author} {\bibfnamefont {T.}~\bibnamefont
  {Fujita}}, \bibinfo {author} {\bibfnamefont {M.}~\bibnamefont {Kawasaki}},
  \bibinfo {author} {\bibfnamefont {K.}~\bibnamefont {Harigaya}}, \ and\
  \bibinfo {author} {\bibfnamefont {R.}~\bibnamefont {Matsuda}},\ }\href
  {\doibase 10.1103/PhysRevD.89.103501} {\bibfield  {journal} {\bibinfo
  {journal} {Phys. Rev.}\ }\textbf {\bibinfo {volume} {D89}},\ \bibinfo {pages}
  {103501} (\bibinfo {year} {2014})},\ \Eprint {http://arxiv.org/abs/1401.1909}
  {arXiv:1401.1909 [astro-ph.CO]} \BibitemShut {NoStop}%
\bibitem [{\citenamefont {Hook}(2014)}]{Hook:2014mla}%
  \BibitemOpen
  \bibfield  {author} {\bibinfo {author} {\bibfnamefont {A.}~\bibnamefont
  {Hook}},\ }\href {\doibase 10.1103/PhysRevD.90.083535} {\bibfield  {journal}
  {\bibinfo  {journal} {Phys. Rev. D}\ }\textbf {\bibinfo {volume} {90}},\
  \bibinfo {pages} {083535} (\bibinfo {year} {2014})},\ \Eprint
  {http://arxiv.org/abs/1404.0113} {arXiv:1404.0113 [hep-ph]} \BibitemShut
  {NoStop}%
\bibitem [{\citenamefont {Hamada}\ and\ \citenamefont
  {Iso}(2017)}]{Hamada:2016jnq}%
  \BibitemOpen
  \bibfield  {author} {\bibinfo {author} {\bibfnamefont {Y.}~\bibnamefont
  {Hamada}}\ and\ \bibinfo {author} {\bibfnamefont {S.}~\bibnamefont {Iso}},\
  }\href {\doibase 10.1093/ptep/ptx011} {\bibfield  {journal} {\bibinfo
  {journal} {PTEP}\ }\textbf {\bibinfo {volume} {2017}},\ \bibinfo {pages}
  {033B02} (\bibinfo {year} {2017})},\ \Eprint
  {http://arxiv.org/abs/1610.02586} {arXiv:1610.02586 [hep-ph]} \BibitemShut
  {NoStop}%
\bibitem [{\citenamefont {Chaudhuri}\ and\ \citenamefont
  {Dolgov}(2020)}]{Chaudhuri:2020wjo}%
  \BibitemOpen
  \bibfield  {author} {\bibinfo {author} {\bibfnamefont {A.}~\bibnamefont
  {Chaudhuri}}\ and\ \bibinfo {author} {\bibfnamefont {A.}~\bibnamefont
  {Dolgov}},\ }\href@noop {} {\  (\bibinfo {year} {2020})},\ \Eprint
  {http://arxiv.org/abs/2001.11219} {arXiv:2001.11219 [astro-ph.CO]}
  \BibitemShut {NoStop}%
\bibitem [{\citenamefont {Hooper}\ and\ \citenamefont
  {Krnjaic}(2021)}]{Hooper:2020otu}%
  \BibitemOpen
  \bibfield  {author} {\bibinfo {author} {\bibfnamefont {D.}~\bibnamefont
  {Hooper}}\ and\ \bibinfo {author} {\bibfnamefont {G.}~\bibnamefont
  {Krnjaic}},\ }\href {\doibase 10.1103/PhysRevD.103.043504} {\bibfield
  {journal} {\bibinfo  {journal} {Phys. Rev. D}\ }\textbf {\bibinfo {volume}
  {103}},\ \bibinfo {pages} {043504} (\bibinfo {year} {2021})},\ \Eprint
  {http://arxiv.org/abs/2010.01134} {arXiv:2010.01134 [hep-ph]} \BibitemShut
  {NoStop}%
\bibitem [{\citenamefont {Perez-Gonzalez}\ and\ \citenamefont
  {Turner}(2021)}]{Perez-Gonzalez:2020vnz}%
  \BibitemOpen
  \bibfield  {author} {\bibinfo {author} {\bibfnamefont {Y.~F.}\ \bibnamefont
  {Perez-Gonzalez}}\ and\ \bibinfo {author} {\bibfnamefont {J.}~\bibnamefont
  {Turner}},\ }\href {\doibase 10.1103/PhysRevD.104.103021} {\bibfield
  {journal} {\bibinfo  {journal} {Phys. Rev. D}\ }\textbf {\bibinfo {volume}
  {104}},\ \bibinfo {pages} {103021} (\bibinfo {year} {2021})},\ \Eprint
  {http://arxiv.org/abs/2010.03565} {arXiv:2010.03565 [hep-ph]} \BibitemShut
  {NoStop}%
\bibitem [{\citenamefont {Datta}\ \emph {et~al.}(2020)\citenamefont {Datta},
  \citenamefont {Ghosal},\ and\ \citenamefont {Samanta}}]{Datta:2020bht}%
  \BibitemOpen
  \bibfield  {author} {\bibinfo {author} {\bibfnamefont {S.}~\bibnamefont
  {Datta}}, \bibinfo {author} {\bibfnamefont {A.}~\bibnamefont {Ghosal}}, \
  and\ \bibinfo {author} {\bibfnamefont {R.}~\bibnamefont {Samanta}},\
  }\href@noop {} {\  (\bibinfo {year} {2020})},\ \Eprint
  {http://arxiv.org/abs/2012.14981} {arXiv:2012.14981 [hep-ph]} \BibitemShut
  {NoStop}%
\bibitem [{\citenamefont {Jyoti~Das}\ \emph {et~al.}(2021)\citenamefont
  {Jyoti~Das}, \citenamefont {Mahanta},\ and\ \citenamefont
  {Borah}}]{JyotiDas:2021shi}%
  \BibitemOpen
  \bibfield  {author} {\bibinfo {author} {\bibfnamefont {S.}~\bibnamefont
  {Jyoti~Das}}, \bibinfo {author} {\bibfnamefont {D.}~\bibnamefont {Mahanta}},
  \ and\ \bibinfo {author} {\bibfnamefont {D.}~\bibnamefont {Borah}},\
  }\href@noop {} {\  (\bibinfo {year} {2021})},\ \Eprint
  {http://arxiv.org/abs/2104.14496} {arXiv:2104.14496 [hep-ph]} \BibitemShut
  {NoStop}%
\bibitem [{\citenamefont {Matsas}\ \emph {et~al.}(1998)\citenamefont {Matsas},
  \citenamefont {Montero}, \citenamefont {Pleitez},\ and\ \citenamefont
  {Vanzella}}]{Matsas:1998zm}%
  \BibitemOpen
  \bibfield  {author} {\bibinfo {author} {\bibfnamefont {G.~E.~A.}\
  \bibnamefont {Matsas}}, \bibinfo {author} {\bibfnamefont {J.~C.}\
  \bibnamefont {Montero}}, \bibinfo {author} {\bibfnamefont {V.}~\bibnamefont
  {Pleitez}}, \ and\ \bibinfo {author} {\bibfnamefont {D.~A.~T.}\ \bibnamefont
  {Vanzella}},\ }in\ \href@noop {} {\emph {\bibinfo {booktitle} {{Conference on
  Topics in Theoretical Physics II: Festschrift for A.H. Zimerman Sao Paulo,
  Brazil, November 20, 1998}}}}\ (\bibinfo {year} {1998})\ \Eprint
  {http://arxiv.org/abs/hep-ph/9810456} {arXiv:hep-ph/9810456 [hep-ph]}
  \BibitemShut {NoStop}%
\bibitem [{\citenamefont {Bell}\ and\ \citenamefont
  {Volkas}(1999)}]{Bell:1998jk}%
  \BibitemOpen
  \bibfield  {author} {\bibinfo {author} {\bibfnamefont {N.~F.}\ \bibnamefont
  {Bell}}\ and\ \bibinfo {author} {\bibfnamefont {R.~R.}\ \bibnamefont
  {Volkas}},\ }\href {\doibase 10.1103/PhysRevD.59.107301} {\bibfield
  {journal} {\bibinfo  {journal} {Phys. Rev.}\ }\textbf {\bibinfo {volume}
  {D59}},\ \bibinfo {pages} {107301} (\bibinfo {year} {1999})},\ \Eprint
  {http://arxiv.org/abs/astro-ph/9812301} {arXiv:astro-ph/9812301 [astro-ph]}
  \BibitemShut {NoStop}%
\bibitem [{\citenamefont {Green}(1999)}]{Green:1999yh}%
  \BibitemOpen
  \bibfield  {author} {\bibinfo {author} {\bibfnamefont {A.~M.}\ \bibnamefont
  {Green}},\ }\href {\doibase 10.1103/PhysRevD.60.063516} {\bibfield  {journal}
  {\bibinfo  {journal} {Phys. Rev. D}\ }\textbf {\bibinfo {volume} {60}},\
  \bibinfo {pages} {063516} (\bibinfo {year} {1999})},\ \Eprint
  {http://arxiv.org/abs/astro-ph/9903484} {arXiv:astro-ph/9903484} \BibitemShut
  {NoStop}%
\bibitem [{\citenamefont {Arbey}\ \emph {et~al.}(2021)\citenamefont {Arbey},
  \citenamefont {Auffinger}, \citenamefont {Sandick}, \citenamefont {Shams
  Es~Haghi},\ and\ \citenamefont {Sinha}}]{Arbey:2021ysg}%
  \BibitemOpen
  \bibfield  {author} {\bibinfo {author} {\bibfnamefont {A.}~\bibnamefont
  {Arbey}}, \bibinfo {author} {\bibfnamefont {J.}~\bibnamefont {Auffinger}},
  \bibinfo {author} {\bibfnamefont {P.}~\bibnamefont {Sandick}}, \bibinfo
  {author} {\bibfnamefont {B.}~\bibnamefont {Shams Es~Haghi}}, \ and\ \bibinfo
  {author} {\bibfnamefont {K.}~\bibnamefont {Sinha}},\ }\href@noop {} {\
  (\bibinfo {year} {2021})},\ \Eprint {http://arxiv.org/abs/2104.04051}
  {arXiv:2104.04051 [astro-ph.CO]} \BibitemShut {NoStop}%
\bibitem [{\citenamefont {Khlopov}\ \emph {et~al.}(2006)\citenamefont
  {Khlopov}, \citenamefont {Barrau},\ and\ \citenamefont
  {Grain}}]{Khlopov:2004tn}%
  \BibitemOpen
  \bibfield  {author} {\bibinfo {author} {\bibfnamefont {M.~{\relax Yu}.}\
  \bibnamefont {Khlopov}}, \bibinfo {author} {\bibfnamefont {A.}~\bibnamefont
  {Barrau}}, \ and\ \bibinfo {author} {\bibfnamefont {J.}~\bibnamefont
  {Grain}},\ }\href {\doibase 10.1088/0264-9381/23/6/004} {\bibfield  {journal}
  {\bibinfo  {journal} {Class. Quant. Grav.}\ }\textbf {\bibinfo {volume}
  {23}},\ \bibinfo {pages} {1875} (\bibinfo {year} {2006})},\ \Eprint
  {http://arxiv.org/abs/astro-ph/0406621} {arXiv:astro-ph/0406621 [astro-ph]}
  \BibitemShut {NoStop}%
\bibitem [{\citenamefont {Allahverdi}\ \emph {et~al.}(2018)\citenamefont
  {Allahverdi}, \citenamefont {Dent},\ and\ \citenamefont
  {Osinski}}]{Allahverdi:2017sks}%
  \BibitemOpen
  \bibfield  {author} {\bibinfo {author} {\bibfnamefont {R.}~\bibnamefont
  {Allahverdi}}, \bibinfo {author} {\bibfnamefont {J.}~\bibnamefont {Dent}}, \
  and\ \bibinfo {author} {\bibfnamefont {J.}~\bibnamefont {Osinski}},\ }\href
  {\doibase 10.1103/PhysRevD.97.055013} {\bibfield  {journal} {\bibinfo
  {journal} {Phys. Rev.}\ }\textbf {\bibinfo {volume} {D97}},\ \bibinfo {pages}
  {055013} (\bibinfo {year} {2018})},\ \Eprint
  {http://arxiv.org/abs/1711.10511} {arXiv:1711.10511 [astro-ph.CO]}
  \BibitemShut {NoStop}%
\bibitem [{\citenamefont {Lennon}\ \emph {et~al.}(2018)\citenamefont {Lennon},
  \citenamefont {March-Russell}, \citenamefont {Petrossian-Byrne},\ and\
  \citenamefont {Tillim}}]{Lennon:2017tqq}%
  \BibitemOpen
  \bibfield  {author} {\bibinfo {author} {\bibfnamefont {O.}~\bibnamefont
  {Lennon}}, \bibinfo {author} {\bibfnamefont {J.}~\bibnamefont
  {March-Russell}}, \bibinfo {author} {\bibfnamefont {R.}~\bibnamefont
  {Petrossian-Byrne}}, \ and\ \bibinfo {author} {\bibfnamefont
  {H.}~\bibnamefont {Tillim}},\ }\href {\doibase 10.1088/1475-7516/2018/04/009}
  {\bibfield  {journal} {\bibinfo  {journal} {JCAP}\ }\textbf {\bibinfo
  {volume} {1804}},\ \bibinfo {pages} {009} (\bibinfo {year} {2018})},\ \Eprint
  {http://arxiv.org/abs/1712.07664} {arXiv:1712.07664 [hep-ph]} \BibitemShut
  {NoStop}%
\bibitem [{\citenamefont {Morrison}\ \emph {et~al.}(2019)\citenamefont
  {Morrison}, \citenamefont {Profumo},\ and\ \citenamefont
  {Yu}}]{Morrison:2018xla}%
  \BibitemOpen
  \bibfield  {author} {\bibinfo {author} {\bibfnamefont {L.}~\bibnamefont
  {Morrison}}, \bibinfo {author} {\bibfnamefont {S.}~\bibnamefont {Profumo}}, \
  and\ \bibinfo {author} {\bibfnamefont {Y.}~\bibnamefont {Yu}},\ }\href
  {\doibase 10.1088/1475-7516/2019/05/005} {\bibfield  {journal} {\bibinfo
  {journal} {JCAP}\ }\textbf {\bibinfo {volume} {1905}},\ \bibinfo {pages}
  {005} (\bibinfo {year} {2019})},\ \Eprint {http://arxiv.org/abs/1812.10606}
  {arXiv:1812.10606 [astro-ph.CO]} \BibitemShut {NoStop}%
\bibitem [{\citenamefont {Gondolo}\ \emph {et~al.}(2020)\citenamefont
  {Gondolo}, \citenamefont {Sandick},\ and\ \citenamefont {Shams
  Es~Haghi}}]{Gondolo:2020uqv}%
  \BibitemOpen
  \bibfield  {author} {\bibinfo {author} {\bibfnamefont {P.}~\bibnamefont
  {Gondolo}}, \bibinfo {author} {\bibfnamefont {P.}~\bibnamefont {Sandick}}, \
  and\ \bibinfo {author} {\bibfnamefont {B.}~\bibnamefont {Shams Es~Haghi}},\
  }\href {\doibase 10.1103/PhysRevD.102.095018} {\bibfield  {journal} {\bibinfo
   {journal} {Phys. Rev. D}\ }\textbf {\bibinfo {volume} {102}},\ \bibinfo
  {pages} {095018} (\bibinfo {year} {2020})},\ \Eprint
  {http://arxiv.org/abs/2009.02424} {arXiv:2009.02424 [hep-ph]} \BibitemShut
  {NoStop}%
\bibitem [{\citenamefont {Bernal}\ and\ \citenamefont
  {Zapata}(2021{\natexlab{a}})}]{Bernal:2020kse}%
  \BibitemOpen
  \bibfield  {author} {\bibinfo {author} {\bibfnamefont {N.}~\bibnamefont
  {Bernal}}\ and\ \bibinfo {author} {\bibfnamefont {O.}~\bibnamefont
  {Zapata}},\ }\href {\doibase 10.1088/1475-7516/2021/03/007} {\bibfield
  {journal} {\bibinfo  {journal} {JCAP}\ }\textbf {\bibinfo {volume} {03}},\
  \bibinfo {pages} {007} (\bibinfo {year} {2021}{\natexlab{a}})},\ \Eprint
  {http://arxiv.org/abs/2010.09725} {arXiv:2010.09725 [hep-ph]} \BibitemShut
  {NoStop}%
\bibitem [{\citenamefont {Bernal}\ and\ \citenamefont
  {Zapata}(2021{\natexlab{b}})}]{Bernal:2020bjf}%
  \BibitemOpen
  \bibfield  {author} {\bibinfo {author} {\bibfnamefont {N.}~\bibnamefont
  {Bernal}}\ and\ \bibinfo {author} {\bibfnamefont {O.}~\bibnamefont
  {Zapata}},\ }\href {\doibase 10.1088/1475-7516/2021/03/015} {\bibfield
  {journal} {\bibinfo  {journal} {JCAP}\ }\textbf {\bibinfo {volume} {03}},\
  \bibinfo {pages} {015} (\bibinfo {year} {2021}{\natexlab{b}})},\ \Eprint
  {http://arxiv.org/abs/2011.12306} {arXiv:2011.12306 [astro-ph.CO]}
  \BibitemShut {NoStop}%
\bibitem [{\citenamefont {Bernal}\ and\ \citenamefont
  {Zapata}(2021{\natexlab{c}})}]{Bernal:2020ili}%
  \BibitemOpen
  \bibfield  {author} {\bibinfo {author} {\bibfnamefont {N.}~\bibnamefont
  {Bernal}}\ and\ \bibinfo {author} {\bibfnamefont {O.}~\bibnamefont
  {Zapata}},\ }\href {\doibase 10.1016/j.physletb.2021.136129} {\bibfield
  {journal} {\bibinfo  {journal} {Phys. Lett. B}\ }\textbf {\bibinfo {volume}
  {815}},\ \bibinfo {pages} {136129} (\bibinfo {year} {2021}{\natexlab{c}})},\
  \Eprint {http://arxiv.org/abs/2011.02510} {arXiv:2011.02510 [hep-ph]}
  \BibitemShut {NoStop}%
\bibitem [{\citenamefont {Kitabayashi}(2021)}]{Kitabayashi:2021hox}%
  \BibitemOpen
  \bibfield  {author} {\bibinfo {author} {\bibfnamefont {T.}~\bibnamefont
  {Kitabayashi}},\ }\href@noop {} {\  (\bibinfo {year} {2021})},\ \Eprint
  {http://arxiv.org/abs/2101.01921} {arXiv:2101.01921 [hep-ph]} \BibitemShut
  {NoStop}%
\bibitem [{\citenamefont {Baldes}\ \emph {et~al.}(2020)\citenamefont {Baldes},
  \citenamefont {Decant}, \citenamefont {Hooper},\ and\ \citenamefont
  {Lopez-Honorez}}]{Baldes:2020nuv}%
  \BibitemOpen
  \bibfield  {author} {\bibinfo {author} {\bibfnamefont {I.}~\bibnamefont
  {Baldes}}, \bibinfo {author} {\bibfnamefont {Q.}~\bibnamefont {Decant}},
  \bibinfo {author} {\bibfnamefont {D.~C.}\ \bibnamefont {Hooper}}, \ and\
  \bibinfo {author} {\bibfnamefont {L.}~\bibnamefont {Lopez-Honorez}},\ }\href
  {\doibase 10.1088/1475-7516/2020/08/045} {\bibfield  {journal} {\bibinfo
  {journal} {JCAP}\ }\textbf {\bibinfo {volume} {08}},\ \bibinfo {pages} {045}
  (\bibinfo {year} {2020})},\ \Eprint {http://arxiv.org/abs/2004.14773}
  {arXiv:2004.14773 [astro-ph.CO]} \BibitemShut {NoStop}%
\bibitem [{\citenamefont {Auffinger}\ \emph {et~al.}(2021)\citenamefont
  {Auffinger}, \citenamefont {Masina},\ and\ \citenamefont
  {Orlando}}]{Auffinger:2020afu}%
  \BibitemOpen
  \bibfield  {author} {\bibinfo {author} {\bibfnamefont {J.}~\bibnamefont
  {Auffinger}}, \bibinfo {author} {\bibfnamefont {I.}~\bibnamefont {Masina}}, \
  and\ \bibinfo {author} {\bibfnamefont {G.}~\bibnamefont {Orlando}},\ }\href
  {\doibase 10.1140/epjp/s13360-021-01247-9} {\bibfield  {journal} {\bibinfo
  {journal} {Eur. Phys. J. Plus}\ }\textbf {\bibinfo {volume} {136}},\ \bibinfo
  {pages} {261} (\bibinfo {year} {2021})},\ \Eprint
  {http://arxiv.org/abs/2012.09867} {arXiv:2012.09867 [hep-ph]} \BibitemShut
  {NoStop}%
\bibitem [{\citenamefont {Cheek}\ \emph {et~al.}(2022)\citenamefont {Cheek},
  \citenamefont {Heurtier}, \citenamefont {Perez-Gonzalez},\ and\ \citenamefont
  {Turner}}]{paperB}%
  \BibitemOpen
  \bibfield  {author} {\bibinfo {author} {\bibfnamefont {A.}~\bibnamefont
  {Cheek}}, \bibinfo {author} {\bibfnamefont {L.}~\bibnamefont {Heurtier}},
  \bibinfo {author} {\bibfnamefont {Y.~F.}\ \bibnamefont {Perez-Gonzalez}}, \
  and\ \bibinfo {author} {\bibfnamefont {J.}~\bibnamefont {Turner}},\ }\href
  {\doibase 10.1103/PhysRevD.105.015023} {\bibfield  {journal} {\bibinfo
  {journal} {Phys. Rev. D}\ }\textbf {\bibinfo {volume} {105}},\ \bibinfo
  {pages} {015023} (\bibinfo {year} {2022})},\ \Eprint
  {http://arxiv.org/abs/2107.00016} {arXiv:2107.00016 [hep-ph]} \BibitemShut
  {NoStop}%
\bibitem [{\citenamefont {Granelli}\ \emph {et~al.}(2021)\citenamefont
  {Granelli}, \citenamefont {Moffat}, \citenamefont {Perez-Gonzalez},
  \citenamefont {Schulz},\ and\ \citenamefont {Turner}}]{Granelli:2020pim}%
  \BibitemOpen
  \bibfield  {author} {\bibinfo {author} {\bibfnamefont {A.}~\bibnamefont
  {Granelli}}, \bibinfo {author} {\bibfnamefont {K.}~\bibnamefont {Moffat}},
  \bibinfo {author} {\bibfnamefont {Y.~F.}\ \bibnamefont {Perez-Gonzalez}},
  \bibinfo {author} {\bibfnamefont {H.}~\bibnamefont {Schulz}}, \ and\ \bibinfo
  {author} {\bibfnamefont {J.}~\bibnamefont {Turner}},\ }\href {\doibase
  10.1016/j.cpc.2020.107813} {\bibfield  {journal} {\bibinfo  {journal}
  {Comput. Phys. Commun.}\ }\textbf {\bibinfo {volume} {262}},\ \bibinfo
  {pages} {107813} (\bibinfo {year} {2021})},\ \Eprint
  {http://arxiv.org/abs/2007.09150} {arXiv:2007.09150 [hep-ph]} \BibitemShut
  {NoStop}%
\bibitem [{\citenamefont {Page}(1976)}]{Page:1976df}%
  \BibitemOpen
  \bibfield  {author} {\bibinfo {author} {\bibfnamefont {D.~N.}\ \bibnamefont
  {Page}},\ }\href {\doibase 10.1103/PhysRevD.13.198} {\bibfield  {journal}
  {\bibinfo  {journal} {Phys. Rev. D}\ }\textbf {\bibinfo {volume} {13}},\
  \bibinfo {pages} {198} (\bibinfo {year} {1976})}\BibitemShut {NoStop}%
\bibitem [{\citenamefont {Page}(1977)}]{Page:1977um}%
  \BibitemOpen
  \bibfield  {author} {\bibinfo {author} {\bibfnamefont {D.~N.}\ \bibnamefont
  {Page}},\ }\href {\doibase 10.1103/PhysRevD.16.2402} {\bibfield  {journal}
  {\bibinfo  {journal} {Phys. Rev. D}\ }\textbf {\bibinfo {volume} {16}},\
  \bibinfo {pages} {2402} (\bibinfo {year} {1977})}\BibitemShut {NoStop}%
\bibitem [{\citenamefont {Unruh}(1976)}]{Unruh:1976fm}%
  \BibitemOpen
  \bibfield  {author} {\bibinfo {author} {\bibfnamefont {W.~G.}\ \bibnamefont
  {Unruh}},\ }\href {\doibase 10.1103/PhysRevD.14.3251} {\bibfield  {journal}
  {\bibinfo  {journal} {Phys. Rev. D}\ }\textbf {\bibinfo {volume} {14}},\
  \bibinfo {pages} {3251} (\bibinfo {year} {1976})}\BibitemShut {NoStop}%
\bibitem [{\citenamefont {Doran}\ \emph {et~al.}(2005)\citenamefont {Doran},
  \citenamefont {Lasenby}, \citenamefont {Dolan},\ and\ \citenamefont
  {Hinder}}]{Doran:2005vm}%
  \BibitemOpen
  \bibfield  {author} {\bibinfo {author} {\bibfnamefont {C.}~\bibnamefont
  {Doran}}, \bibinfo {author} {\bibfnamefont {A.}~\bibnamefont {Lasenby}},
  \bibinfo {author} {\bibfnamefont {S.}~\bibnamefont {Dolan}}, \ and\ \bibinfo
  {author} {\bibfnamefont {I.}~\bibnamefont {Hinder}},\ }\href {\doibase
  10.1103/PhysRevD.71.124020} {\bibfield  {journal} {\bibinfo  {journal} {Phys.
  Rev. D}\ }\textbf {\bibinfo {volume} {71}},\ \bibinfo {pages} {124020}
  (\bibinfo {year} {2005})},\ \Eprint {http://arxiv.org/abs/gr-qc/0503019}
  {arXiv:gr-qc/0503019} \BibitemShut {NoStop}%
\bibitem [{\citenamefont {MacGibbon}\ and\ \citenamefont
  {Webber}(1990{\natexlab{a}})}]{MacGibbon:1990zk}%
  \BibitemOpen
  \bibfield  {author} {\bibinfo {author} {\bibfnamefont {J.~H.}\ \bibnamefont
  {MacGibbon}}\ and\ \bibinfo {author} {\bibfnamefont {B.~R.}\ \bibnamefont
  {Webber}},\ }\href {\doibase 10.1103/PhysRevD.41.3052} {\bibfield  {journal}
  {\bibinfo  {journal} {Phys. Rev.}\ }\textbf {\bibinfo {volume} {D41}},\
  \bibinfo {pages} {3052} (\bibinfo {year} {1990}{\natexlab{a}})}\BibitemShut
  {NoStop}%
\bibitem [{\citenamefont {MacGibbon}(1991{\natexlab{a}})}]{MacGibbon:1991tj}%
  \BibitemOpen
  \bibfield  {author} {\bibinfo {author} {\bibfnamefont {J.~H.}\ \bibnamefont
  {MacGibbon}},\ }\href {\doibase 10.1103/PhysRevD.44.376} {\bibfield
  {journal} {\bibinfo  {journal} {Phys. Rev.}\ }\textbf {\bibinfo {volume}
  {D44}},\ \bibinfo {pages} {376} (\bibinfo {year}
  {1991}{\natexlab{a}})}\BibitemShut {NoStop}%
\bibitem [{\citenamefont {Ukwatta}\ \emph {et~al.}(2016)\citenamefont
  {Ukwatta}, \citenamefont {Stump}, \citenamefont {Linnemann}, \citenamefont
  {MacGibbon}, \citenamefont {Marinelli}, \citenamefont {Yapici},\ and\
  \citenamefont {Tollefson}}]{Ukwatta:2015iba}%
  \BibitemOpen
  \bibfield  {author} {\bibinfo {author} {\bibfnamefont {T.~N.}\ \bibnamefont
  {Ukwatta}}, \bibinfo {author} {\bibfnamefont {D.~R.}\ \bibnamefont {Stump}},
  \bibinfo {author} {\bibfnamefont {J.~T.}\ \bibnamefont {Linnemann}}, \bibinfo
  {author} {\bibfnamefont {J.~H.}\ \bibnamefont {MacGibbon}}, \bibinfo {author}
  {\bibfnamefont {S.~S.}\ \bibnamefont {Marinelli}}, \bibinfo {author}
  {\bibfnamefont {T.}~\bibnamefont {Yapici}}, \ and\ \bibinfo {author}
  {\bibfnamefont {K.}~\bibnamefont {Tollefson}},\ }\href {\doibase
  10.1016/j.astropartphys.2016.03.007} {\bibfield  {journal} {\bibinfo
  {journal} {Astropart. Phys.}\ }\textbf {\bibinfo {volume} {80}},\ \bibinfo
  {pages} {90} (\bibinfo {year} {2016})},\ \Eprint
  {http://arxiv.org/abs/1510.04372} {arXiv:1510.04372 [astro-ph.HE]}
  \BibitemShut {NoStop}%
\bibitem [{\citenamefont {MacGibbon}\ and\ \citenamefont
  {Webber}(1990{\natexlab{b}})}]{PhysRevD.41.3052}%
  \BibitemOpen
  \bibfield  {author} {\bibinfo {author} {\bibfnamefont {J.~H.}\ \bibnamefont
  {MacGibbon}}\ and\ \bibinfo {author} {\bibfnamefont {B.~R.}\ \bibnamefont
  {Webber}},\ }\href {\doibase 10.1103/PhysRevD.41.3052} {\bibfield  {journal}
  {\bibinfo  {journal} {Phys. Rev. D}\ }\textbf {\bibinfo {volume} {41}},\
  \bibinfo {pages} {3052} (\bibinfo {year} {1990}{\natexlab{b}})}\BibitemShut
  {NoStop}%
\bibitem [{\citenamefont {MacGibbon}(1991{\natexlab{b}})}]{PhysRevD.44.376}%
  \BibitemOpen
  \bibfield  {author} {\bibinfo {author} {\bibfnamefont {J.~H.}\ \bibnamefont
  {MacGibbon}},\ }\href {\doibase 10.1103/PhysRevD.44.376} {\bibfield
  {journal} {\bibinfo  {journal} {Phys. Rev. D}\ }\textbf {\bibinfo {volume}
  {44}},\ \bibinfo {pages} {376} (\bibinfo {year}
  {1991}{\natexlab{b}})}\BibitemShut {NoStop}%
\bibitem [{\citenamefont {Buonanno}\ \emph {et~al.}(2008)\citenamefont
  {Buonanno}, \citenamefont {Kidder},\ and\ \citenamefont
  {Lehner}}]{Buonanno:2007sv}%
  \BibitemOpen
  \bibfield  {author} {\bibinfo {author} {\bibfnamefont {A.}~\bibnamefont
  {Buonanno}}, \bibinfo {author} {\bibfnamefont {L.~E.}\ \bibnamefont
  {Kidder}}, \ and\ \bibinfo {author} {\bibfnamefont {L.}~\bibnamefont
  {Lehner}},\ }\href {\doibase 10.1103/PhysRevD.77.026004} {\bibfield
  {journal} {\bibinfo  {journal} {Phys. Rev.}\ }\textbf {\bibinfo {volume}
  {D77}},\ \bibinfo {pages} {026004} (\bibinfo {year} {2008})},\ \Eprint
  {http://arxiv.org/abs/0709.3839} {arXiv:0709.3839 [astro-ph]} \BibitemShut
  {NoStop}%
\bibitem [{\citenamefont {Kesden}(2008)}]{Kesden:2008ga}%
  \BibitemOpen
  \bibfield  {author} {\bibinfo {author} {\bibfnamefont {M.}~\bibnamefont
  {Kesden}},\ }\href {\doibase 10.1103/PhysRevD.78.084030} {\bibfield
  {journal} {\bibinfo  {journal} {Phys. Rev.}\ }\textbf {\bibinfo {volume}
  {D78}},\ \bibinfo {pages} {084030} (\bibinfo {year} {2008})},\ \Eprint
  {http://arxiv.org/abs/0807.3043} {arXiv:0807.3043 [astro-ph]} \BibitemShut
  {NoStop}%
\bibitem [{\citenamefont {Tichy}\ and\ \citenamefont
  {Marronetti}(2008)}]{Tichy:2008du}%
  \BibitemOpen
  \bibfield  {author} {\bibinfo {author} {\bibfnamefont {W.}~\bibnamefont
  {Tichy}}\ and\ \bibinfo {author} {\bibfnamefont {P.}~\bibnamefont
  {Marronetti}},\ }\href {\doibase 10.1103/PhysRevD.78.081501} {\bibfield
  {journal} {\bibinfo  {journal} {Phys. Rev.}\ }\textbf {\bibinfo {volume}
  {D78}},\ \bibinfo {pages} {081501} (\bibinfo {year} {2008})},\ \Eprint
  {http://arxiv.org/abs/0807.2985} {arXiv:0807.2985 [gr-qc]} \BibitemShut
  {NoStop}%
\bibitem [{\citenamefont {Chandrasekhar}\ and\ \citenamefont
  {Detweiler}(1975)}]{Chandrasekhar:1975zz}%
  \BibitemOpen
  \bibfield  {author} {\bibinfo {author} {\bibfnamefont {S.}~\bibnamefont
  {Chandrasekhar}}\ and\ \bibinfo {author} {\bibfnamefont {S.~L.}\ \bibnamefont
  {Detweiler}},\ }\href {\doibase 10.1098/rspa.1975.0130} {\bibfield  {journal}
  {\bibinfo  {journal} {Proc. Roy. Soc. Lond. A}\ }\textbf {\bibinfo {volume}
  {345}},\ \bibinfo {pages} {145} (\bibinfo {year} {1975})}\BibitemShut
  {NoStop}%
\bibitem [{\citenamefont {Chandrasekhar}(1976)}]{10.2307/79115}%
  \BibitemOpen
  \bibfield  {author} {\bibinfo {author} {\bibfnamefont {S.}~\bibnamefont
  {Chandrasekhar}},\ }\href {http://www.jstor.org/stable/79115} {\bibfield
  {journal} {\bibinfo  {journal} {Proc. Roy. Soc. Lond. A}\ }\textbf {\bibinfo
  {volume} {348}},\ \bibinfo {pages} {39} (\bibinfo {year} {1976})}\BibitemShut
  {NoStop}%
\bibitem [{\citenamefont {Chandrasekhar}\ and\ \citenamefont
  {Detweiler}(1977)}]{Chandrasekhar:1977kf}%
  \BibitemOpen
  \bibfield  {author} {\bibinfo {author} {\bibfnamefont {S.}~\bibnamefont
  {Chandrasekhar}}\ and\ \bibinfo {author} {\bibfnamefont {S.~L.}\ \bibnamefont
  {Detweiler}},\ }\href {\doibase 10.1098/rspa.1977.0002} {\bibfield  {journal}
  {\bibinfo  {journal} {Proc. Roy. Soc. Lond. A}\ }\textbf {\bibinfo {volume}
  {352}},\ \bibinfo {pages} {325} (\bibinfo {year} {1977})}\BibitemShut
  {NoStop}%
\bibitem [{\citenamefont {Arbey}\ and\ \citenamefont
  {Auffinger}(2019)}]{Arbey:2019mbc}%
  \BibitemOpen
  \bibfield  {author} {\bibinfo {author} {\bibfnamefont {A.}~\bibnamefont
  {Arbey}}\ and\ \bibinfo {author} {\bibfnamefont {J.}~\bibnamefont
  {Auffinger}},\ }\href {\doibase 10.1140/epjc/s10052-019-7161-1} {\bibfield
  {journal} {\bibinfo  {journal} {Eur. Phys. J. C}\ }\textbf {\bibinfo {volume}
  {79}},\ \bibinfo {pages} {693} (\bibinfo {year} {2019})},\ \Eprint
  {http://arxiv.org/abs/1905.04268} {arXiv:1905.04268 [gr-qc]} \BibitemShut
  {NoStop}%
\bibitem [{\citenamefont {Chandrasekhar}\ and\ \citenamefont
  {Detweiler}(1976)}]{Chandrasekhar:1976zz}%
  \BibitemOpen
  \bibfield  {author} {\bibinfo {author} {\bibfnamefont {S.}~\bibnamefont
  {Chandrasekhar}}\ and\ \bibinfo {author} {\bibfnamefont {S.~L.}\ \bibnamefont
  {Detweiler}},\ }\href {\doibase 10.1098/rspa.1976.0101} {\bibfield  {journal}
  {\bibinfo  {journal} {Proc. Roy. Soc. Lond. A}\ }\textbf {\bibinfo {volume}
  {350}},\ \bibinfo {pages} {165} (\bibinfo {year} {1976})}\BibitemShut
  {NoStop}%
\bibitem [{\citenamefont {Bode}\ \emph {et~al.}(2001)\citenamefont {Bode},
  \citenamefont {Ostriker},\ and\ \citenamefont {Turok}}]{Bode:2000gq}%
  \BibitemOpen
  \bibfield  {author} {\bibinfo {author} {\bibfnamefont {P.}~\bibnamefont
  {Bode}}, \bibinfo {author} {\bibfnamefont {J.~P.}\ \bibnamefont {Ostriker}},
  \ and\ \bibinfo {author} {\bibfnamefont {N.}~\bibnamefont {Turok}},\ }\href
  {\doibase 10.1086/321541} {\bibfield  {journal} {\bibinfo  {journal}
  {Astrophys. J.}\ }\textbf {\bibinfo {volume} {556}},\ \bibinfo {pages} {93}
  (\bibinfo {year} {2001})},\ \Eprint {http://arxiv.org/abs/astro-ph/0010389}
  {arXiv:astro-ph/0010389} \BibitemShut {NoStop}%
\bibitem [{\citenamefont {Boyarsky}\ \emph {et~al.}(2009)\citenamefont
  {Boyarsky}, \citenamefont {Lesgourgues}, \citenamefont {Ruchayskiy},\ and\
  \citenamefont {Viel}}]{Boyarsky:2008xj}%
  \BibitemOpen
  \bibfield  {author} {\bibinfo {author} {\bibfnamefont {A.}~\bibnamefont
  {Boyarsky}}, \bibinfo {author} {\bibfnamefont {J.}~\bibnamefont
  {Lesgourgues}}, \bibinfo {author} {\bibfnamefont {O.}~\bibnamefont
  {Ruchayskiy}}, \ and\ \bibinfo {author} {\bibfnamefont {M.}~\bibnamefont
  {Viel}},\ }\href {\doibase 10.1088/1475-7516/2009/05/012} {\bibfield
  {journal} {\bibinfo  {journal} {JCAP}\ }\textbf {\bibinfo {volume} {05}},\
  \bibinfo {pages} {012} (\bibinfo {year} {2009})},\ \Eprint
  {http://arxiv.org/abs/0812.0010} {arXiv:0812.0010 [astro-ph]} \BibitemShut
  {NoStop}%
\bibitem [{\citenamefont {Baur}\ \emph {et~al.}(2017)\citenamefont {Baur},
  \citenamefont {Palanque-Delabrouille}, \citenamefont {Yeche}, \citenamefont
  {Boyarsky}, \citenamefont {Ruchayskiy}, \citenamefont {Armengaud},\ and\
  \citenamefont {Lesgourgues}}]{Baur:2017stq}%
  \BibitemOpen
  \bibfield  {author} {\bibinfo {author} {\bibfnamefont {J.}~\bibnamefont
  {Baur}}, \bibinfo {author} {\bibfnamefont {N.}~\bibnamefont
  {Palanque-Delabrouille}}, \bibinfo {author} {\bibfnamefont {C.}~\bibnamefont
  {Yeche}}, \bibinfo {author} {\bibfnamefont {A.}~\bibnamefont {Boyarsky}},
  \bibinfo {author} {\bibfnamefont {O.}~\bibnamefont {Ruchayskiy}}, \bibinfo
  {author} {\bibfnamefont {E.}~\bibnamefont {Armengaud}}, \ and\ \bibinfo
  {author} {\bibfnamefont {J.}~\bibnamefont {Lesgourgues}},\ }\href {\doibase
  10.1088/1475-7516/2017/12/013} {\bibfield  {journal} {\bibinfo  {journal}
  {JCAP}\ }\textbf {\bibinfo {volume} {12}},\ \bibinfo {pages} {013} (\bibinfo
  {year} {2017})},\ \Eprint {http://arxiv.org/abs/1706.03118} {arXiv:1706.03118
  [astro-ph.CO]} \BibitemShut {NoStop}%
\bibitem [{\citenamefont {{Lesgourgues}}(2011)}]{CLASSI}%
  \BibitemOpen
  \bibfield  {author} {\bibinfo {author} {\bibfnamefont {J.}~\bibnamefont
  {{Lesgourgues}}},\ }\href@noop {} {\bibfield  {journal} {\bibinfo  {journal}
  {arXiv e-prints}\ ,\ \bibinfo {eid} {arXiv:1104.2932}} (\bibinfo {year}
  {2011})},\ \Eprint {http://arxiv.org/abs/1104.2932} {arXiv:1104.2932
  [astro-ph.IM]} \BibitemShut {NoStop}%
\bibitem [{\citenamefont {{Blas}}\ \emph {et~al.}(2011)\citenamefont {{Blas}},
  \citenamefont {{Lesgourgues}},\ and\ \citenamefont {{Tram}}}]{CLASSII}%
  \BibitemOpen
  \bibfield  {author} {\bibinfo {author} {\bibfnamefont {D.}~\bibnamefont
  {{Blas}}}, \bibinfo {author} {\bibfnamefont {J.}~\bibnamefont
  {{Lesgourgues}}}, \ and\ \bibinfo {author} {\bibfnamefont {T.}~\bibnamefont
  {{Tram}}},\ }\href {\doibase 10.1088/1475-7516/2011/07/034} {\bibfield
  {journal} {\bibinfo  {journal} {\it{JCAP}}\ }\textbf {\bibinfo {volume}
  {2011}},\ \bibinfo {eid} {034} (\bibinfo {year} {2011})},\ \Eprint
  {http://arxiv.org/abs/1104.2933} {arXiv:1104.2933 [astro-ph.CO]} \BibitemShut
  {NoStop}%
\bibitem [{\citenamefont {{Lesgourgues}}\ and\ \citenamefont
  {{Tram}}(2011)}]{CLASSIV}%
  \BibitemOpen
  \bibfield  {author} {\bibinfo {author} {\bibfnamefont {J.}~\bibnamefont
  {{Lesgourgues}}}\ and\ \bibinfo {author} {\bibfnamefont {T.}~\bibnamefont
  {{Tram}}},\ }\href {\doibase 10.1088/1475-7516/2011/09/032} {\bibfield
  {journal} {\bibinfo  {journal} {\it{JCAP}}\ }\textbf {\bibinfo {volume}
  {2011}},\ \bibinfo {eid} {032} (\bibinfo {year} {2011})},\ \Eprint
  {http://arxiv.org/abs/1104.2935} {arXiv:1104.2935 [astro-ph.CO]} \BibitemShut
  {NoStop}%
\bibitem [{\citenamefont {Khlopov}(2010)}]{Khlopov:2008qy}%
  \BibitemOpen
  \bibfield  {author} {\bibinfo {author} {\bibfnamefont {M.~Y.}\ \bibnamefont
  {Khlopov}},\ }\href {\doibase 10.1088/1674-4527/10/6/001} {\bibfield
  {journal} {\bibinfo  {journal} {Res. Astron. Astrophys.}\ }\textbf {\bibinfo
  {volume} {10}},\ \bibinfo {pages} {495} (\bibinfo {year} {2010})},\ \Eprint
  {http://arxiv.org/abs/0801.0116} {arXiv:0801.0116 [astro-ph]} \BibitemShut
  {NoStop}%
\bibitem [{\citenamefont {Hooper}\ \emph {et~al.}(2020)\citenamefont {Hooper},
  \citenamefont {Krnjaic}, \citenamefont {March-Russell}, \citenamefont
  {McDermott},\ and\ \citenamefont {Petrossian-Byrne}}]{Hooper:2020evu}%
  \BibitemOpen
  \bibfield  {author} {\bibinfo {author} {\bibfnamefont {D.}~\bibnamefont
  {Hooper}}, \bibinfo {author} {\bibfnamefont {G.}~\bibnamefont {Krnjaic}},
  \bibinfo {author} {\bibfnamefont {J.}~\bibnamefont {March-Russell}}, \bibinfo
  {author} {\bibfnamefont {S.~D.}\ \bibnamefont {McDermott}}, \ and\ \bibinfo
  {author} {\bibfnamefont {R.}~\bibnamefont {Petrossian-Byrne}},\ }\href@noop
  {} {\  (\bibinfo {year} {2020})},\ \Eprint {http://arxiv.org/abs/2004.00618}
  {arXiv:2004.00618 [astro-ph.CO]} \BibitemShut {NoStop}%
\bibitem [{\citenamefont {Flores}\ and\ \citenamefont
  {Kusenko}(2021)}]{Flores:2021tmc}%
  \BibitemOpen
  \bibfield  {author} {\bibinfo {author} {\bibfnamefont {M.~M.}\ \bibnamefont
  {Flores}}\ and\ \bibinfo {author} {\bibfnamefont {A.}~\bibnamefont
  {Kusenko}},\ }\href@noop {} {\  (\bibinfo {year} {2021})},\ \Eprint
  {http://arxiv.org/abs/2106.03237} {arXiv:2106.03237 [astro-ph.CO]}
  \BibitemShut {NoStop}%
\bibitem [{\citenamefont {Dong}\ \emph {et~al.}(2016)\citenamefont {Dong},
  \citenamefont {Kinney},\ and\ \citenamefont {Stojkovic}}]{Dong:2015yjs}%
  \BibitemOpen
  \bibfield  {author} {\bibinfo {author} {\bibfnamefont {R.}~\bibnamefont
  {Dong}}, \bibinfo {author} {\bibfnamefont {W.~H.}\ \bibnamefont {Kinney}}, \
  and\ \bibinfo {author} {\bibfnamefont {D.}~\bibnamefont {Stojkovic}},\ }\href
  {\doibase 10.1088/1475-7516/2016/10/034} {\bibfield  {journal} {\bibinfo
  {journal} {JCAP}\ }\textbf {\bibinfo {volume} {10}},\ \bibinfo {pages} {034}
  (\bibinfo {year} {2016})},\ \Eprint {http://arxiv.org/abs/1511.05642}
  {arXiv:1511.05642 [astro-ph.CO]} \BibitemShut {NoStop}%
\bibitem [{\citenamefont {Han}\ \emph {et~al.}(1999)\citenamefont {Han},
  \citenamefont {Lykken},\ and\ \citenamefont {Zhang}}]{Han:1998sg}%
  \BibitemOpen
  \bibfield  {author} {\bibinfo {author} {\bibfnamefont {T.}~\bibnamefont
  {Han}}, \bibinfo {author} {\bibfnamefont {J.~D.}\ \bibnamefont {Lykken}}, \
  and\ \bibinfo {author} {\bibfnamefont {R.-J.}\ \bibnamefont {Zhang}},\ }\href
  {\doibase 10.1103/PhysRevD.59.105006} {\bibfield  {journal} {\bibinfo
  {journal} {Phys. Rev. D}\ }\textbf {\bibinfo {volume} {59}},\ \bibinfo
  {pages} {105006} (\bibinfo {year} {1999})},\ \Eprint
  {http://arxiv.org/abs/hep-ph/9811350} {arXiv:hep-ph/9811350} \BibitemShut
  {NoStop}%
\bibitem [{\citenamefont {Lee}\ \emph {et~al.}(2014)\citenamefont {Lee},
  \citenamefont {Park},\ and\ \citenamefont {Sanz}}]{Lee:2013bua}%
  \BibitemOpen
  \bibfield  {author} {\bibinfo {author} {\bibfnamefont {H.~M.}\ \bibnamefont
  {Lee}}, \bibinfo {author} {\bibfnamefont {M.}~\bibnamefont {Park}}, \ and\
  \bibinfo {author} {\bibfnamefont {V.}~\bibnamefont {Sanz}},\ }\href {\doibase
  10.1140/epjc/s10052-014-2715-8} {\bibfield  {journal} {\bibinfo  {journal}
  {Eur. Phys. J. C}\ }\textbf {\bibinfo {volume} {74}},\ \bibinfo {pages}
  {2715} (\bibinfo {year} {2014})},\ \Eprint {http://arxiv.org/abs/1306.4107}
  {arXiv:1306.4107 [hep-ph]} \BibitemShut {NoStop}%
\bibitem [{\citenamefont {Falkowski}\ and\ \citenamefont
  {Kamenik}(2016)}]{Falkowski:2016glr}%
  \BibitemOpen
  \bibfield  {author} {\bibinfo {author} {\bibfnamefont {A.}~\bibnamefont
  {Falkowski}}\ and\ \bibinfo {author} {\bibfnamefont {J.~F.}\ \bibnamefont
  {Kamenik}},\ }\href {\doibase 10.1103/PhysRevD.94.015008} {\bibfield
  {journal} {\bibinfo  {journal} {Phys. Rev. D}\ }\textbf {\bibinfo {volume}
  {94}},\ \bibinfo {pages} {015008} (\bibinfo {year} {2016})},\ \Eprint
  {http://arxiv.org/abs/1603.06980} {arXiv:1603.06980 [hep-ph]} \BibitemShut
  {NoStop}%
\end{thebibliography}%

\end{document}